\def \inparg{\leftskip = 40pt\rightskip = 40pt}
\def \outparg{\leftskip = 0 pt\rightskip = 0pt}

\def\npb{{Nucl.\ Phys.\ }{\bf B}}
\def\plb{{Phys.\ Lett.\ }{ \bf B}}

\def\prd{{Phys.\ Rev.\ }{\bf D}}

\def\prl{Phys.\ Rev.\ Lett.\ }

\def\tZ{\tilde Z}
\def\tz{\tilde z}
\def\ov #1{\overline{#1}}
\def\wt #1{\widetilde{#1}}
\def\Tr{\mathop{\rm Tr}}

\def\th{{\theta}}
\def\bth{{\overline{\theta}}}

\def\frak#1#2{{\textstyle{{#1}\over{#2}}}}
\def\nhalf{${\cal N} = \frak{1}{2}$}
\def\none{${\cal N} = 1$}
\def\ntwo{${\cal N} = 2$}

\def\ybar{\overline y}
\def\mbar{\overline m}

\def\Ahat{\hat A}
\def\Chat{\hat C}
\def\mbar{\overline{m}}
\def\mubar{\overline{\mu}}
\def\deltabar{\overline\delta}
\def\alphabar{\overline\alpha}

\def\frak#1#2{{\textstyle{{#1}\over{#2}}}}

\def\go{\rightarrow}
\def\lambdabar{\overline\lambda}
\def\lambdahatbar{\overline{\hat\lambda}}
\def\lambdahat{\hat\lambda}
\def\Dtil{\tilde D}
\def\Dhat{\hat D}
\def\Ftil{\tilde F}
\def\Fhat{\hat F}

\def\sigmabar{\overline\sigma}
\def\phibar{\overline\phi}
\def\phitilbar{\overline{\tilde{\phi}}}
\def\psitilbar{\overline{\tilde{\psi}}}

\def\psibar{\overline\psi}
\def\Fbar{\overline F}

\def\Dtil{\tilde D}

\def\phitil{\tilde\phi}
\def\psitil{\tilde\psi}
\def\Ncal{{\cal N}}
\def\Ftil{\tilde F}
\def\alphadot{\dot\alpha}
\def\betadot{\dot\beta}

\def\pa{\partial}

\input harvmac
\input epsf
%
\newbox\hdbox%
\newcount\hdrows%
\newcount\multispancount%
\newcount\ncase%
\newcount\ncols
\newcount\nrows%
\newcount\nspan%
\newcount\ntemp%
\newdimen\hdsize%
\newdimen\newhdsize%
\newdimen\parasize%
\newdimen\spreadwidth%
\newdimen\thicksize%
\newdimen\thinsize%
\newdimen\tablewidth%
\newif\ifcentertables%
\newif\ifendsize%
\newif\iffirstrow%
\newif\iftableinfo%
\newtoks\dbt%
\newtoks\hdtks%
\newtoks\savetks%
\newtoks\tableLETtokens%
\newtoks\tabletokens%
\newtoks\widthspec%
%
%
%
%
\tableinfotrue%
\catcode`\@=11
%
%
\def\tstrut{\vrule height3.1ex depth1.2ex width0pt}%
\def\and{\char`\&}
\def\tablerule{\noalign{\hrule height\thinsize depth0pt}}%
\thicksize=1.5pt
\thinsize=0.6pt
\def\thickrule{\noalign{\hrule height\thicksize depth0pt}}%
\def\ctr#1{\hfil\ #1\hfil}%
%
%
%
%
\tablewidth=-\maxdimen%
\spreadwidth=-\maxdimen%
\def\tabskipglue{0pt plus 1fil minus 1fil}%
%
%
\centertablestrue%
%
%
%
%
\parasize=4in%
\gdef\ARGS{########}
\gdef\headerARGS{####}
\def\@mpersand{&}
{\catcode`\|=13
\gdef\letbarzero{\let|0}
\gdef\letbartab{\def|{&&}}%
\gdef\letvbbar{\let\vb|}%
}
{\catcode`\&=4
\def\ampskip{&\omit\hfil&}
\catcode`\&=13
\let&0
\xdef\letampskip{\def&{\ampskip}}%
\gdef\letnovbamp{\let\novb&\let\tab&}
}
\def\begintable{
   \begingroup%
   \catcode`\|=13\letbartab\letvbbar%
   \catcode`\&=13\letampskip\letnovbamp%
   \def\multispan##1{
      \omit \mscount##1%
      \multiply\mscount\tw@\advance\mscount\m@ne%
      \loop\ifnum\mscount>\@ne \sp@n\repeat%
   }
   \def\|{%
      &\omit\widevline&%
   }%
   \ruledtable
}
\long\def\ruledtable#1\endtable{%
%
%
%
   \offinterlineskip
   \tabskip 0pt
   \def\widevline{\vrule width\thicksize}
   \def\endrow{\@mpersand\omit\hfil\crnorm\@mpersand}%
   \def\crthick{\@mpersand\crnorm\thickrule\@mpersand}%
   \def\crthickneg##1{\@mpersand\crnorm\thickrule
          \noalign{{\skip0=##1\vskip-\skip0}}\@mpersand}%
   \def\crnorule{\@mpersand\crnorm\@mpersand}%
   \def\crnoruleneg##1{\@mpersand\crnorm
          \noalign{{\skip0=##1\vskip-\skip0}}\@mpersand}%
   \let\nr=\crnorule
   \def\endtable{\@mpersand\crnorm\thickrule}%
   \let\crnorm=\cr
%
%
   \edef\cr{\@mpersand\crnorm\tablerule\@mpersand}%
   \def\crneg##1{\@mpersand\crnorm\tablerule
          \noalign{{\skip0=##1\vskip-\skip0}}\@mpersand}%
   \let\ctneg=\crthickneg
   \let\nrneg=\crnoruleneg
   \the\tableLETtokens
%
%
   \tabletokens={&#1}
%
%
   \countROWS\tabletokens\into\nrows%
   \countCOLS\tabletokens\into\ncols%
%
%
   \advance\ncols by -1%
   \divide\ncols by 2%
   \advance\nrows by 1%
%
%
   \iftableinfo %
      \immediate\write16{[Nrows=\the\nrows, Ncols=\the\ncols]}%
   \fi%
%
%
   \ifcentertables
      \ifhmode \par\fi
      \line{
      \hss
   \else %
      \hbox{%
   \fi
      \vbox{%
         \makePREAMBLE{\the\ncols}
         \edef\next{\preamble}
         \let\preamble=\next
         \makeTABLE{\preamble}{\tabletokens}
      }
      \ifcentertables \hss}\else }\fi
   \endgroup
   \tablewidth=-\maxdimen
   \spreadwidth=-\maxdimen
}
\def\makeTABLE#1#2{
   {
   \let\ifmath0
   \let\header0
   \let\multispan0
%
%
   \ncase=0%
   \ifdim\tablewidth>-\maxdimen \ncase=1\fi%
   \ifdim\spreadwidth>-\maxdimen \ncase=2\fi%
   \relax
%
   \ifcase\ncase %
      \widthspec={}%
   \or %
      \widthspec=\expandafter{\expandafter t\expandafter o%
                 \the\tablewidth}%
   \else %
      \widthspec=\expandafter{\expandafter s\expandafter p\expandafter r%
                 \expandafter e\expandafter a\expandafter d%
                 \the\spreadwidth}%
   \fi %
   \xdef\next{
      \halign\the\widthspec{%
      #1
      \noalign{\hrule height\thicksize depth0pt}
      \the#2\endtable
%
      }
   }
   }
   \next
}
\def\makePREAMBLE#1{
   \ncols=#1
   \begingroup
   \let\ARGS=0
   \edef\xtp{\widevline\ARGS\tabskip\tabskipglue%
   &\ctr{\ARGS}\tstrut}
   \advance\ncols by -1
   \loop
      \ifnum\ncols>0 %
      \advance\ncols by -1%
      \edef\xtp{\xtp&\vrule width\thinsize\ARGS&\ctr{\ARGS}}%
   \repeat
   \xdef\preamble{\xtp&\widevline\ARGS\tabskip0pt%
   \crnorm}
   \endgroup
}
\def\countROWS#1\into#2{
   \let\countREGISTER=#2%
   \countREGISTER=0%
   \expandafter\ROWcount\the#1\endcount%
}%
\def\ROWcount{%
   \afterassignment\subROWcount\let\next= %
}%
\def\subROWcount{%
   \ifx\next\endcount %
      \let\next=\relax%
   \else%
      \ncase=0%
      \ifx\next\cr %
         \global\advance\countREGISTER by 1%
         \ncase=0%
      \fi%
      \ifx\next\endrow %
         \global\advance\countREGISTER by 1%
         \ncase=0%
      \fi%
      \ifx\next\crthick %
         \global\advance\countREGISTER by 1%
         \ncase=0%
      \fi%
      \ifx\next\crnorule %
         \global\advance\countREGISTER by 1%
         \ncase=0%
      \fi%
      \ifx\next\crthickneg %
         \global\advance\countREGISTER by 1%
         \ncase=0%
      \fi%
      \ifx\next\crnoruleneg %
         \global\advance\countREGISTER by 1%
         \ncase=0%
      \fi%
      \ifx\next\crneg %
         \global\advance\countREGISTER by 1%
         \ncase=0%
      \fi%
      \ifx\next\header %
         \ncase=1%
      \fi%
      \relax%
      \ifcase\ncase %
         \let\next\ROWcount%
      \or %
         \let\next\argROWskip%
      \else %
      \fi%
   \fi%
   \next%
}
\def\counthdROWS#1\into#2{%
\dvr{10}%
   \let\countREGISTER=#2%
   \countREGISTER=0%
\dvr{11}%
\dvr{13}%
   \expandafter\hdROWcount\the#1\endcount%
\dvr{12}%
}%
\def\hdROWcount{%
   \afterassignment\subhdROWcount\let\next= %
}%
\def\subhdROWcount{%
   \ifx\next\endcount %
      \let\next=\relax%
   \else%
      \ncase=0%
      \ifx\next\cr %
         \global\advance\countREGISTER by 1%
         \ncase=0%
      \fi%
      \ifx\next\endrow %
         \global\advance\countREGISTER by 1%
         \ncase=0%
      \fi%
      \ifx\next\crthick %
         \global\advance\countREGISTER by 1%
         \ncase=0%
      \fi%
      \ifx\next\crnorule %
         \global\advance\countREGISTER by 1%
         \ncase=0%
      \fi%
      \ifx\next\header %
         \ncase=1%
      \fi%
\relax%
      \ifcase\ncase %
         \let\next\hdROWcount%
      \or%
         \let\next\arghdROWskip%
      \else %
      \fi%
   \fi%
   \next%
}%
{\catcode`\|=13\letbartab
\gdef\countCOLS#1\into#2{%
   \let\countREGISTER=#2%
   \global\countREGISTER=0%
   \global\multispancount=0%
   \global\firstrowtrue
   \expandafter\COLcount\the#1\endcount%
   \global\advance\countREGISTER by 3%
   \global\advance\countREGISTER by -\multispancount
}%
\gdef\COLcount{%
   \afterassignment\subCOLcount\let\next= %
}%
{\catcode`\&=13%
\gdef\subCOLcount{%
   \ifx\next\endcount %
      \let\next=\relax%
   \else%
      \ncase=0%
      \iffirstrow
         \ifx\next& %
            \global\advance\countREGISTER by 2%
            \ncase=0%
         \fi%
         \ifx\next\span %
            \global\advance\countREGISTER by 1%
            \ncase=0%
         \fi%
         \ifx\next| %
            \global\advance\countREGISTER by 2%
            \ncase=0%
         \fi
         \ifx\next\|
            \global\advance\countREGISTER by 2%
            \ncase=0%
         \fi
         \ifx\next\multispan
            \ncase=1%
            \global\advance\multispancount by 1%
         \fi
         \ifx\next\header
            \ncase=2%
         \fi
         \ifx\next\cr       \global\firstrowfalse \fi
         \ifx\next\endrow   \global\firstrowfalse \fi
         \ifx\next\crthick  \global\firstrowfalse \fi
         \ifx\next\crnorule \global\firstrowfalse \fi
         \ifx\next\crnoruleneg \global\firstrowfalse \fi
         \ifx\next\crthickneg  \global\firstrowfalse \fi
         \ifx\next\crneg       \global\firstrowfalse \fi
      \fi
\relax
      \ifcase\ncase %
         \let\next\COLcount%
      \or %
         \let\next\spancount%
      \or %
         \let\next\argCOLskip%
      \else %
      \fi %
   \fi%
   \next%
}%
\gdef\argROWskip#1{%
   \let\next\ROWcount \next%
}
\gdef\arghdROWskip#1{%
   \let\next\ROWcount \next%
}
\gdef\argCOLskip#1{%
   \let\next\COLcount \next%
}
}
}
\def\spancount#1{
   \nspan=#1\multiply\nspan by 2\advance\nspan by -1%
   \global\advance \countREGISTER by \nspan
   \let\next\COLcount \next}%
\def\dvr#1{\relax}%
\def\header#1{%
\dvr{1}{\let\cr=\@mpersand%
\hdtks={#1}%
\counthdROWS\hdtks\into\hdrows%
\advance\hdrows by 1%
\ifnum\hdrows=0 \hdrows=1 \fi%
\dvr{5}\makehdPREAMBLE{\the\hdrows}%
\dvr{6}\getHDdimen{#1}%
{\parindent=0pt\hsize=\hdsize{\let\ifmath0%
\xdef\next{\valign{\headerpreamble #1\crnorm}}}\dvr{7}\next\dvr{8}%
}%
}\dvr{2}}
\def\makehdPREAMBLE#1{
\dvr{3}%
\hdrows=#1
{
\let\headerARGS=0%
\let\cr=\crnorm%
\edef\xtp{\vfil\hfil\hbox{\headerARGS}\hfil\vfil}%
\advance\hdrows by -1
\loop
\ifnum\hdrows>0%
\advance\hdrows by -1%
\edef\xtp{\xtp&\vfil\hfil\hbox{\headerARGS}\hfil\vfil}%
\repeat%
\xdef\headerpreamble{\xtp\crcr}%
}
\dvr{4}}
\def\getHDdimen#1{%
\hdsize=0pt%
\getsize#1\cr\end\cr%
}
\def\getsize#1\cr{%
\endsizefalse\savetks={#1}%
\expandafter\lookend\the\savetks\cr%
\relax \ifendsize \let\next\relax \else%
\setbox\hdbox=\hbox{#1}\newhdsize=1.0\wd\hdbox%
\ifdim\newhdsize>\hdsize \hdsize=\newhdsize \fi%
\let\next\getsize \fi%
\next%
}%
\def\lookend{\afterassignment\sublookend\let\looknext= }%
\def\sublookend{\relax%
\ifx\looknext\cr %
\let\looknext\relax \else %
   \relax
   \ifx\looknext\end \global\endsizetrue \fi%
   \let\looknext=\lookend%
    \fi \looknext%
}%
%
%
\def\tablelet#1{%
   \tableLETtokens=\expandafter{\the\tableLETtokens #1}%
}%
\catcode`\@=12

{\nopagenumbers
\line{\hfil LTH 709}
\line{\hfil hep-th/0701096}
\vskip .5in
\centerline{\titlefont One-loop renormalisation of }
\centerline{\titlefont $\Ncal=\frak12$ supersymmetric gauge theory}
\centerline{\titlefont with a superpotential}
\vskip 1in
\centerline{\bf I.~Jack, D.R.T.~Jones and L.A.~Worthy}
\bigskip
\centerline{\it Department of Mathematical Sciences,  
University of Liverpool, Liverpool L69 3BX, U.K.}
\vskip .3in
We construct a superpotential for the general $\Ncal=\frak12$ 
supersymmetric gauge theory coupled to chiral matter in the fundamental
and adjoint representations, and investigate the one-loop renormalisability 
of the theories.  
\Date{July 2006}}

\newsec{Introduction} $\Ncal=\frak12$ supersymmetric theories (i.e.
theories defined on non-anticommutative superspace) have recently attracted
much attention\ref\ferr{S.~Ferrara and
M.A.~Lledo, JHEP 0005 (2000) 008}\nref\klemm{D.~Klemm, S.~Penati and
L.~Tamassia, Class. Quant. Grav.
20 (2003) 2905}\nref\seib{N.~Seiberg, JHEP
{\bf 0306} (2003) 010}--\ref\araki{T.~Araki, K.~Ito and  A. Ohtsuka,
\plb573 (2003) 209}. Such theories are non-hermitian and
only have half the supersymmetry of the corresponding $\Ncal=1$
theory.
These theories are not power-counting  renormalisable\foot{See
Refs.~\ref\britto{R.~Britto, B.~Feng and S.-J.~Rey,
JHEP 0307 (2003) 067; JHEP 0308 (2003) 001}\ref\terash{S. Terashima
and J-T Yee, JHEP {\bf 0312}
(2003) 053} for other discussions of the ultra-violet properties of these
theories.} but it has been
argued\ref\gris{M.T.~Grisaru, S.~Penati and  A.~Romagnoni, JHEP {\bf
0308} (2003) 003\semi
R.~Britto and B.~Feng, \prl91 (2003) 201601\semi
A.~Romagnoni, JHEP {\bf 0310} (2003) 016}\nref\lunin{O.~Lunin
and S.-J. Rey, JHEP  {\bf 0309}
(2003) 045}\nref\alish{
M.~Alishahiha, A.~Ghodsi and N.~Sadooghi, \npb691
(2004) 111}--\ref\berrey{D.~Berenstein and S.-J.~Rey, \prd68 (2003) 121701}
 that they are in  fact nevertheless
renormalisable, in other words only a finite number of additional
terms need to be added to the Lagrangian to absorb divergences to all
orders. Generally speaking, in a non-renormalisable theory, the dimensionality
of Green's functions suffering from logarithmical divergences increases as 
the number of loops increases, leading inevitably to an infinite 
number of potential counter-terms; that this does not happen here is due to a set 
of discrete symmetries whose origin is linked to the non-hermitian nature 
of the relevant actions. (An elegant analysis of this appearing in Ref.~\lunin{} is 
straightforward to generalise to incorporate Yukawa couplings in the 
case of adjoint matter that we consider here. We present this analysis in Appendix C, 
showing in fact that Ref.~\lunin{} omitted a few relevant terms).
In previous work we have shown that although divergent 
gauge non-invariant terms are generated at the one-loop level, they can be 
removed by  
divergent field redefinitions leading to a renormalisable theory
in which $\Ncal=\frak12$ supersymmetry is preserved at the one-loop level
in both the pure gauge case\ref\jjw{I.~Jack, D.R.T.~Jones
and L.A.~Worthy, \plb611 (2005) 199}\ and in the presence of 
chiral matter in the fundamental 
representation\ref\jjwa{I.~Jack, D.R.T.~Jones
and L.A.~Worthy, \prd72 (2005) 065002}.  In the latter case, the joint
requirements of
renormalisability and $\Ncal=\frak12$ supersymmetry impose the choice of
gauge group $SU(N)\otimes U(1)$ (rather than $U(N)$ or $SU(N)$).
It is interesting to compare our results with those obtained using   
superfields. The authors of Ref.~\ref\penrom{
S.~Penati and A.~Romagnoni, JHEP {\bf 0502} (2005) 064}
obtained the one loop effective action for pure $\Ncal=\frak12$
supersymmetry using a superfield formalism. Although they found
divergent contributions which broke supergauge invariance, their final
result was gauge-invariant without the need for any redefinition. In 
subsequent work\ref\gpr{M.T.~Grisaru, S.~Penati and A.~Romagnoni, 
JHEP {\bf 0602} (2006) 043} it was shown that the $\Ncal=\frak12$
superfield action requires modification to ensure renormalisability, which
is consistent with our findings in the component formulation\jjwa.

It was pointed out in Ref.~\araki\ that an $\Ncal=\frak12$ supersymmetric 
theory can also be constructed with matter in the adjoint representation (at 
the classical level).
Our purpose here is to repeat the analysis of Ref.~\jjwa\ for the
adjoint case, then proceed to consider the addition of superpotential
terms for both the adjoint and fundamental cases,
which will turn out to be a non-trivial task. Our goal is to construct
a renormalisable $\Ncal=\frak12$ supersymmetric theory. Renormalisability
may require the addition of new terms and associated couplings to the
original classical $\Ncal=\frak12$ supersymmetric theory in order that
all divergences may be removed by adding counterterms to the couplings;\ but the
hope is that this can be done while preserving $\Ncal=\frak12$ invariance.
We shall find that this can be achieved in the fundamental case, where
only mass terms are possible in the superpotential, but not in the adjoint 
case where trilinear superpotential terms are also allowed. (In the 
$\Ncal=\frak12$ case, these trilinear superpotential terms are accompanied
by additional terms with gauge fields; we shall refer to the full set of these
terms as the ``Yukawa'' superpotential.)  
\newsec{The classical adjoint action without superpotential}
The adjoint action of Ref.~\araki\ was written for the gauge group $U(N)$.
As we noted in Refs.~\jjw, \jjwa, at the quantum level the $U(N)$ gauge 
invariance cannot be retained. As mentioned earlier, in
the case of chiral matter in the 
fundamental representation we were obliged to consider a modified theory 
with the gauge group $SU(N)\otimes U(1)$. In the adjoint case with a 
Yukawa superpotential, 
it will turn out that the matter fields must also be in a 
representation of $SU(N)\otimes U(1)$. 
However, for simplicity of exposition we 
shall
start by considering the adjoint case without a superpotential, in other
words adapting the calculations of Ref.~\jjwa\ to the adjoint case.
The classical action without a superpotential may be written
\eqn\Sadj{\eqalign{
S_0=&\int d^4x
\Bigl[-\frak14F^{\mu\nu A}F^A_{\mu\nu}-i\lambdabar^A\sigmabar^{\mu}
(D_{\mu}\lambda)^A+\frak12D^AD^A\cr
&-\frak12iC^{\mu\nu}d^{ABC}e^{ABC}F^A_{\mu\nu}\lambdabar^B\lambdabar^C\cr
&+\frak18g^2|C|^2d^{abe}d^{cde}(\lambdabar^a\lambdabar^b)
(\lambdabar^c\lambdabar^d)
+\frak{1}{4N}\frak{g^4}{g_0^2}|C|^2(\lambdabar^a\lambdabar^a)
(\lambdabar^b\lambdabar^b)\cr
&+
\Fbar F -i\psibar\sigmabar^{\mu}D_{\mu}\psi-D^{\mu}\phibar D_{\mu}\phi\cr
&+g\phibar D^F\phi +
ig\sqrt2(\phibar \lambda^F\psi-\psibar\lambdabar{}^F\phi)\cr
&+d^{abc}gC^{\mu\nu}\left(
\sqrt2D_{\mu}\phibar^a\lambdabar{}^b\sigmabar_{\nu}\psi^c
+i\phibar^a F^b_{\mu\nu}F^c\right)\cr
&+d^{ab0}g_0C^{\mu\nu}\left(
\sqrt2D_{\mu}\phibar^a\lambdabar{}^0\sigmabar_{\nu}\psi^b   
+i\phibar^a F^0_{\mu\nu}F^b\right)\cr
&+d^{000}g_0C^{\mu\nu}\left(
\sqrt2\pa_{\mu}\phibar^0\lambdabar{}^0\sigmabar_{\nu}\psi^0          
+i\phibar^0 F^0_{\mu\nu}F^0\right)\cr      
&+d^{ab0}gC_1^{\mu\nu}\left(\sqrt2\pa_{\mu}\phibar^0\lambdabar{}^a
\sigmabar_{\nu}\psi^b
+i\phibar^0 F^a_{\mu\nu}F^b\right)\cr
&+d^{ab0}gC_2^{\mu\nu}\left(\sqrt2D_{\mu}\phibar^a\lambdabar{}^b
\sigmabar_{\nu}\psi^0
+i\phibar^a F^b_{\mu\nu}F^0\right)\cr
&-\frak{1}{4}g^2|C|^2\phibar \lambdabar^F\lambdabar^FF
\Bigr].\cr}}
Here
\eqn\lamdef{
\lambda^F=\lambda^a\Ftil^a,\quad (\Ftil^A)^{BC}=if^{BAC},}
(similarly for $D^F$), and we have
\eqn\Dmudef{\eqalign{
D_{\mu}\phi=&\pa_{\mu}\phi+igA^F_{\mu}\phi,\cr
F_{\mu\nu}^A=&\pa_{\mu}A_{\nu}^A-\pa_{\nu}A_{\mu}^A-gf^{ABC}
A_{\mu}^BA_{\nu}^C,\cr}}
with similar definitions for $D_{\mu}\psi$, $D_{\mu}\lambda$.
If one decomposes $U(N)$ as
$SU(N)\otimes U(1)$ then our convention is that $\phi^a$ (for example) 
are the $SU(N)$
components and $\phi^0$ the $U(1)$ component. (For later convenience 
we also define $g_A$ similarly to encompass both $g_a=g$ and $g_0$.)
Of course then $f^{ABC}=0$
unless all indices are $SU(N)$. 
We note that $d^{ab0}=\sqrt{\frak2N}\delta^{ab}$, $d^{000}=\sqrt{\frak2N}$.
(Useful identities for $U(N)$ are listed in Appendix D.)
We also have
\eqn\etensor{
e^{abc}=g,\quad e^{a0b}=e^{ab0}=e^{000}=g_0,\quad e^{0ab}={g^2\over{g_0}}.}
We have written the $\phibar\lambdabar\lambdabar F$ term as it is 
given starting from the superspace formalism. We note that it has the 
opposite sign from 
that given in Ref.~\araki. This term is $\Ncal=\frak12$ supersymmetric on
its own and so the exact form chosen should not affect the renormalisability
of the theory.
It is easy to show that Eq.~\Sadj\ is invariant under
\eqn\newsusy{
\eqalign{
\delta A^A_{\mu}=& -i\lambdabar^A\sigmabar_{\mu}\epsilon\cr  
\delta \lambda^A_{\alpha}=& i\epsilon_{\alpha}D^A+\left(\sigma^{\mu\nu}\epsilon
\right)_{\alpha}\left[F^A_{\mu\nu}
+\frak12iC_{\mu\nu}e^{ABC}d^{ABC}\lambdabar^B\lambdabar^C\right],\quad
\delta\lambdabar^A_{\alphadot}=0,\cr
\delta D^A=& -\epsilon\sigma^{\mu}D_{\mu}\lambdabar^A,\cr
\delta\phi=& \sqrt2\epsilon\psi,\quad\delta\phibar=0,\cr
\delta\psi^{\alpha}=& \sqrt2\epsilon^{\alpha} F,\quad
\delta\psibar_{\alphadot}=-i\sqrt2(D_{\mu}\phibar)
(\epsilon\sigma^{\mu})_{\alphadot},\cr
\delta F^A=& 0,\cr
\delta \Fbar^a=& -i\sqrt2D_{\mu}\psibar^a\sigmabar^{\mu}\epsilon
-2ig(\phibar\epsilon\lambda^F)^a\cr
& +2gC^{\mu\nu}D_{\mu}(\phibar^b\epsilon\sigma_{\nu}   
\lambdabar^cd^{bca}+\phibar^b\epsilon\sigma_{\nu}
\lambdabar^0d^{b0a})
+2gC_1^{\mu\nu}D_{\mu}(\phibar^0\epsilon\sigma_{\nu}   
\lambdabar^bd^{0ba}),\cr
\delta \Fbar^0=& -i\sqrt2D_{\mu}\psibar^0\sigmabar^{\mu}\epsilon\cr
& +2gC_2^{\mu\nu}D_{\mu}(\phibar^a\epsilon\sigma_{\nu}   
\lambdabar^bd^{ab0})
+2g_0C^{\mu\nu}D_{\mu}(\phibar^0\epsilon\sigma_{\nu}   
\lambdabar^0d^{000}).\cr
}}
In Eq.~\Sadj, $C^{\mu\nu}$ is related to the non-anti-commutativity 
parameter $C^{\alpha\beta}$ by  
\eqn\Cmunu{
C^{\mu\nu}=C^{\alpha\beta}\epsilon_{\beta\gamma}
\sigma^{\mu\nu}_{\alpha}{}^{\gamma},} 
where 
\eqn\sigmunu{\eqalign{
\sigma^{\mu\nu}=&\frak14(\sigma^{\mu}\sigmabar^{\nu}-
\sigma^{\nu}\sigmabar^{\mu}),\cr
\sigmabar^{\mu\nu}=&\frak14(\sigmabar^{\mu}\sigma^{\nu}-
\sigmabar^{\nu}\sigma^{\mu}),\cr }} 
and 
\eqn\Csquar{
|C|^2=C^{\mu\nu}C_{\mu\nu}.} 
Our conventions are in accord with \seib; in particular, 
\eqn\sigid{
\sigma^{\mu}\sigmabar^{\nu}=-\eta^{\mu\nu}+2\sigma^{\mu\nu}.}
Properties of $C$ which follow from
Eq.~\Cmunu\ are  
\eqna\cprop$$\eqalignno{
C^{\alpha\beta}&=\frak12\epsilon^{\alpha\gamma}
\left(\sigma^{\mu\nu}\right)_\gamma{}^{\beta}C_{\mu\nu},
& \cprop a\cr
C^{\mu\nu}\sigma_{\nu\alpha\betadot}&=C_{\alpha}{}^{\gamma}
\sigma^{\mu}{}_{\gamma\betadot},&\cprop b\cr
C^{\mu\nu}\sigmabar_{\nu}^{\alphadot\beta}&=-C^{\beta}{}_{\gamma}
\sigmabar^{\mu\alphadot\gamma}.&\cprop c\cr}$$ 

In Eqs.~\Sadj, $C^{\mu\nu}_{1,2}$ will be identical to $C^{\mu\nu}$ at the
classical level; but we have distinguished them to allow for the
possibility of different renormalisations (in practice an overall numerical
factor) at the quantum level; so that $C^{\mu\nu}_{1,2}$ will obey 
properties analogous to Eqs.~\Cmunu, \Csquar\ and \cprop{}.
It is important to note that this is only 
compatible with $\Ncal=\frak12$ supersymmetry due to the fact that the 
$d^{ab0}\pa_{\mu}\phibar^0\lambdabar^a
\sigmabar_{\nu}\psi^b$ term in Eq.~\Sadj\ contains no gauge field; and  
the variation of the gauge field in $d^{ab0} D_{\mu}\phibar^a\lambdabar^b
\sigmabar_{\nu}\psi^0$ gives zero. This implies that the variations of the
terms containing either $C^{\mu\nu}_1$ or $C^{\mu\nu}_2$ 
respectively are self-contained. (By contrast, 
the variation of the gauge field in the $d^{abc}D_{\mu}\phibar^a\lambdabar^b
\sigmabar_{\nu}\psi^c$ term is cancelled by the $C^{\mu\nu}$ term in the
variation of the $\lambda$ in the 
$\phibar\lambda\psi$ term, which forces the $C^{\mu\nu}$ in the 6th line of
Eq.~\Sadj\ to be equal to that in the pure gauge terms, and similarly
for that in the 7th line; the terms in the 8th line do not get renormalised
at all.)

We use the standard gauge-fixing term 
\eqn\gafix{
S_{\rm{gf}}={1\over{2\alpha}}\int d^4x (\pa.A)^2} 
with its associated
ghost terms.  The gauge propagators for $SU(N)$ and $U(1)$ are both given by  
\eqn\gprop{
\Delta_{\mu\nu}=-{1\over{p^2}}\left(\eta_{\mu\nu}
+(\alpha-1){p_{\mu}p_{\nu}\over{p^2}}\right)}
(omitting group factors) and the fermion propagator is  
\eqn\fprop{
\Delta_{\alpha\alphadot}={p_\mu\sigma^{\mu}_{\alpha\alphadot}\over{p^2}},}
where the momentum enters at the end of the propagator with the undotted 
index.  
The one-loop graphs contributing
to the ``standard'' terms in the Lagrangian (those without a
$C^{\mu\nu}$) are the same as in the ordinary $\Ncal=1$ case, so 
anomalous dimensions and gauge $\beta$-functions are as for
$\Ncal=1$. Since our gauge-fixing term in Eq.~\gafix\ does not preserve 
supersymmetry, the anomalous dimensions for $A_{\mu}$ and $\lambda$
are
different (and moreover gauge-parameter dependent), as are those for
$\phi$ and $\psi$. However, the 
gauge $\beta$-functions are of course gauge-independent. 
The one-loop one-particle-irreducible (1PI) 
graphs contributing to the new terms (those
containing $C$) are depicted in Figs.~1--6. With the exception of Fig.~6 
(which gives zero contributions in the case of chiral fields in the fundamental
representation) these diagrams are the same as those
considered in Ref.~\jjwa. The divergent contributions from these and other
diagrams considered later are listed (for the adjoint case) in Appendix A. 

\newsec{Renormalisation of the adjoint $SU(N)$ action}
The renormalisation of $\Ncal=\frak12$
supersymmetric gauge theory presents certain subtleties.
The bare action is given by 
\eqn\Sbare{\eqalign{
S_B=&S_{0B}\cr
&+\frak{1}{N}\kappa_1g_0^2|C|^2(\lambdabar^a\lambdabar^a)
(\lambdabar^0\lambdabar^0)\cr
&-gd^{ab0}\kappa_2C_2^{\mu\nu}\left(\sqrt2D_{\mu}\phibar^a\lambdabar^b
\sigmabar_{\nu}\psi^0+\sqrt2\phibar^a\lambdabar^b
\sigmabar_{\nu}\pa_{\mu}\psi^0
+i\phibar^a F^b_{\mu\nu}F^0\right)\cr
&-g_0d^{ab0}\kappa_3C^{\mu\nu}\left(\sqrt2D_{\mu}\phibar^a\lambdabar^0
\sigmabar_{\nu}\psi^b+\sqrt2\phibar^a\lambdabar^0
\sigmabar_{\nu}D_{\mu}\psi^b
+i\phibar^a F^0_{\mu\nu}F^b\right)\cr  
}}
where $S_{0B}$ is obtained by replacing all fields and couplings in $S_0$
(in Eq.~\Sadj) by their bare versions, given below.
The terms involving 
$\kappa_{1-3}$ 
 are separately invariant under $\Ncal=\frak12$ 
supersymmetry. Those with $\kappa_1$, $\kappa_2$
 must be included at this stage to obtain a 
renormalisable Lagrangian; those with $\kappa_3$ will be required when we 
introduce a superpotential but could be omitted at present. 

We found in Refs.~\jjw, \jjwa\ that non-linear renormalisations of $\lambda$ 
and $\Fbar$ were required; and in a subsequent
paper\ref\jjwb{I.~Jack, D.R.T.~Jones
and L.A.~Worthy, \prd72 (2005) 107701}\ we pointed out that non-linear 
renormalisations of $F$, $\Fbar$ are required even in ordinary $\Ncal=1$ 
supersymmetric gauge theory when working in the uneliminated formalism.
Note that in the $\Ncal=\frak12$ supersymmetric case, fields and their
conjugates may renormalise differently. 
The renormalisations of the remaining fields and couplings are linear as 
usual and given by
\eqn\bare{\eqalign{ 
\lambdabar^a_B=Z_{\lambda}^{\frak12}\lambdabar^a,
\quad
A^{a}_{\mu B}=Z_A^{\frak12}A^{a}_{\mu}, \quad& D^a_B=Z_D^{\frak12}D^a,
\quad \phi^a_B=Z_{\phi}^{\frak12}\phi^a,\cr
\psi^a_B=Z_{\psi}^{\frak12}\psi^a,\quad
\phibar^a_B=Z_{\phi}^{\frak12}\phibar^a,
\quad \psibar^a_B=&Z_{\psi}^{\frak12}\psibar^a, \quad F_B=Z_FF,\quad
g_B=Z_gg,\cr  
C_B^{\mu\nu}=Z_CC^{\mu\nu}, \quad |C|_B^2=Z_{|C|^2}|C|^2,\quad&
C_{1,2B}^{\mu\nu}=Z_{C_{1,2}}C_{1,2}^{\mu\nu},\quad \kappa_{1-3B}=
Z_{1-3}.
\cr}}
The corresponding $U(1)$ gauge multiplet fields 
$\lambdabar^0$ etc are unrenormalised 
(as are the $U(1)$ chiral fields $\phi^0$ etc in the case with no 
superpotential); so is $g_0$. The auxiliary field $F$ is also unrenormalised,
i.e. $Z_F=1$
(though again this will no longer be the case when we later introduce a 
superpotential). In Eq.~\bare, $Z_{1-3}$ are divergent
contributions, in other words we have set the renormalised couplings
$\kappa_{1-3}$ to zero for simplicity.   
The other renormalisation constants start with 
tree-level values of 1. As we mentioned before,
the renormalisation constants for the fields
and for the gauge coupling $g$ are the same as in the ordinary $\Ncal=1$
supersymmetric theory (for a gauge theory coupled to an adjoint
chiral field) and are therefore given up to one loop 
by\ref\timj{
D.~Gross and F.~Wilcek, \prd8 (1973) 3633\semi
D.R.T.~Jones, \npb87 (1975) 127}: 
\eqn\Zgg{\eqalign{
Z_{\lambda}=&1-g^2NL(2\alpha +2),\cr
Z_A=&1+g^2NL(1-\alpha)\cr
Z_D=&1-2NLg^2,\cr
Z_g=&1-2g^2NL,\cr
Z_{\phi}=&1+2g^2(1-\alpha)LN,\cr
Z_{\psi}=&1-2g^2(1+\alpha)LN,\cr}}
where (using dimensional regularisation with $d=4-\epsilon$)
$L={1\over{16\pi^2\epsilon}}$.
The renormalisation of $\lambda^A$ is given by 
\eqn\lchange{\eqalign{
\lambda_B^a=&Z_{\lambda}^{\frak12}\lambda^a
-\frak12NLg^3C^{\mu\nu}d^{abc}
\sigma_{\mu}\lambdabar^cA_{\nu}^b
-NLg^2g_0C^{\mu\nu}d^{ab0}         
\sigma_{\mu}\lambdabar^0A_{\nu}^b\cr
&+i\sqrt2\tau_4NLg^3d^{abc}(C\psi)^b\phibar^c
+i\sqrt2\tau_5NLg^3d^{ab0}(C\psi)^0\phibar^b,\cr
\lambda_B^0=&\lambda^0i\sqrt2\tau_6NLg^2g_0d^{0ab}
(C\psi)^a\phibar^b,\cr 
}}
where $(C\psi)^{\alpha}=C^{\alpha}{}_{\beta}\psi^{\beta}$. The replacement
of $\lambda$ by $\lambda_B$ produces a change in the action given (to
first order) by 
\eqn\Schlamb{\eqalign{
S_0(\lambda_B)-S_0(\lambda)=&NLg^2\int d^4x\Bigl\{
\tau_4g\bigl[igd^{abe}f^{cde}\phibar^a\phibar^b\psi^c(C\psi^d)\cr
&+\sqrt2C^{\mu\nu}d^{abc}\phibar^a\lambdabar^b\sigmabar_{\nu}D_{\mu}\psi^c
+\sqrt2C^{\mu\nu}d^{abc}D_{\mu}\phibar^a\lambdabar^b\sigmabar_{\nu}
\psi^c\bigr]\cr
&+\tau_5\sqrt2gC^{\mu\nu}d^{ab0}(\phibar^a\lambdabar^b\sigmabar_{\nu}
\pa_{\mu}\psi^0
+D_{\mu}\phibar^a\lambdabar^b\sigmabar_{\nu}\psi^0)\cr
&+\tau_6\sqrt2g_0C^{\mu\nu}d^{0ab}(\phibar^a\lambdabar^0\sigmabar_{\nu}
D_{\mu}\psi^b
+D_{\mu}\phibar^a\lambdabar^0\sigmabar_{\nu}\psi^b)+\ldots\Bigr\},\cr}}
where the ellipsis indicates the terms not involving $\tau_{4-6}$ 
(which were given previously in Ref.~\jjwa). 
The value of $\tau_4$ will be chosen
so as to cancel the divergent contributions from Fig.~6; $\tau_{5,6}$
will be specified later when we renormalise the theory with a 
superpotential. 

We now find that to render finite the contributions linear in $F$ we require
\eqn\fbarredef{\eqalign{
\Fbar^a_B=&Z_F\Fbar^a+
iC^{\mu\nu}Lg^2\Bigl\{gN\Bigl[(5+2\alpha)
\pa_{\mu}A_{\nu}^b-\frak14(11+4\alpha)
gf^{bde}A_{\mu}^dA_{\nu}^e\Bigr]\phibar^c d^{abc}\cr
&+\sqrt{2N}g\Bigl[2\left((4+\alpha)-z_{C_1}\right)\pa_{\mu}A_{\nu}^a
-\left(\frak12(9+2\alpha)-z_{C_1}\right)
gf^{abc}A_{\mu}^bA_{\nu}^c\Bigr]\phibar^0\cr
&+2\sqrt{2N}g_0\left(-(1-\alpha)+z_3\right)
\pa_{\mu}A_{\nu}^0\phibar^a\Bigr\}\cr
&+\frak{1}{8}Lg^4|C|^2\Bigl[2(1-\alpha)Nf^{ace}f^{bde}
-11Nd^{abe}d^{cde}+4(\delta^{ab}\delta^{cd}+\delta^{ac}\delta^{bd})\Bigr]
\phibar^b\lambdabar^c\lambdabar^d\cr
&-Lg^3|C|^2\Bigl\{d^{abc}\sqrt{2N}\Bigl[g\phibar^0\lambdabar^b\lambdabar^c
+3g_0\phibar^b\lambdabar^c\lambdabar^0\Bigr]
+4g_0\phibar^0\lambdabar^0\lambdabar^a\Bigr\},\cr
\Fbar^0_B=&Z_F\Fbar^0
+i\sqrt{2N}L\left(2+z_2-z_{C_2}\right)C^{\mu\nu}F^a_{\mu\nu}\phibar^a\cr
&-2d^{abc}g^4L|C|^2\sqrt{2N}\phibar^a\lambdabar^b\lambdabar^c
-8g^3g_0L|C|^2\phibar^a\lambdabar^a\lambdabar^0.\cr}}
Writing $Z_C^{(n)}$ for the $n$-loop
contribution to $Z_C$ we set
\eqn\zformb{
Z_C^{(1)}=z_CNLg^2}
with similar definitions for $Z_{|C|^2}$, $Z_{C_{1,2}}$, $Z_{1-3}$.
We now find that with
\eqn\zforms{
z_C=z_{|C|^2}=0,\quad z_{C_1}=-z_{C_2}=2,\quad
z_1=-3,\quad \tau_4=1,\quad \tau_5=z_2-z_{C_2}, \quad \tau_6=z_3,}
the one-loop effective action is finite, for arbitrary $z_2$, $z_3$. It would 
appear that we do not have enough information yet to specify the 
renormalisation constants $\kappa_2$, $\kappa_3$ in Eq.~\Sbare. This 
apparent arbitrariness is due to the possibility of making 
non-linear renormalisations of $\lambda$ as in Eq.~\lchange\ so that changes
in $z_2$ and $z_3$ can be compensated by changes in $\tau_4$, $\tau_5$.
We shall however find 
ourselves obliged to pick certain values for these constants when we introduce
a superpotential. This behaviour is unexpected but we shall find similar 
features in 
the case of the fundamental representation later. It would be more satisfying
to find some underlying reason for making these choices {\it before} 
introducing the superpotential. 

\newsec{The superpotential in the adjoint case}
We now consider the problem of adding superpotential terms to the 
Lagrangian Eq.~\Sadj. The following potential terms are $\Ncal=\frak12$ 
invariant at the classical level:  
\eqn\lagmass{\eqalign{
S_{\rm{int}}=&\int d^4x\tr\Bigl\{y\bigl[
\phi^2 F-\psi^2\phi\cr
&+\phibar^2\Fbar-\psibar^2\phibar+\frak43ig
C^{\mu\nu}\phibar^3\Fhat_{\mu\nu}+\frak23C^{\mu\nu}D_{\mu}\phibar D_{\nu}\phibar
\phibar\bigr]\cr
&+m\bigl[\phi F -\frak12\psi\psi+\phibar\Fbar-\frak12\psibar\psibar
+iC^{\mu\nu}\phibar \Fhat_{\mu\nu}\phibar
-\frak18g^2|C|^2\phibar \phibar\lambdabar^F
\lambdabar^F\bigr]\Bigr\}.\cr}}
Here in the interests of conciseness we have written the superpotential
in index-free form, so that 
\eqn\hatdef{
\phi=\phi^A R^A, \quad 
\psi=\psi^AR^A,\quad \Ahat_{\mu}=gA_{\mu}^aR^a+g_0A_{\mu}^0R^0;}
it then follows that $\Fhat_{\mu\nu}=g_AF_{\mu\nu}^AR^A$, with 
$F_{\mu\nu}$ defined as in Eq.~\Dmudef. 
We denote the group matrices for the fundamental representation 
of $SU(N)\otimes U(1)$ by $R^A$ where
our convention is that $R^a$ are the $SU(N)$
generators and $R^0$ the $U(1)$ generator.
The matrices are normalised so that
$\Tr[R^AR^B]=\frak12\delta^{AB}$. In particular, $R^0=\sqrt{\frak{1}{2N}}1$.
It is easy to check that $S_{\rm{int}}$ is $\Ncal=\frak12$ invariant.
Except for the last mass term, this superpotential is most readily  
derived directly from the superspace formalism.
Denoting an adjoint chiral superfield as 
$\Phi_A$, 
we have that under a gauge transformation 
$$
\Phi_A \to \Omega * \Phi_A *  \Omega^{-1}, \quad 
\ov \Phi_A \to \ov \Omega * \ov \Phi_A *  \ov \Omega^{-1}, 
$$
so that the gauge interactions are written in superfield form as 
$$
\int d^4\th \,\tr\left[\ov \Phi_A * e^V * \Phi_A * e^{-V}\right]  .
$$
The following superpotential terms are manifestly also invariant:
\eqn\super{\eqalign{
&\int d^2 \theta 
\,\tr\left[\frak12m \Phi_A * \Phi_A + \frak13y\Phi_A * \Phi_A * \Phi_A  
\right]  
\quad \cr
+&  
\int d^2 \bth \,\tr\left[\frak13m \ov \Phi_A * \ov \Phi_A 
+ \frak13y\ov \Phi_A * \ov \Phi_A * \ov \Phi_A  \right].
}}

Expanded in component fields we have 
\eqna\comps$$\eqalignno{
\Phi_A (y, \th) &= \phi (y) + \sqrt{2} \th \psi (y) + \th\th
F (y) & \comps a\cr
\ov \Phi_A (\ov y, \bth) &= \ov \phi (\ov y) + \sqrt{2} \bth \ov \psi (\ov y)\cr
&+ \bth\bth \Big( \ov F (\ov y) + i gC^{\mu\nu} \partial_{\mu}
\{\ov \phi, A_{\nu}\}(\ov y) 
- \frak{g^2}{2} C^{\mu\nu} \left[ A_{\mu},\{ A_{\nu},  
\ov \phi\}\right](\ov y) \Big), & 
\comps b\cr
}$$
where $\ov y^{\mu} = y^{\mu} - 2 i \th \sigma^{\mu} \bth$. Note the 
modification of the $\bth\bth$-term\araki.

If we substitute Eq.~\comps{}\ in Eq.~\super\ we obtain Eq.~\lagmass\
except for the 
last term. (This can also
be expressed in superfields but in a more unwieldy form). 
The coefficient 
of this final term is arbitrary since it is separately 
$\Ncal=\frak12$ invariant; the reason for our particular choice will be
explained later (after Eq.~(A.18) in Appendix A).
A similar set of mass terms is admissible in the case of the 
fundamental representation, with mass terms coupling the fundamental and
anti-fundamental representation fields, as we show later. 
However, no Yukawa terms are possible
in the $\Ncal=\frak12$ case for the fundamental representation.
If we have both adjoint and fundamental (anti-fundamental) representations 
$\Phi (\wt \Phi)$ we can construct \ntwo-type invariants, of the form 
\eqn\supertwo{
y \left[\int d^2 \theta \,\wt \Phi * \Phi_A * \Phi \quad + \quad  
\int d^2 \bth \,\ov \Phi * \ov \Phi_A * \ov {\wt \Phi} \right].
}

At the classical level $\phi$ may be 
considered as forming a representation of $U(N)$. However, just as we saw in
Ref.~\jjwa\ for the gauge group, the $U(N)$ structure is not preserved at the
quantum level. The $\phi^a$ renormalise differently from the $\phi^0$ and this 
means that, for instance, there must be a different mass parameter ($m$, say)
for the $\phi^a F^a$, $\psi^a\psi^a$ terms than for the $\phi^0 F^0$, 
$\psi^0\psi^0$ terms ($m_0$, say). In the case of the mass terms this does not 
present serious difficulty since we can separate the mass terms in 
Eq.~\lagmass\ into separately $\Ncal=\frak12$ invariant sets of terms
involving either $m$ or $m_0$. However, in the case of the Yukawa 
superpotential terms, we need to invoke three separate couplings, one
($y$, say) for $\phi^a\phi^b F^c$ terms, one ($y_1$, say)
for $\phi^a\phi^b F^0$, $\phi^a\phi^0 F^b$ etc and one ($y_2$, say)
for $\phi^0\phi^0 F^0$. In the $\Ncal=1$ case the theory would, of course,
be renormalisable, with each of $y$, $y_{1,2}$ 
renormalising differently. By contrast, in the $\Ncal=\frak12$ case many of the 
$\phibar^3 A_{\mu}$ terms are linked by $\Ncal=\frak12$ transformations 
to more than one of these groups of terms and so cannot be assigned a unique
coupling out of $y$, $y_{1,2}$. So in the presence of Yukawa 
superpotential terms, the $\Ncal=\frak12$ invariance cannot be maintained at 
the quantum level. It is this linking of different groups of terms, 
specifically those corresponding purely to $SU(N)$ with those containing
$U(1)$ fields, which implies that we cannot have an $\Ncal=\frak12$ theory
with a superpotential if the chiral fields belong to $SU(N)$ alone. 
\vfill
\eject
\newsec{The renormalised action with superpotential}
As we explained in the previous section, many of the individual terms
with couplings $m$ or $y$ 
in Eq.~\lagmass\ will renormalise differently and hence need to
be assigned their own separate couplings. 
For renormalisability, Eq.~\lagmass\ needs to be replaced by
\eqn\lagmassren{\eqalign{
S_{\rm{int}}=&\int d^4x\Bigl\{\frak14y d^{abc}(\phi^a\phi^b F^c
-\psi^a\psi^b\phi^c)\cr
&+\frak14y_1d^{ab0}(\phi^a\phi^b F^0
+2\phi^a\phi^0F^b-\psi^a\psi^b\phi^0-2\psi^a\psi^0\phi^b)\cr
&+\frak14y_2d^{000}(\phi^0\phi^0F^0-\psi^0\psi^0\phi^0)
\cr
&+\frak14y d^{abc}(\phibar^a\phibar^b \Fbar^c
-\psibar^a\psibar^b\phibar^c)\cr          
&+\frak14y_1d^{ab0}(\phibar^a\phibar^b \Fbar^0+2\phibar^a\phibar^0\Fbar^b
-\psibar^a\psibar^b\phibar^0 
-2\psibar^a\psibar^0\phibar^b)\cr
&+\frak14y_2d^{000}(\phibar^0\phibar^0\Fbar^0
-\psibar^0\psibar^0\phibar^0)
\cr
&+igC^{\mu\nu}F^d_{\mu\nu}\Bigl(
\frak16y d^{abe}d^{cde}\phibar^a\phibar^b\phibar^c
+\frak13y_3\frak1N\delta^{ab}\delta^{cd}\phibar^a\phibar^b\phibar^c\cr
&+\frak12y_4\sqrt{\frak2N}d^{abd}\phibar^0\phibar^a\phibar^b
+y_5\frak1N\phibar^0\phibar^0\phibar^d\Bigr)\cr
&+ig_0C^{\mu\nu}F_{\mu\nu}^0\Bigl(\frak16y\sqrt{\frak2N}
d^{abc}\phibar^a\phibar^b\phibar^c
+y_1\frak1N\phibar^a\phibar^a\phibar^0
+\frak13y_2\frak1N\phibar^0\phibar^0\phibar^0\Bigr)\cr
&+\frak16iy C^{\mu\nu}f^{abc}(D_{\mu}\phibar)^a (D_{\nu}\phibar)^b
\phibar^c\cr
&+m\Bigl[\phi^a F^a -\frak12\psi^a\psi^a+\phibar^a\Fbar^a
-\frak12\psibar^a\psibar^a\cr
&+\frak12igC^{\mu\nu}d^{abc}F^c_{\mu\nu}\phibar^a \phibar^b
+\frak12ig
C_1^{\mu\nu}d^{0ab}F^a_{\mu\nu}\phibar^0 \phibar^b
+\frak12ig
C^{\mu\nu}d^{0ab}F^0_{\mu\nu}\phibar^a\phibar^b
\Bigr]\cr
&+m_0\Bigl[\phi^0 F^0 -\frak12\psi^0\psi^0+\phibar^0\Fbar^0
-\frak12\psibar^0\psibar^0\cr
&+\frak12igC_2^{\mu\nu}d^{0ab}F^a_{\mu\nu}\phibar^0 \phibar^b
+\frak12ig
C^{\mu\nu}d^{000}F^0_{\mu\nu}\phibar^0 \phibar^0\Bigr]\cr
&+|C|^2\bigl[-\frak18
\mu_1f^{ace}f^{bde}+\mu_2\frak2N\delta^{ab}\delta^{cd}\bigr]
g^2\phibar^a\phibar^b\lambdabar^c\lambdabar^d\cr
&+gd^{abc}\sqrt{2N}|C|^2\phibar^a\lambdabar^b\left(\mu_3g_0\phibar^c\lambdabar^0
+g\mu_4\phibar^0\lambdabar^c\right)
+\frak2N\mu_5gg_0|C|^2\phibar^a\phibar^0\lambdabar^a\lambdabar^0\cr
&+\kappa_4myC^{\mu\nu}[\sqrt2\lambdabar^a\sigmabar_{\nu}D_{\mu}\psi^a
+iF^a_{\mu\nu}F^a]+\kappa_5yf^{abc}F^a(C\psi)^b\psi^c
\Bigr\}.\cr}}
Each of the coefficients $m$, $y$, etc above will renormalise 
separately. However, for simplicity when we quote the results for
Feynman diagrams, we will use the values of the coefficients as implied by
Eq.~\lagmass, i.e. $y_{1-5}=y$, $m_0=\mu_1=m$, $\mu_{2-5}=0$,
so that these are effectively        
the renormalised values of these couplings.
Note that the $g_0C^{\mu\nu}F_{\mu\nu}^0\sqrt{\frak2N}
d^{abc}\phibar^a\phibar^b\phibar^c$ and 
$\frak1Ng_0C^{\mu\nu}F_{\mu\nu}^0\phibar^a\phibar^a\phibar^0$
terms only mix with the $d^{abc}\phibar^a\phibar^b\Fbar^c$ or 
$d^{ab0}\phibar^a\phibar^b\Fbar^0$ terms respectively and hence
can be assigned the coupling $y$ or $y_1$ respectively. We emphasise that the 
mass terms in Eq.~\lagmassren\ are 
$\Ncal=\frak12$ invariant although the 
Yukawa terms are not.

The terms with $\kappa_4$ and that with $\kappa_5$ 
are separately $\Ncal=\frak12$ invariant. As with
$\kappa_{1-3}$, we set the renormalised couplings to zero so that 
$\kappa_4=Z_4$, $\kappa_5=Z_5$ 
with $Z_4$, $Z_5$ divergent.
The renormalisation constants $Z_{\phi,\psi}$, $Z_F$
now acquire $y$-dependent contributions, so we have 
\eqn\newZ{
\eqalign{
Z_{\phi}=&1+\left[-\frak14y^2+2g^2(1-\alpha)\right]LN,\cr
Z_{\psi}=&1+\left[-\frak14y^2-2g^2(1+\alpha)\right]LN,\cr
Z_{\phi^0}=&Z_{\psi^0}=1-\frak14y^2LN,\cr
Z_F=&1-\frak14y^2LN.\cr}}
Here we write $\phi^0_B=Z_{\phi^0}^{\frak12}\phi^0$, etc, since the
$U(1)$ chiral fields are now renormalised. There are now several new one-loop
diagrams giving $y^2$-dependent divergent contributions to terms in the
action without superpotential, Eq.~\Sadj\ which are cancelled by 
the $y^2$ terms in 
Eq.~\newZ, but we have not calculated them here; the process of accounting
for these divergences would be similar to that elucidated in Ref.~\jjwa, and
we have preferred to concentrate on the renormalisation of the new terms
in Eq.~\lagmass. Moreover, we have not computed divergent contributions
to terms purely involving 
$\phibar$, $\lambdabar$ and/or $F$, which are individually 
$\Ncal=\frak12$ invariant, nor have we explicitly displayed such terms in the
action Eq.~\lagmassren; these contributions 
would not give any more information about the preservation
of $\Ncal=\frak12$ supersymmetry.

Now for the bare action we also need to replace $m_B=Z_m m$, 
$y_B=Z_{y} y$ etc in addition to the replacements given
earlier. These renormalisation constants are given according to the
non-renormalisation theorem by
\eqn\zlam{\eqalign{
Z_m=&Z_{\Phi}^{-1},\cr
Z_{m_0}=&Z_{\Phi_0}^{-1},\cr
Z_{y}=&Z_{\Phi}^{-\frak32},\cr
Z_{y_1}=&Z_{\Phi}^{-1}Z_{\Phi_0}^{-\frak12},\cr
Z_{y_2}=&Z_{\Phi_0}^{-\frak32},\cr}}
where $Z_{\Phi}$ , $Z_{\Phi_0}$ are the
renormalisation constants for the chiral superfield $\Phi$ given by
\eqn\zPhi{
\eqalign{
Z_{\Phi}=&1+\left[-\frak14y^2+4g^2\right]LN,\cr
Z_{\Phi^0}=&1-\frak14y^2LN.\cr}}
 
The redefinitions of $F$ and $\Fbar$ found in Ref.~\jjw\ need to be
modified in the presence of mass terms and the $U(1)$ gauge group. 
This is easily done following the
arguments of Ref.~\jjwb; there are no one-loop diagrams giving divergent 
contributions to $m\phi F$ or $m\phibar\Fbar$ although there are counterterm
contributions from $m_B\phi_B F$, $m_B\phibar_B\Fbar$. 
At one loop we have 
\eqn\newF{\eqalign{
\Fbar^{a\prime}_B=&\Fbar^{a}_B+(\alpha+3)g^2NL\left(m\phibar^a
+\frak14y d^{abc}\phibar^b\phibar^c\right) 
+\frak12(\alpha+3)y g^2NLd^{ab0}\phibar^b\phibar^0\cr
&+\tau_7f^{abc}(C\psi)^b\psi^c,\cr
\Fbar_B^{0\prime}=&\Fbar^0_B,\cr
F^{a\prime}_B=&Z_FF^a+(\alpha+3)g^2NL\left(m\phi^a
 +\frak14y d^{abc}\phi^b\phi^c\right) 
+\frak12(\alpha+3)y g^2NLd^{ab0}\phi^b\phi^0,\cr
F^{0\prime}_B=&Z_FF^0.\cr}}
Here $\Fbar^{a}_B$ etc are as given in Eq.~\fbarredef, though of course using 
the non-zero $Z_F$ given in Eq.~\newZ. We have also included the term with 
$\tau_7$ which is needed to cancel divergences from Figs.~15 and 16.  
The new $C$-dependent diagrams in the presence of a superpotential 
are depicted in Figs.~7--11, and their divergent contributions in the
corresponding Tables. Note that we do not show Figs. 11(e)--(v) explicitly; 
instead they are described in Appendix A.
We omit diagrams giving contributions of the form 
$A_{\mu}A_{\nu}\phibar^3$ which complete the $F_{\mu\nu}$ in 
$F_{\mu\nu}\phibar^3$
contributions; we already have ample evidence that gauge invariance, even
when apparently violated, can be restored by making divergent field 
redefinitions. 

We now choose the renormalisation constants at our disposal to
ensure finiteness. In order to ensure renormalisability of the action in 
Eq.~\lagmassren, we find we now need to impose specific values for the hitherto
arbitrary coefficients $z_2$, $z_3$, namely
\eqn\zformb{
z_2=-4,\quad z_3=4.}
In other words the effective action 
is rendered finite by adding counterterms
to the fields and parameters in Eq.~\lagmassren, without the need for 
further couplings.
Different choices for $z_2$, $z_3$ would require introducing
additional 
parameters into Eq.~\lagmassren, for instance separate couplings (other than
$m$, $m_0$ respectively)
for the 
$C_1^{\mu\nu}d^{0ab}F_{\mu\nu}^0\phibar^a\phibar^b$ 
and $C_2^{\mu\nu}d^{0ab}F_{\mu\nu}^a\phibar^0\phibar^b$ terms, spoiling the 
$\Ncal=\frak12$ invariance of the mass terms.

We find moreover
\eqn\moreZ{\eqalign{
Z_{y_3}=&1-6LNg^2,\cr
Z_{y_4}=&1-4LNg^2,\cr
Z_{y_5}=&1-2LNg^2,\cr
Z_{\mu_1}=&1+\frak{32}{N}Lg^2\left(1-\frak{g^2}{g_0^2}\right),\cr
Z_{\mu_2}=&-\frak{g^4}{g_0^2}L,\cr
Z_{\mu_3}=&0,\cr
Z_{\mu_4}=&2LNg^2,\cr
Z_{\mu_5}=&4LNg^2,\cr
Z_4=&\frak12LNg^2,\cr
Z_5=&-3LNg^2,\cr
\tau_7=&NLg^2.\cr
}}
We note that $Z_4$ and $Z_5$ are chosen to cancel the divergences from 
Figs.~13, 14 and 12. (We have only computed those terms involving $\kappa_4$ 
in Eq.~\lagmassren\ which contain a derivative; we assume that the others
will be as implied by gauge invariance.)
\vfill
\eject
\newsec{The eliminated formalism in the adjoint case}
It is instructive and also provides a useful check to perform the calculation 
in the eliminated formalism. In the eliminated case Eq.~\lagmassren\ is replaced
by
\eqn\lagmassa{\eqalign{
\tilde S_{\rm{mass}}=&\int d^4x\Bigl\{
-\frak14y d^{abc}\psi^a\psi^b\phi^c
-\frak14y_1d^{ab0}(\psi^a\psi^b\phi^0+2\psi^a\psi^0\phi^b)
-\frak14y_2d^{000}\psi^0\psi^0\phi^0\cr
&-\frak14y d^{abc}\psibar^a\psibar^b\phibar^c       
-\frak14y_1d^{ab0}(\psibar^a\psibar^b\phibar^0
+2\psibar^a\psibar^0\phibar^b)
-\frak14y_2d^{000}\psibar^0\psibar^0\phibar^0                  
\cr
&-m^2\phibar^a\phi^a
-\frak12m\psi^a\psi^a-\frak12m\psibar^a\psibar^a
-m_0^2\phibar^0\phi^0      
-\frak12m_0\psi^0\psi^0-\frak12m_0\psibar^0\psibar^0
\cr
&-\left(\frak14y d^{eab}\phi^a\phi^b+\frak12y_1 d^{ea0}\phi^a\phi^0
+m\phi^e\right)
\left(\frak14y d^{eab}\phibar^a\phibar^b+\frak12y_1 
d^{ea0}\phibar^a\phibar^0+m\phibar^e\right)\cr
&-\left(\frak14y_1 d^{0ab}\phi^a\phi^b+\frak14y_2 d^{000}\phi^0\phi^0
+m_0\phi^0\right)
\left(\frak14y_1 d^{0ab}\phibar^a\phibar^b+
\frak14y_2 d^{000}\phibar^0\phibar^0+m_0\phibar^0\right)\cr
&+\frak16iy C^{\mu\nu}f^{abc}(D_{\mu}\phibar)^a (D_{\nu}\phibar)^b
\phibar^c\cr
&+\frak16igF^d_{\mu\nu}\Bigl(
-\frak12y C^{\mu\nu}d^{abe}d^{cde}\phibar^a\phibar^b\phibar^c
+[2y_3C^{\mu\nu}-3y_1(1-Z_2)C_2^{\mu\nu}]\frak1N
\delta^{ab}\delta^{cd}\phibar^a\phibar^b\phibar^c\cr
&+3[(y_4-y_1)C^{\mu\nu}-\frak12y C_1^{\mu\nu}]
\sqrt{\frak2N}d^{abd}\phibar^0\phibar^a\phibar^b\cr
&+6[y_5C^{\mu\nu}-y_1C_1^{\mu\nu}
-\frak12y_2(1-Z_2)C_2^{\mu\nu}]\frak1N\phibar^0\phibar^0\phibar^d\Bigr)\cr
&-\frak{1}{12}ig_0C^{\mu\nu}F_{\mu\nu}^0\Bigl(y
(1-3Z_3)\sqrt{\frak2N}
d^{abc}\phibar^a\phibar^b\phibar^c
+6y_1(1-2Z_3)
\frak1N\phibar^a\phibar^a\phibar^0\cr
&+2y_2\frak1N\phibar^0\phibar^0\phibar^0\Bigr)\cr
&-\frak12i\Bigl\{gmd^{abc}C^{\mu\nu}F^c_{\mu\nu}\phibar^a \phibar^b
+g_0m_0(1-2Z_3)C^{\mu\nu}d^{ab0}F^0_{\mu\nu}\phibar^a \phibar^b\cr
&+g\left[mC_1^{\mu\nu}
+m_0(1-2Z_2)C_2^{\mu\nu}\right]
d^{ab0}F^b_{\mu\nu}\phibar^a \phibar^0\Bigr\}\cr
&+g^2|C|^2\bigl[-\frak18(\mu_1-2m)f^{ace}f^{bde}
+\mu_2\frak2N\delta^{ab}\delta^{cd}\bigr]
\phibar^a\phibar^b\lambdabar^c\lambdabar^d\cr
&+gd^{abc}\sqrt{2N}|C|^2\phibar^a\lambdabar^b\left(\mu_3g_0\phibar^c\lambdabar^0
+g\mu_4\phibar^0\lambdabar^c\right)
+\frak2N\mu_5gg_0|C|^2\phibar^a\phibar^0\lambdabar^a\lambdabar^0\cr
&+\kappa_4myC^{\mu\nu}[\sqrt2\lambdabar^a\sigmabar_{\nu}D_{\mu}\psi^a
-imF^a_{\mu\nu}\phibar^a]-\kappa_5f^{abc}y(m\phibar^a
+\frak14yd^{acd}\phibar^c\phibar^d)(C\psi)^b\psi^c
\Bigr\}\cr
}}
while we simply strike out the terms involving $F$, $\Fbar$ in Eq.~\Sadj.
Once again note that in quoting diagrammatic results we set
$y_{1-5}=y$, $m_0=\mu_1=m$, $\mu_{2-5}=0$, so that these are effectively
the renormalised values of these couplings. In Table~7, the contributions from 
Figs.~7(f-k) are now absent while those from
Figs.~7(l-r) change sign. Similarly, in Table~8, the contributions from 
Figs.~8(e-p) are now absent while those from Figs.~8(q-dd) change sign.
In Table~9, the contributions from Figs.~9(f-o) are now absent while
those from Figs.~9(p-z) change sign. In Table~10, the contribution from 
Fig.~10(d) is now absent. In Table~11, the contributions from 
Figs.~11(j-o) are now absent while those from Figs.~11(p-v) which contain 
two factors of $d^{abc}$ acquire an additional factor of 
$\left(-\frak12\right)$. 
Again, note that we do not show Figs. 11(e)--(v) explicitly,
instead describing them in Appendix A.
Figs.~12 and 13 are no longer present, of course,
while the result of Fig.~14 is still given by Eq.~(A.27).
In Table~15, the contribution from Fig.~15(b) is absent while that from
15(c) changes sign, and 15(d) must now also be considered.
 Likewise in Table~16, the contribution from Fig.~16(b) is 
absent while that from
16(c) changes sign, and 16(e) must now also be considered.
 The net divergent contribution from Figs.~15 and 16 
is therefore unchanged. 
The results from Figs.~7--11 and Figs.~14--16 now add to
\eqn\sumsixelim{\eqalign{
\Gamma_{7\rm{1PI elim}}^{(1)\rm{pole}}=
&iLg^2C^{\mu\nu}m\Bigl[-\frak12(7+5\alpha)Ngd^{abc}\pa_{\mu}A_{\nu}^a\phibar^b
\phibar^c\cr
&+3(1-\alpha)g\sqrt{2N}\pa_{\mu}A_{\nu}^a\phibar^a\phibar^0
-2(5+\alpha)g_0\sqrt{2N}\pa_{\mu}A_{\nu}^0\phibar^a\phibar^a\Bigr],\cr
\Gamma_{8\rm{1PI elim}}^{(1)\rm{pole}}=&
iLg^4C^{\mu\nu}mf^{abe}A_{\mu}^aA_{\nu}^b\Bigl[
\frak12(5+3\alpha)Nd^{cde}\phibar^c
\phibar^d+2\alpha\sqrt{2N}\phibar^e\phibar^0\Bigr],\cr
\Gamma_{9\rm{1PIelim}}^{(1)\rm{pole}}=&
|C|^2mL\Bigl\{\bigl[
2\frak{g^2}{g_0^2}\delta^{ab}\delta^{cd}
+\left\{\frak12N(3+\alpha)+\frak4N\left(1-\frak{g^2}{g_0^2}\right)\right\}
f^{ace}f^{bde}\bigr]
g^4\phibar^a\phibar^b\lambdabar^c\lambdabar^d\cr
&-2g^4d^{abc}\sqrt{2N}\phibar^a\lambdabar^b\phibar^0\lambdabar^c
-8g^3g_0\phibar^a\phibar^0\lambdabar^a\lambdabar^0\Bigr\},\cr
\Gamma_{10\rm{1PI elim}}^{(1)\rm{pole}}=&
\Gamma_{10\rm{1PI}}^{(1)\rm{pole}},\cr
\Gamma_{11\rm{1PI elim}}^{(1)\rm{pole}}=&iC^{\mu\nu}yg^2L\Bigl(
-\frak12g\left(3+\frak73\alpha\right)Nf^{abe}f^{cde}\pa_{\mu}\phibar^a
\phibar^b\phibar^cA_{\nu}^d\cr
&+\left[-\left(\frak34+\frak{7}{12}\alpha\right)d^{abe}d^{cde}
+\left(\frak52-\frak76\alpha\right)\delta^{ab}\delta^{cd}\right]
g\phibar^a\phibar^b\phibar^c\pa_{\mu}A_{\nu}^d\cr
&-\frak14(7+5\alpha)g\sqrt{2N}d^{abc}\phibar^0\phibar^a\phibar^b
\pa_{\mu}A_{\nu}^c+\frak32(1-\alpha)g\phibar^0\phibar^0\phibar^a 
\pa_{\mu}A_{\nu}^a\cr
&-\frak12(5+\alpha)g_0\sqrt{2N}d^{abc}\phibar^a\phibar^b\phibar^c\pa_{\mu}A_{\nu}^0
-2(5+\alpha)g_0\phibar^a\phibar^a\phibar^0\pa_{\mu}A_{\nu}^0\Bigr),\cr
\Gamma_{14\rm{1PI elim}}^{(1)\rm{pole}}=&
\Gamma_{14\rm{1PI}}^{(1)\rm{pole}},\cr
\Gamma_{15\rm{1PI elim}}^{(1)\rm{pole}}=&
-3iLNg^2myf^{abc}\phibar^a                      
(C\psi)^b\psi^c,\cr
\Gamma_{16\rm{1PI elim}}^{(1)\rm{pole}}=&
-\frak34LNg^2y^2f^{abe}d^{cde}(C\psi)^a\psi^b\phibar^c\phibar^d,\cr
\Gamma_{17\rm{1PI elim}}^{(1)\rm{pole}}=&\frak12im^2ygLC^{\mu\nu}F_{\mu\nu}^a
\phibar^a.}}
respectively. The results in Eq.~\moreZ\ are unchanged, which is a very good
check on the calculation.

\newsec{$\Ncal=\frak12$ supersymmetric theory with chiral matter
in the fundamental representation}
We now turn to the case of the $\Ncal=\frak12$ supersymmetric theory with 
chiral matter in the fundamental representation. As we saw
in Ref.~\jjwa, in this case renormalisability combined with
$\Ncal=\frak12$ supersymmetry requires us to consider 
an $SU(N)\otimes U(1)$ gauge 
theory. The action (with no superpotential) is given by\jjwa
\eqn\lagranb{\eqalign{
S_0=&\int d^4x
\Bigl[-\frak14F^{\mu\nu A}F^A_{\mu\nu}-i\lambdabar^A\sigmabar^{\mu}
(D_{\mu}\lambda)^A+\frak12D^AD^A\cr
&-\frak12iC^{\mu\nu}d^{ABC}e^{ABC}F^A_{\mu\nu}\lambdabar^B\lambdabar^C\cr
&+\frak18g^2|C|^2d^{abe}d^{cde}(\lambdabar^a\lambdabar^b)
(\lambdabar^c\lambdabar^d)
+\frak{1}{4N}\frak{g^4}{g_0^2}|C|^2(\lambdabar^a\lambdabar^a)
(\lambdabar^b\lambdabar^b)\cr
&+\Bigl\{
\Fbar F -i\psibar\sigmabar^{\mu}D_{\mu}\psi-D^{\mu}\phibar D_{\mu}\phi\cr
&+\phibar \Dhat\phi +
i\sqrt2(\phibar \lambdahat\psi-\psibar\lambdahatbar\phi)\cr
&+\sqrt2C^{\mu\nu}D_{\mu}\phibar\lambdahatbar\sigmabar_{\nu}\psi
+iC^{\mu\nu}\phibar \Fhat_{\mu\nu}F
-\frak{1}{4}|C|^2\phibar \lambdahatbar\lambdahatbar
F\cr
&+\frak{1}{N}\gamma_1g_0^2|C|^2(\lambdabar^a\lambdabar^a)
(\lambdabar^0\lambdabar^0)\cr
&-\gamma_2 C^{\mu\nu}g\left(
\sqrt2D_{\mu}\phibar\lambdabar^aR^a\sigmabar_{\nu}\psi+\sqrt2\phibar
\lambdabar^aR^a
\sigmabar_{\nu}D_{\mu}\psi+i\phibar F^a_{\mu\nu}R^aF\right)\cr
&-\gamma_3 C^{\mu\nu}g_0\left(
\sqrt2D_{\mu}\phibar\lambdabar^0R^0\sigmabar_{\nu}\psi+\sqrt2\phibar
\lambdabar^0R^0
\sigmabar_{\nu}D_{\mu}\psi+i\phibar F^0_{\mu\nu}R^0F\right)\cr
&+(\phi\go\phitil,\psi\go\psitil,F\go\Ftil,R^A\go -(R^A)^*,C^{\mu\nu}\go
-C^{\mu\nu})\Bigr\}\Bigr],\cr}} 
where $\gamma_{1-3}$ are constants, and 
\eqn\afdef{\eqalign{
D_{\mu}\phi=&\pa_{\mu}\phi+i\Ahat_{\mu}\phi,\cr
(D_{\mu}\lambda)^A=&\pa_{\mu}\lambda^A-gf^{ABC}A_{\mu}^B\lambda^C,\cr
\quad F_{\mu\nu}^A=&\pa_{\mu}A_{\nu}^A-\pa_{\nu}A_{\mu}^A-gf^{ABC}
A_{\mu}^BA_{\nu}^C,\cr}}
(with similar expressions for $D_{\mu}\phitil$, $D_{\mu}\psi$,
$D_{\mu}\psitil$). Here
\eqn\hatdefs{
\Ahat_{\mu}=\Ahat_{\mu}^AR^A=gA_{\mu}^aR^a+g_0A_{\mu}^0R^0,}
with similar definitions for $\lambdahat$, $\Dhat$, $\Fhat_{\mu\nu}$.
We also have
\eqn\etensor{
e^{abc}=g,\quad e^{a0b}=e^{ab0}=e^{000}=g_0,\quad e^{0ab}={g^2\over{g_0}}.}
We include a multiplet $\{\phi,\psi,F\}$ transforming according to the
fundamental representation of $SU(N)\otimes U(1)$ and, 
to ensure anomaly cancellation, a 
multiplet $\{\phitil,\psitil,\Ftil\}$ transforming according to its conjugate.
The change $C^{\mu\nu}\go-C^{\mu\nu}$ for the conjugate representation is due 
to the fact that the anticommutation relations for the conjugate fundamental
representation 
differ by a sign from those for the fundamental representation. 
The group matrices $R^A$ 
for the fundamental representation of $SU(N)\otimes U(1)$ are as defined in 
Section 4.  
They satisfy
\eqn\comrel{ [R^A,R^B]=if^{ABC}R^C,\quad
\{R^A,R^B\}=d^{ABC}R^C,}
where $d^{ABC}$ is totally symmetric. 
We note that $d^{ab0}=\sqrt{\frak2N}\delta^{ab}$, $d^{000}=\sqrt{\frak2N}$. 

It is easy to show that Eq.~\lagranb\ is invariant under 
\eqn\newsusy{
\eqalign{
\delta A^A_{\mu}=&-i\lambdabar^A\sigmabar_{\mu}\epsilon\cr
\delta \lambda^A_{\alpha}=&i\epsilon_{\alpha}D^A+\left(\sigma^{\mu\nu}\epsilon
\right)_{\alpha}\left[F^A_{\mu\nu}
+\frak12iC_{\mu\nu}e^{ABC}d^{ABC}\lambdabar^B\lambdabar^C\right],\quad
\delta\lambdabar^A_{\alphadot}=0,\cr
\delta D^A=&-\epsilon\sigma^{\mu}D_{\mu}\lambdabar^A,\cr
\delta\phi=&\sqrt2\epsilon\psi,\quad\delta\phibar=0,\cr
\delta\psi^{\alpha}=&\sqrt2\epsilon^{\alpha} F,\quad 
\delta\psibar_{\alphadot}=-i\sqrt2(D_{\mu}\phibar)
(\epsilon\sigma^{\mu})_{\alphadot},\cr
\delta F=&0,\quad
\delta \Fbar=-i\sqrt2D_{\mu}\psibar\sigmabar^{\mu}\epsilon
-2i\phibar\epsilon\lambdahat+2C^{\mu\nu}D_{\mu}(\phibar\epsilon\sigma_{\nu}
\lambdahatbar).\cr}}
The terms involving $\gamma_{1-3}$ are 
separately invariant under $\Ncal=\frak12$ supersymmetry and must
be included to obtain a renormalisable Lagrangian. In fact only the 
$\gamma_{1,2}$ terms were required in the case without a superpotential\jjwa;
to ensure renormalisability in the massive case we need to include
the $\gamma_3$ terms and also modify $\gamma_2$, with a corresponding change to
the bare gaugino $\lambda_B$ (see later).

We now consider the problem of adding superpotential terms to the 
Lagrangian Eq.~\lagranb. Again, this problem is most succinctly addressed by
returning to the superfield formalism whence the $\Ncal=\frak12$ action was
originally derived.
Denoting fundamental (anti-fundamental) chiral superfield representations as 
$\Phi$ ($\wt \Phi$)
it is simple to see that
$$
\int d^2 \theta \, \wt \Phi * \Phi \quad + \quad  
\int d^2 \bth\,\ov \Phi * \ov{\wt \Phi} 
$$
is gauge invariant, since under a gauge transformation we have 
$$
\Phi \to \Omega * \Phi, \quad \wt \Phi \to \wt \Phi *  \Omega^{-1}.
$$

In the \none\ case an interaction term is possible for the group $SU(3)$, 
i.e.
$$
\int d^2\theta\,\epsilon_{a b c} \Phi_1^a \Phi_2^b \Phi_3^c  
\quad + \quad\hbox{c.c.} 
$$
This construction does not, however, generalise to the \nhalf\ case, because 
of the non-anticommutative product. 

We may express the superfields in terms of component fields as follows:

\eqna\comps$$\eqalignno{
\Phi (y, \th) &= \phi (y) + \sqrt{2} \th \psi (y) + \th\th
F (y) & \comps a\cr
\wt \Phi (y, \th) &= \wt \phi  (y) + \sqrt{2} \th
\wt \psi (y) + \th\th \wt F  (y) & \comps b\cr
\ov \Phi (\ov y, \bth) &= \ov \phi (\ov y) + \sqrt{2} \bth \ov
\psi (\ov y) \cr
&\quad+ \bth\bth \Big( \ov F (\ov y) + i C^{\mu\nu} \partial_{\mu}
(\ov \phi  A_{\nu})(\ov y) -
\frak{1}{4} C^{\mu\nu} \ov \phi  A_{\mu} A_{\nu} (\ov y) \Big) & \comps c\cr
\ov {\wt \Phi}  (\ov y, \bth) &= \ov {\wt \phi}  (\ov y)
+ \sqrt{2} \bth \ov {\wt \psi}  (\ov y) \cr
&\quad+ \bth\bth \Big( \ov
{\wt F}  (\ov y) + i C^{\mu\nu} \partial_{\mu} (\ov {\wt \phi} 
A_{\nu})(\ov y) - \frak{1}{4} C^{\mu\nu} \ov {\wt \phi}  A_{\mu} A_{\nu} (\ov y)
\Big), & \comps d \cr}$$
where $\ov y^{\mu} = y^{\mu} - 2 i \th \sigma^{\mu} \bth$. Note the 
modification of the $\bth\bth$-term\araki.

We thus obtain

\eqna\expans$$\eqalignno{
m\int d^2 \theta \,\Phi * \wt \Phi  
&=  m\left[\phi \wt F + F \wt \phi  -\psi \wt\psi      \right] & \expans a \cr
m\int  d^2 \bth \,\ov \Phi * \ov{\wt \Phi} &=  m\left[\ov \phi \ov{\wt F} + 
\ov F \ov{\wt \phi} -\ov \psi \ov{\wt\psi}  
+ i C^{\mu\nu} \ov \phi {\hat F}_{\mu\nu}\ov{\wt\phi}\right] & \expans b \cr
}$$
In fact, the most general
mass term is in components
\eqn\lagmassb{\eqalign{
S_{\rm{mass}}=&m\int d^4x\Bigl[(\phi\Ftil+F\phitil 
-\psi\psitil)
+\hbox{h.c.}+iC^{\mu\nu}\phibar \Fhat_{\mu\nu}\phitilbar\cr
&-\frak{1}{8}|C|^2d^{ABC}\phibar R^A\lambdahatbar^B\lambdahatbar^C
\phitilbar\Bigr].\cr}}
As in the adjoint case, the coefficient
of the final term in Eq.~\lagmassb\ is arbitrary since it is separately
$\Ncal=\frak12$ invariant; we make a particular choice for similar 
reasons, as 
explained after Eq.~(B.6) in Appendix B.
This final term can also
be expressed in superfields but in a more  unwieldy form. 

%

The one-loop one-particle-irreducible (1PI) 
graphs contributing to the new terms (those
containing $C$) in the absence of a superpotential were 
given in Ref.~\jjwa; the new diagrams in the presence of the 
mass terms are depicted in Figs.~7--9. The divergent contributions from these
diagrams are listed in Appendix B.

\newsec{Renormalisation of the action in the fundamental case}
The renormalisations
of $\lambda$, $F$ and $\Fbar$, which are non-linear as in the adjoint case,
will be given later. (Note that
$F$ is unrenormalised in the absence of Yukawa superpotential terms.)
The renormalisations of the remaining fields and couplings are linear as 
usual and given by
\eqn\barea{\eqalign{
\lambdabar^a_B=Z_{\lambda}^{\frak12}\lambdabar^a,
\quad \lambdabar^0_B=Z_{\lambda^0}^{\frak12}\lambdabar^0,\quad&
A^{a}_{\mu B}=Z_A^{\frak12}A^{a}_{\mu},\quad A^{0}_{\mu B}
=Z_{A^0}^{\frak12}A^0_{\mu},\cr
D^a_B=Z_D^{\frak12}D^a,\quad& D^0_B=Z_{D^0}^{\frak12}D^0,\cr  
\phi_B=Z_{\phi}^{\frak12}\phi,
\quad \psi_B=Z_{\psi}^{\frak12}\psi,
\quad &\phibar_B=Z_{\phi}^{\frak12}\phibar,
\quad \psibar_B=Z_{\psi}^{\frak12}\psibar,\cr
\quad g_B=Z_gg,\quad g_{0B}=&Z_{g_0}g_0,
\quad m_B=Z_m m,\cr
\gamma_{1-3B}=\tZ_{1-3}, \quad 
C_B^{\mu\nu}=&Z_CC^{\mu\nu}, \quad |C|_B^2=Z_{|C|^2}|C|^2,
\cr}} 
with similar expressions for $\phitil_B$, $\psitil_B$ etc. 
In Eq.~\barea, $\tZ_{1-3}$ are divergent contributions, in other words
we have set the renormalised couplings $\gamma_{1-3}$ to zero for
simplicity. The other renormalisation constants start with 
tree-level values of 1. As we mentioned before,
the renormalisation constants for the fields
and for the gauge couplings $g$, $g_0$ are the same as in the ordinary $\Ncal=1$
supersymmetric theory and are therefore given up to one loop 
by \ref\timj{
D.~Gross and F.~Wilcek, \prd8 (1973) 3633\semi
D.R.T.~Jones, \npb87 (1975) 127}: 
\eqn\Zgga{\eqalign{
Z_{\lambda}=&1-g^2L(2\alpha N +2),\cr
Z_A=&1+g^2L[(3-\alpha)N-2]\cr
Z_D=&1-2g^2L,\cr
Z_g=&1+g^2L\left(1-3N\right),\cr
Z_{\phi}=&1+2(1-\alpha)L\Chat_2,\cr
Z_{\psi}=&1-2(1+\alpha)L\Chat_2,\cr
Z_m=&Z_{\Phi}^{-1},\cr
Z_{\Phi}=&1+4L\Chat_2,\cr}}
where 
\eqn\ctildef{
\Chat_2=g^2R^aR^a+g_0^2R^0R^0=\frak12\left(Ng^2+\frak1N\Delta\right)}
with
\eqn\deltadefn{
\Delta=g_0^2-g^2.}
(For the gauge multiplet, 
the renormalisation constants given in Eq.~\Zgga\ are those corresponding to the
$SU(N)$ sector of the $U(N)$ theory; those for the $U(1)$ sector, namely
$Z_{\lambda^0}$, $Z_{A^0}$, $Z_{D^0}$
 and $Z_{g_0}$, are given by omitting the terms in
$N$ and replacing $g$ by $g_0$.) 
In Eq.~\Zgga, $Z_{\Phi}$ is the
renormalisation constant for the chiral superfield $\Phi$ so that the   
result for $m_B$ is the consequence of the non-renormalisation theorem.
For later convenience we write
(denoting for instance the $n$-loop
contribution to $\tZ_1$ by $\tZ_1^{(n)}$) 
\eqn\zforma{
\tZ^{(1)}_1=\tz_1L}
with similar expressions for $\tZ_{2,3}$.
The renormalisation of $\lambda^a$ is given by 
\eqn\lchangea{\eqalign{
\lambda_B^a=&Z_{\lambda}^{\frak12}\lambda^a 
-\frak12NLg^3C^{\mu\nu}d^{abc}
\sigma_{\mu}\lambdabar^cA_{\nu}^b
-NLg^2g_0C^{\mu\nu}d^{ab0}
\sigma_{\mu}\lambdabar^0A_{\nu}^b\cr
&+i\sqrt2Lg\rho_1[\phibar R^a(C\psi)
+(C\psitil)R^a\bar{\tilde\phi}],\cr
\lambda_B^0=&iZ_{\lambda^0}^{\frak12}\lambda^0+
i\sqrt2Lg_0\rho_2[\phibar R^0(C\psi)+(C\psitil)R^0\bar{\tilde\phi}],\cr
}}
where $(C\psi)^{\alpha}=C^{\alpha}{}_\beta\psi^{\beta}$. Here $\rho_{1,2}$
are divergent parameters to be defined later. Note that the
renormalisation of $\lambda^a$ required in the case of the 
fundamental representation is different from that required in the 
adjoint case. 
The replacement
of $\lambda$ by $\lambda_B$ produces a change in the action given (to
first order) by
\eqn\Schlamb{\eqalign{
S_0(\lambda_B)-S_0(\lambda)=&L\int d^4x\Bigl\{
\rho_1\sqrt2gC^{\mu\nu}(\phibar\lambda^aR^a\sigmabar_{\nu}D_{\mu}\psi
+D_{\mu}\phibar\lambda^aR^a\sigmabar_{\nu}\psi)\cr
&+\rho_2\sqrt2g_0C^{\mu\nu}(\phibar\lambda^0R^0\sigmabar_{\nu}D_{\mu}\psi
+D_{\mu}\phibar\lambda^0R^0\sigmabar_{\nu}\psi)\cr
&+(\phi\go\phitil,\psi\go\psitil,R^A\go -(R^A)^*,C^{\mu\nu}\go
-C^{\mu\nu})+\ldots\Bigr\},\cr}}
where the ellipsis indicates the terms not involving $\rho_1$, $\rho_2$
(which were given previously in Ref.~\jjwa).

The final term in Eq.~\lagmassb\ may be decomposed into four terms 
each of which are separately gauge and $\Ncal=\frak12$ invariant and
hence can (and do) renormalise separately. Consequently, in order to 
consider the renormalisation of the theory we need to replace Eq.~\lagmassb\
by 
\eqn\mlagmassnew{\eqalign{
S_{\rm{mass}}=&\int d^4x\Bigl\{m(\phi\Ftil+F\phitil
-\psi\psitil)
+\hbox{h.c.}+imC^{\mu\nu}\phibar \Fhat_{\mu\nu}\phitilbar\cr
&-\frak14|C|^2\phibar \Bigl(\frak12\mu_1g^2d^{abc}R^c\lambdabar^a\lambdabar^b
+\frak{1}{2N}\mu_2g^2\lambdabar^a\lambdabar^a\cr
&+2\mu_3gg_0R^aR^0\lambdabar^a\lambdabar^0
+\mu_4g_0^2R^0R^0\lambdabar^0\lambdabar^0\Bigr)
\phitilbar\Bigr\},\cr}}
where each of $\mu_{1-4}$ will renormalise separately.
However, for simplicity when we quote results for Feynman diagrams, we
use the values of the coefficients as implied by Eq.~\lagmassb, i.e. $\mu_{1-4}
=m$; so that we are setting the renormalised values of $\mu_{1-4}$ to be $m$.
In contrast to the adjoint case, where it was impossible to
maintain $\Ncal=\frak12$ invariance for the Yukawa terms at the quantum 
level, in the fundamental case where only mass terms are allowed we shall
find that $\Ncal=\frak12$ invariance can be preserved.

The redefinitions of $F$ and $\Fbar$ found in Ref.~\jjw\ need to be
modified in the presence of mass terms. As in the adjoint case
this is readily done following the
arguments of Ref.~\jjwb. However, note that due 
to the afore-mentioned change in sign for the $\phibar\lambdabar\lambdabar F$
term, the result for Fig.~8 in Ref.~\jjwb\ is modified to
\eqn\sumeight{\eqalign{
\Gamma_{8\rm{1PI}}^{(1)\rm{pole}}=
&L|C|^2\phibar\bigl\{g^2\left[\frak18(13-2\alpha)Ng^2-2\Chat_2\right]
\lambdabar^a\lambdabar^bd^{abc}R^c\cr
&+gg_0\left[\frak12(13-\alpha)Ng^2-8\Chat_2\right]
\lambdabar^0\lambdabar^aR^0R^a\cr
&-\left[2\Chat_2+\frak14\alpha Ng^2\right]
g^2d^{0bc}R^0\lambdabar^b\lambdabar^c
-4g_0^2\Chat_2\lambdabar^0\lambdabar^0R^0R^0\bigr\}F.
\cr}}      
We find
\eqn\fbarredef{\eqalign{
\Fbar_B=&\Fbar+(\alpha+3)mL\Chat_2\phitil+
L\Bigl\{\Bigl[\left(7Ng^2+2(1+\alpha) \Chat_2+2\tz_2\right)
g\pa_{\mu}A_{\nu}^a\cr
&-\left(\frak{15}{4}Ng^2+(1+\alpha)\Chat_2+\tz_2\right)
g^2f^{abc}A_{\mu}^bA_{\nu}^c\Bigr]iC^{\mu\nu}\phibar R^a\cr
&+2\left((1+\alpha)\Chat_2+\tz_3\right)
g_0\pa_{\mu}A_{\nu}^0iC^{\mu\nu}\phibar R^0\cr
&+\frak{1}{8}|C^2|\Bigl[\left(-19Ng^2+(17-\alpha)\Chat_2\right)
g^2d^{abc}\phibar R^c\lambdabar^a\lambdabar^b\cr
&+4\left(-16Ng^2+(17-\alpha)\Chat_2\right)
gg_0\phibar \lambdabar^0\lambdabar^aR^0R^a\cr
&+2(17-\alpha)\Chat_2
g_0^2\phibar \lambdabar^0\lambdabar^0R^0R^0
+\left(-6Ng^2+(17-\alpha)\Chat_2\right)g^2
d^{ab0}\phibar R^0\lambdabar^a\lambdabar^b\Bigr]\Bigr\},\cr
F_B=&F+(\alpha+3)mL\Chat_2\phitilbar.\cr}}
Again, note that these are different from the corresponding results in
the adjoint representation (Eq.~\newF).
We now find that with
\eqn\zformsa{\eqalign{
Z^{(1)}_C=&Z_{|C|^2}^{(1)}=0,\quad \tz_1=-3Ng^2, \quad
\tz_2=8(2\Chat_2-Ng^2), \cr
\tz_3=&4\left(\left[4-2\frak{\Delta}{g_0^2}\right]
\Chat_2-Ng^2\right),
\quad \rho_1=1+\tz_2,\quad  \rho_2=\tz_3, 
\cr
Z_{\mu_1}=&1+\left[\left(44+64\frak{g^2}{g_0^2}\right)\Chat_2
-\left(28+32\frak{g^2}{g_0^2}\right)Ng^2\right]L,\cr
Z_{\mu_2}=&1+\left[\left(44+128\frak{g^2}{g_0^2}\right)\Chat_2
-\left(28+32\frak{g^2}{g_0^2}\right)Ng^2\right]L,\cr
Z_{\mu_3}=&1+(44\Chat_2-30Ng^2)L,\cr
Z_{\mu_4}=&1+44\Chat_2L,\cr
\cr
}}
the one-loop effective action is finite. In fact the massless theory
is finite for
arbitrary choices of $\tz_2$, $\tz_3$; the particular values chosen are 
necessary to ensure renormalisability of the mass terms in Eq.~\mlagmassnew,
in analogy to the adjoint case. 
As explained earlier, by renormalisability we mean that
the massive theory is rendered finite by adding counterterms
to the fields and parameters in Eq.~\mlagmassnew, without the need for
further parameters. The $\Ncal=\frak12$
invariance is thereby retained. Different choices for $\tz_2$, $\tz_3$ would 
require introducing additional 
parameters into Eq.~\mlagmassnew, specifically a separate coupling (other
than $m$) for the 
$C^{\mu\nu}\phibar \Fhat_{\mu\nu}\phitilbar$ term, spoiling the 
$\Ncal=\frak12$ invariance.

\newsec{The eliminated formalism in the fundamental case}
Once again it is a useful check to perform the calculation 
in the eliminated formalism. In the eliminated case Eq.~\mlagmassnew\ is replaced
by
\eqn\lagmassab{\eqalign{
\tilde S_{\rm{mass}}=&\int d^4x\Bigl\{-m^2(\phibar\phi+\phitil\phitilbar)
-m(\psi\psitil+\psitilbar\psibar)\cr
&
-imC^{\mu\nu}\phibar \left[(1-2\gamma_2)gF^a_{\mu\nu}R^a
+(1-2\gamma_3)g_0F^0_{\mu\nu}R^0\right]\phitilbar\cr
&-\frak14|C|^2\phibar 
\Bigl(\frak12(\mu_1-2m)g^2d^{abc}R^c\lambdabar^a\lambdabar^b 
+\frak{1}{2N}(\mu_2-2m)g^2\lambdabar^a\lambdabar^a\cr
&+2(\mu_3-2m)gg_0R^aR^0\lambdabar^a\lambdabar^0
+(\mu_4-2m)g_0^2R^0R^0\lambdabar^0\lambdabar^0\Bigr)
\phitilbar\Bigr\}.\cr}}
while we simply strike out the terms involving $F$, $\Fbar$ in Eq.~\lagranb.
In Table~17, the contributions from Figs.~7(f-k) are now absent while those from
Figs.~7(l-r) change sign. Similarly, in Table~18, the contributions from
Figs.~8(e-p) are now absent while those from Figs.~8(q-dd) change sign.
In Table~19, the contributions from Figs.~9(f-o) are now absent while
those from Figs.~9(p-z) change sign. 
The results from Figs.~7, 8 and 9 now add to
\eqn\sumelim{\eqalign{
\Gamma_{B7\rm{1PIelim}}^{(1)\rm{pole}}=&
iLgC^{\mu\nu}\pa_{\mu}A_{\nu}^A\phibar R^A\Bigl[\left\{-(68+4\alpha)
+32\frak{\Delta}{g_0^2}\delta^{A0}\right\}
\Chat_2\cr
&+\left\{(29-\alpha)c^A+16\delta^{A0}\right\}Ng^2\Bigr]
\phitilbar,\cr
\Gamma_{B8\rm{1PIelim}}^{(1)\rm{pole}}=&
iLNg^2C^{\mu\nu}f^{abc}A_{\mu}^aA_{\nu}^b\phibar 
\left[2(17+\alpha)\Chat_2-(13-\alpha)Ng^2\right]R^c\phitilbar,\cr
\Gamma_{B9\rm{1PIelim}}^{(1)\rm{pole}}=&\phibar\Bigl(
\left\{\left[\frak14(25+\alpha)+8\frak{g^2}{g_0^2}
\right]\Chat_2-\left[\frak14(11-\alpha)
+4\frak{g^2}{g_0^2}\right]Ng^2\right\}g^2d^{abc}R^c\lambdabar^a\lambdabar^b\cr
+&\left\{\left[\frak14(25+\alpha)+16\frak{g^2}{g_0^2}
\right]\Chat_2-\left[\frak14(11-\alpha)
+4\frak{g^2}{g_0^2}\right]Ng^2\right\}
\frak1Ng^2\lambdabar^a\lambdabar^a\cr
&+\left\{(25+\alpha)\Chat_2-\frak12(27-\alpha)g^2N\right\}gg_0R^aR^0
\lambdabar^a\lambdabar^0\cr
&+\frak12(25+\alpha)g_0^2\Chat_2R^0R^0\lambdabar^0\lambdabar^0\Bigr)
\phitilbar,\cr
}}
where $c^A=1-\delta^{A0}$.
The results in Eq.~\zformsa\ are again unchanged, giving a convincing
check on the calculation.
 
\newsec{Conclusions}
We have repeated our earlier one-loop analysis of $\Ncal=\frak12$ supersymmetry
for the case of chiral matter in the adjoint representation.
We have constructed an $\Ncal=\frak12$ invariant set of mass terms and an
$\Ncal=\frak12$ invariant set of Yukawa terms 
for this case. The $\Ncal=\frak12$ invariance of the Yukawa terms
requires that the chiral matter be in the adjoint representation of 
$U(N)$ rather than $SU(N)$ at the classical level. However, once we consider
quantum corrections, the $U(1)$ chiral fields will renormalise differently
from the $SU(N)$ fields and so at the quantum level we are obliged to
consider $SU(N)\otimes U(1)$ rather than $U(N)$.  On the other hand, the
$\Ncal=\frak12$ transformations mix superpotential terms with different
kinds of field ($SU(N)$ or $U(1)$) and so it is clear that the 
$\Ncal=\frak12$ invariance of the Yukawa terms cannot be preserved 
at the quantum level. This is because separate couplings must be introduced
for most of the $C$-dependent superpotential terms. The only remaining vestige
of $\Ncal=\frak12$ supersymmetry is that since  
the $g_0C^{\mu\nu}F_{\mu\nu}^0\sqrt{\frak2N}
d^{abc}\phibar^a\phibar^b\phibar^c$ and
$\frak1Ng_0C^{\mu\nu}F_{\mu\nu}^0\phibar^a\phibar^a\phibar^0$
terms only mix with the $d^{abc}\phibar^a\phibar^b\Fbar^c$ or
$d^{ab0}\phibar^a\phibar^b\Fbar^0$ fields respectively, they 
can be assigned the coupling $y$ or $y_1$ respectively which are already in the 
$\Ncal=1$ part of the theory.
In contrast, we have shown
that the $\Ncal=\frak12$ invariance of the mass terms is preserved  
at the one-loop level. 
However the invariance is assured by making a
particular choice of the parameters $\kappa_2$, $\kappa_3$
(in Eq.~\Sbare), as determined by Eq.~\zformb. 
This also implies (through Eq.~\zforms) a particular choice of renormalisation
for the gaugino $\lambda$, parametrised by $\tau_5$  (in Eq.~\lchange). 

We have also 
constructed a set of mass terms for the $\Ncal=\frak12$ supersymmetric
theory with chiral matter in the fundamental representation, and we have shown
that the one-loop renormalisation presents similar features to the adjoint case.
The $\Ncal=\frak12$ invariance is preserved
at one loop since the Yukawa terms
which presented difficulties in the adjoint case are absent, leaving only the 
mass terms.
Once again the invariance is assured by making a
particular parameter choice, in this case of
the parameters $\gamma_2$, $\gamma_3$
(in Eq.~\lagranb) combined with a particular choice of renormalisations
for the gaugino
$\lambda$, parametrised by $\rho_1$, $\rho_2$ (in Eq.~\lchangea). These choices
were listed in Eq.~\zformsa.

The necessity for the above choices in both the fundamental and adjoint 
cases seems somewhat counterintuitive as these
renormalisations are all present in the theory without 
superpotential and yet there
appeared to be nothing in the theory without       
superpotential to enforce these choices. The fact that the same feature 
appears in both cases is at least an indication that this really is a generic
property of the theory. However 
it would be reassuring if some independent confirmation could be found for
these particular values. Presumably the necessity for the non-linear
renormalisations we are compelled to make lies in our use of a
non-supersymmetric gauge (the obvious choice when working in components, of
course). So the answer to this puzzle might lie in a close scrutiny of the
gauge-invariance Ward identities. Of course a calculation in superspace
would also be illuminating. It is always tempting to investigate whether
the behaviour at one loop persists to higher orders but the proliferation
of diagrams in this case would almost certainly be prohibitive.

\centerline{{\bf Acknowledgements}}\nobreak

LAW was supported by PPARC through a Graduate Studentship. 

\appendix{A}{Results for one-loop diagrams}
In this Appendix we list the divergent contributions from the one-loop
diagrams. 

The contributions from the graphs shown in Fig.~1 are of the form
\eqn\formone{
\sqrt2Ng^2g_BLC^{\mu\nu}d^{ABC}\left(\pa_{\mu}\phibar^A X_1^{ABC}\lambdabar^B 
\sigmabar_{\nu}\psi^C
+\phibar^A Y_1^{ABC}\lambdabar^B \sigmabar_{\nu}\pa_{\mu}\psi^C\right)  }
where 
$X_1^{ABC}$ and $Y_1^{ABC}$ consist of a number $X_1$, $Y_1$ multiplying a 
tensor structure formed of a product of terms like $c^A$ or $d^A$,
where $c^A=1-\delta^{A0}$, $d^A=1+\delta^{A0}$.
 The 
$X_1$, $Y_1$ and the tensor structures are given separately in Table~1. 
(The contributions from Figs.~2--4, 7, 8 also involve tensors 
$X^{ABC\ldots}_i$, $Y_i^{ABC\ldots}$ etc (for Fig.~$i$)
which can be decomposed similarly and will be similarly presented.)
\vbox{
\begintable
Fig.| $X_1$| $Y_1$|Tensor\cr
1a|$\frak32$|$-\alpha $|$c^Ac^Bd^C$\cr
1b|$\alpha $|$\alpha $|$c^Ac^Bd^C$\cr
1c|$\alpha $|$0$|$d^Ac^Bc^C$\cr
1d|$1$|$-1$|$c^Ac^Bd^C$\cr
1e|$1$|$0$|$d^Ac^Bc^C$\cr
1f|$-\frak12 (1-2\alpha)$|$0$|$c^Ad^Bc^C$\cr
1g|$1$|$0$|$c^Ad^Bc^C$\cr
1h|$1$|$-1$|$c^Ad^Bc^C$\cr
1i|$0$|$1$|$c^Ad^Bc^C$\cr
1j|$-3$|$0$|$c^A$
\endtable}
\centerline{{\it Table~1:\/} Contributions from Fig.~1}
\bigskip

The sum of the contributions from Table~1 can be written in the form
\eqn\sumone{\eqalign{  
\Gamma_{1\rm{1PI}}^{(1)\rm{pole}}=&
Ng^2\sqrt2LC^{\mu\nu}\Bigl[(2+3\alpha)gd^{abc}\pa_{\mu}\phibar^a 
\lambdabar^b\sigmabar_{\nu}\psi^c
-gd^{abc}\phibar^a \lambdabar^b \sigmabar_{\nu}\pa_{\mu}\psi^c  \cr
& +2(1+\alpha)gd^{ab0}\pa_{\mu}\phibar^a \lambdabar^b\sigmabar_{\nu}\psi^0
-2gd^{ab0}\phibar^a \lambdabar^b \sigmabar_{\nu}\pa_{\mu}\psi^0\cr
& +2\alpha g_0d^{ab0}\pa_{\mu}\phibar^a \lambdabar^0\sigmabar_{\nu}\psi^b\cr
&+2(1+\alpha)gd^{ab0}\pa_{\mu}\phibar^0 \lambdabar^a\sigmabar_{\nu}\psi^b
\Bigr] \cr}}

The contributions from the graphs shown in Fig.~2 are of the form
\eqn\formtwo{\eqalign{
\sqrt2
&g^3g_CNLC^{\mu\nu}
A_{\mu}^A\phibar^B\lambdabar^C \sigmabar_{\nu}\psi^D
\Bigl(X^{ABCD}_2f^{BAE}d^{CDE}\cr
&+Y^{ABCD}_2f^{DAE}d^{CBE}+Z^{ABCD}_2f^{BDE}d^{CAE}\Bigr)\cr}}
where $g_c\equiv g$. The $X_2$, $Y_2$, $Z_2$ and tensor products in the
decomposition of $X_2^{ABCD}$, $Y_2^{ABCD}$ and $Z_2^{ABCD}$ (as described 
earlier) are shown in Table~2:
\vfill
\eject
\vbox{
\begintable
Fig.| $X_2$| $Y_2$| $Z_2$| Tensor\cr
2a|$\frak12$|$\frak12$|$-\frak12$|$c^Ac^Bc^Cd^D$\cr
2b|$\frak12$|$\frak12$|$\frak12$|$c^Ac^Bc^Cd^D$\cr
2c|$1$|$-1$|$1$|$c^Ac^Bc^Cd^D$\cr
2d|$-1$|$-1$|$-1$|$c^Ac^Bc^Cd^D$\cr
2e|$1$|$0$|$0$|$c^Ac^Bc^Cc^D$\cr
2f|$-\frak14(1-\alpha)$|$\frak14(1-\alpha)$|$-\frak14(1-\alpha)$
|$c^Ac^Bd^Cc^D$\cr
2g|$\frak12$|$\frak12$|$\frak12$|$c^Ac^Bd^Cc^D$\cr
2h|$\frak12\alpha$|$0$|$0$|$c^Ac^B$\cr
2i|$\frak34\alpha$|$0$|$0$|$c^Ac^B$\cr
2j|$-\frak34(3+\alpha)$|$0$|$0$|$c^Ac^B$\cr
2k|$\frak18\alpha$|$-\frak18\alpha$|$\frak18\alpha$|$c^Ac^Bd^Cc^D$\cr
2l|$-\frak38(1-\alpha)$|$-\frak18(1-\alpha)$|$\frak18(1-\alpha)$
|$c^Ac^Bd^Cc^D$\cr
2m|$\frak12\alpha$|$\frak12\alpha$|$-\frak12\alpha$|$c^Ac^Bd^Cc^D$\cr
2n|$\frak12\alpha$|$-\frak12\alpha$|$\frak12\alpha$|$c^Ac^Bc^Cd^D$\cr
2o|$-\frak14\alpha$|$-\frak14\alpha$|$\frak14\alpha$|$c^Ac^Bd^Cc^D$\cr
2p|$\frak38(3+\alpha)$|$-\frak18(3+\alpha)$|$\frak18(3+\alpha)$
|$c^Ac^Bc^Cd^D$\cr
2q|$\alpha$|$0$|$0$|$c^Ac^Bc^Cc^D$\cr
2r|$-\frak14\alpha$|$\frak14\alpha$|$-\frak14\alpha$|$c^Ac^Bc^Cd^D$\cr
2s|$\frak34(1+\alpha)$|$\frak34(1+\alpha)$|$-\frak34(1+\alpha)$|$c^Ac^Bc^Cd^D$\cr
2t|$-\frak12\alpha$|$-\frak12\alpha$|$-\frak12\alpha$|$c^Ac^Bc^Cd^D$\cr
2u|$\frak12\alpha$|$\frak12\alpha$|$\frak12\alpha$|$c^Ac^Bc^Cd^D$\cr
2v|$-\frak38\alpha$|$-\frak38\alpha$|$\frak38\alpha$|$c^Ac^Bc^Cd^D$\cr
2w|$-\frak14(3+\alpha)$|$-\frak14(3+\alpha)$|$-\frak14(3+\alpha)$
|$c^Ac^Bd^Cc^D$\cr
2x|$\frak12$|$\frak12$|$-\frak12$|$c^Ac^Bd^Cc^D$\cr
2y|$1$|$-1$|$1$|$c^Ac^Bd^Cc^D$
\endtable}
\centerline{{\it Table~2:\/} Contributions from Fig.~2}
\vfill
\eject
\vbox{
\begintable
Fig.| $X_2$| $Y_2$| $Z_2$| Tensor\cr
2z|$\frak14(2+\alpha)$|$\frak14(2+\alpha)$|$\frak14(2+\alpha)$
|$c^Ac^Bd^Cc^D$\cr
2aa|$\frak14\alpha$|$\frak14\alpha$|$-\frak14\alpha$|$c^Ac^Bd^Cc^D$\cr
2bb|$\frak34$|$-\frak14$|$\frak14$|$c^Ac^Bd^Cc^D$\cr
2cc|$-\frak14\alpha$|$-\frak14\alpha$|$\frak14\alpha$|$c^Ac^Bd^Cc^D$
\endtable}
\centerline{{\it Table~2:\/} Contributions from Fig.~2 (continued)}
\bigskip

The sum of the contributions from Table~2 can be written in the form
\eqn\sumtwo{\eqalign{
\Gamma_{2\rm{1PI}}^{(1)\rm{pole}}=&\sqrt2
g^3LC^{\mu\nu}A_{\mu}^a\Bigl[
\left(\frak72(1+\alpha)f^{bae}d^{cde}-f^{dae}d^{cbe}+\frak12f^{bde}
d^{cae}\right)Ng
\phibar^b\lambdabar^c \sigmabar_{\nu}\psi^d\cr
&-\frak12(1+5\alpha)\sqrt{2N}g_0
f^{abc}\phibar^b\lambdabar^0 \sigmabar_{\nu}\psi^c
-\frak12(7+5\alpha)\sqrt{2N}gf^{abc}\phibar^b\lambdabar^c \sigmabar_{\nu}\psi^0
\Bigr]\cr
}}

The contributions from Fig.~3 are of the form
\eqn\formthree{
ig^3NLC^{\mu\nu}(\pa_{\mu}A^A_{\nu}
\phibar^B X_3^{ABC}F^C+A^A_{\nu}\pa_{\mu}\phibar^B Y_3^{ABC}F^C)d^{ABC}}
where the $X_3$, $Y_3$ and tensor products in the decomposition of
$X_3^{ABC}$ and $Y_3^{ABC}$ are given in Table 3:
\bigskip
\vbox{
\begintable
Fig.|$X_3$|$Y_3$|Tensor\cr
3a|0|$3$|$c^Ac^Bd^C$\cr
3b|0|$-2$|$c^Ac^Bd^C$\cr
3c|$1$|$1$|$c^Ac^Bd^C$\cr
3d|$-(5+\alpha)$|0|$c^A$\cr
3e|$2\alpha $|$-2$|$c^Ac^Bd^C$   
\endtable}
\centerline{{\it Table~3:\/} Contributions from Fig.~3}

The contributions from Table~3 add to
\eqn\sumthree{\eqalign{
\Gamma_{3\rm{1PI}}^{(1)\rm{pole}}=&
iNg^3LC^{\mu\nu}\Bigl[-(4-\alpha)d^{abc}
\phibar^b \pa_{\mu}A^a_{\nu}F^c\cr
&-3(1-\alpha)d^{ab0}\phibar^a\pa_{\mu}A^b_{\nu}
F^0-
(5+\alpha)d^{ab0}\phibar^0\pa_{\mu}A^a_{\nu}F^b\Bigr].\cr}}

The contributions from Fig.~4 are of the form
\eqn\formfour{
ig^4NLC^{\mu\nu}A^A_{\mu}A^B_{\nu}(X_4^{ABCD}f^{ABE}d^{CDE}
+Y_4^{ABCD}f^{ACE}d^{BDE})\phibar^CF^D}
where the $X_4$ and $Y_4$ and tensor products in the usual decomposition
are given in Table 4:
\bigskip
\vbox{
\begintable
Fig.|$X_4$|$Y_4$|Tensor\cr
4a|$-\frak34\alpha$|$0$|$c^Ac^Bc^Cd^D$\cr
4b|$\frak12\alpha$|$\alpha$|$c^Ac^Bc^Cd^D$\cr
4c|$-\frak12\alpha$|$-\alpha$|$c^Ac^Bc^Cd^D$\cr
4d|$0$|$0$|$c^Ac^Bc^Cd^D$\cr
4e|$\frak14(2+\alpha)$|$2+\alpha$|$c^Ac^Bc^Cd^D$\cr
4f|$-\frak12$|$1$|$c^Ac^Bc^Cd^D$\cr
4g|$-\frak32\alpha$|$0$|$c^Ac^B$\cr
4h|$\frak32(1+\alpha)$|$0$|$c^Ac^B$\cr
4i|$-\frak14(3+\alpha)$|$-(3+\alpha)$|$c^Ac^Bc^Cd^D$\cr
4j|$\frak12\alpha$|$0$|$c^Ac^Bc^Cd^D$\cr
4k|$-\frak34\alpha$|$0$|$c^Ac^Bc^Cd^D$\cr
4l|$0$|$0$|
\endtable}
\centerline{{\it Table~4:\/} Contributions from Fig.~4}

The contributions from Table~4 add to
\eqn\sumfour{\eqalign{
\Gamma_{4\rm{1PI}}^{(1)\rm{pole}}=&ig^4
LC^{\mu\nu}A^a_{\mu}A^b_{\nu}
\Bigl(\frak14(3-4\alpha)Nf^{abe}d^{cde}\phibar^cF^d\cr
&-2\alpha\sqrt{2N}f^{abc}\phibar^cF^0
+\frak32\sqrt{2N}f^{abc}\phibar^0F^c\Bigr).\cr}}

The contributions from Fig.~5 are of the form
\eqn\formfive{
X_5^{ABCD}|C|^2g^2g_Cg_DL\phibar^A\lambdabar^C\lambdabar^DF^B}
where $X_5^{ABCD}$ is given in Table~5.
In Table~5 we have introduced the notation $(\Dtil^A)^{BC}=d^{ABC}$.
Using results from Appendix D, the contributions from Table~5 add to
\eqn\sumfive{\eqalign{
\Gamma_{5\rm{1PI}}^{(1)\rm{pole}}=&
g^4L|C|^2\Bigl[-\frak{1}{2}(3+\alpha)Nf^{ace}f^{bde}
+\frak{11}{8}Nd^{abe}d^{cde}\cr
&-\frak12\delta^{ab}\delta^{cd}-\frak12\delta^{ac}\delta^{bd}\Bigr]
\phibar^a\lambdabar^c\lambdabar^dF^b\cr
&+d^{abc}g^3L|C|^2\sqrt{2N}\Bigl[g\phibar^0\lambdabar^a\lambdabar^bF^c
+3g_0\phibar^a\lambdabar^b\lambdabar^0F^c
+2g\phibar^a\lambdabar^b\lambdabar^cF^0\Bigr]\cr
&+4g^3g_0L|C|^2\left(\phibar^0\lambdabar^0\lambdabar^aF^a
+2\phibar^a\lambdabar^a\lambdabar^0F^0\right).\cr}}
\bigskip
\vbox{
\begintable
Fig.|$X_5^{ABCD}$\cr
5a|0\cr
5b|$4\tr[\Ftil^A\Ftil^C\Dtil^B\Dtil^D]$\cr
5c|$-2\tr[\Ftil^A\Dtil^C\Ftil^D\Dtil^B]$\cr
5d|$-\alpha Nd^Cc^Dc^Xd^{ABX}d^{CDX}$\cr
5e|$(1+\alpha)Nc^Xd^{ABX}d^{CDX}$\cr
5f|$-\frak12N\alpha c^Ad^Bc^Xd^{ABX}d^{CDX}$\cr
5g|$0$\cr
5h|$2\alpha \tr[\Ftil^C\Ftil^A\Dtil^B\Dtil^D]$\cr
5i|$-\frak12(3+\alpha)\tr[\Ftil^A\Ftil^B\Ftil^D\Ftil^C]$\cr
5j|$\frak12\alpha (\tr[\Ftil^C\Ftil^A\Ftil^D\Ftil^B]-\frak12
Nf^{XAC}f^{XBD})$\cr
5k|$\tr[\Ftil^A\Ftil^C\Ftil^D\Ftil^B]+\frak12
Nf^{XAC}f^{XBD}$
\endtable}
\centerline{{\it Table~5:\/} Contributions from Fig.~5}

The divergent contributions
to the effective action from the graphs in Fig. 6 are of the form
\eqn\formsixa{
iLNg^3X_6C^{\alpha\beta}d^{abe}f^{cde}\phibar^a
\phibar^b\psi^c_{\alpha}\psi^d_{\beta}}
where the contributions from the
individual graphs to $X_6$ are given in Table 6:
\bigskip
\vbox{
\begintable
Fig.|$X_6$\cr
6a|$0$\cr
6b|$0$\cr
6c|$\alpha$\cr
6d|$-\alpha$\cr
6e|$-1$
\endtable}
\centerline{{\it Table~6:\/} Contributions from Fig.~6}
\bigskip

The contributions from Table~6 add to
\eqn\sumsixa{
\Gamma_{6\rm{1PI}}^{(1)\rm{pole}}
=-iLNg^3C^{\alpha\beta}d^{abe}f^{cde}\phibar^a
\phibar^b\psi^c_{\alpha}\psi^d_{\beta}.}

The divergent contributions
to the effective action from the graphs in Fig. 7 are of the form
\eqn\formsix{
imLNg^2g_AX^{ABC}_7C^{\mu\nu}d^{ABC}\pa_{\mu}A_{\nu}^A\phibar^B
\phibar^C}
where the contributions from the
individual graphs to $X_7$ and the associated tensors in the usual 
decomposition are given in Table 7:
\bigskip
\vbox{
\begintable
Fig.|$X_7$|Tensor\cr
7a|$2$|$c^Ac^Bd^C$\cr
7b|$-1$|$c^Ac^Bd^C$\cr
7c|$-1$|$c^Ac^Bd^C$\cr
7d|$0$|\cr
7e|$-4$|$d^Ac^Bc^C$\cr
7f|$-2\alpha$|$d^Ac^Bc^C$\cr
7g|$-2$|$d^Ac^Bc^C$\cr
7h|$-\frak12$|$c^Ac^Bd^C$\cr
7i|$-1$|$c^Ac^Bd^C$\cr
7j|$-(1+2\alpha)$|$c^Ac^Bd^C$\cr
7k|$\frak32$|$c^Ac^Bd^C$\cr
7l|$-\frak12(5+\alpha)$|$c^A$\cr
7m|$\alpha$|$d^Ac^Bc^C$\cr
7n|$1$|$d^Ac^Bc^C$\cr
7o|$\frak12$|$c^Ac^Bd^C$\cr
7p|$1$|$c^Ac^Bd^C$\cr
7q|$1+2\alpha$|$c^Ac^Bd^C$\cr
7r|$-\frak32$|$c^Ac^Bd^C$
\endtable}
\centerline{{\it Table~7:\/} Contributions from Fig.~7}
\bigskip
These results add to
\eqn\sumsix{\eqalign{
\Gamma_{7\rm{1PI}}^{(1)\rm{pole}}=
&-\frak12(5+\alpha)iLg^2C^{\mu\nu}m\Bigl[3Ngd^{abc}\pa_{\mu}A_{\nu}^a\phibar^b 
\phibar^c\cr
&+2g\sqrt{2N}\pa_{\mu}A_{\nu}^a\phibar^a\phibar^0
+4g_0\sqrt{2N}\pa_{\mu}A_{\nu}^0\phibar^a\phibar^a\Bigr].\cr}}
(Note that the contributions from Figs.~7(h-k) cancel those from 
Figs.~7(o-r) respectively.)

The divergent contributions
to the effective action from the graphs in Fig. 8 are of the form
\eqn\formseven{
imLNg^4X_8^{ABCD}C^{\mu\nu}f^{ABE}d^{CDE}A_{\mu}^AA_{\nu}^B\phibar^C
\phibar^D}
where the contributions from the
individual graphs to $X_8$ and the associated tensors in the usual 
decomposition are given in Table 8:
\bigskip
\vbox{
\begintable
Fig.|$X_8$|Tensor\cr
8a|$-2$|$c^Ac^Bc^Cc^D$\cr
8b|$1$|$c^Ac^Bc^Cc^D$\cr
8c|$1$|$c^Ac^Bc^Cc^D$\cr
8d|$2$|$c^Ac^Bc^Cc^D$\cr
8e|$\alpha$|$c^Ac^Bc^Cc^D$\cr
8f|$1$|$c^Ac^Bc^Cc^D$\cr
8g|$-\frak14(3+\alpha)$|$c^Ac^Bc^Cd^D$\cr
8h|$0$|\cr
8i|$0$|\cr
8j|$1$|$c^Ac^Bc^Cd^D$\cr
8k|$\frak34\alpha$|$c^Ac^Bc^Cd^D$\cr
8l|$-\frak12\alpha$|$c^Ac^Bc^Cd^D$\cr
8m|$\frak34\alpha$|$c^Ac^Bc^Cd^D$\cr
8n|$\frak14(2+\alpha)$|$c^Ac^Bc^Cd^D$\cr
8o|$0$|\cr
8p|$0$|\cr
8q|$-\frak34\alpha$|$c^Ac^B$\cr
8r|$\frak34(1+\alpha)$|$c^Ac^B$\cr
8s|$-\frak12\alpha$|$c^Ac^Bc^Cc^D$\cr
8t|$-\frak12$|$c^Ac^Bc^Cc^D$
\endtable}
\centerline{{\it Table~8:\/} Contributions from Fig.~8}
\bigskip  
\vbox{
\begintable
Fig.|$X_8$|Tensor\cr
8u|$\frak14(3+\alpha)$|$c^Ac^Bc^Cd^D$\cr
8v|$0$|\cr
8w|$0$|\cr
8x|$-1$|$c^Ac^Bc^Cd^D$\cr
8y|$-\frak34\alpha$|$c^Ac^Bc^Cd^D$\cr
8z|$\frak12\alpha$|$c^Ac^Bc^Cd^D$\cr
8aa|$-\frak34\alpha$|$c^Ac^Bc^Cd^D$\cr
8bb|$-\frak14(2+\alpha)$|$c^Ac^Bc^Cd^D$\cr
8cc|$0$|\cr
8dd|$0$|
\endtable}
\centerline{{\it Table~8:\/} Contributions from Fig.~8 (continued)}
\bigskip
These results add to
\eqn\sumseven{
\Gamma_{8\rm{1PI}}^{(1)\rm{pole}}=
iLg^4C^{\mu\nu}mf^{abe}A_{\mu}^aA_{\nu}^b\Bigl[
\frak14(13+2\alpha)Nd^{cde}\phibar^c 
\phibar^d+\frak32\sqrt{2N}\phibar^e\phibar^0\Bigr]}
(Note that the contributions from Figs.~8(g-p) cancel those from
Figs.~8(u-dd) respectively.)

The contributions from Fig.~9 are of the form
\eqn\formeight{
X_9^{ABCD}g^2g_Cg_DmL|C|^2\phibar^A\phibar^B\lambdabar^C\lambdabar^D.}
The contributions from the individual graphs to 
$X_9^{ABCD}$ are given in Table~9.
The results in Table~9 add to
\eqn\sumeight{\eqalign{
\Gamma_{9\rm{1PI}}^{(1)\rm{pole}}=&
|C|^2mL\Bigl\{\bigl[\frak{11}{8}Nd^{abe}d^{cde}
-\frak12\left(1-4\frak{g^2}{g_0^2}
\right)\delta^{ab}\delta^{cd}
-\frak12\delta^{ad}\delta^{bc}\cr
&+\frak4N\left(1-\frak{g^2}{g_0^2}\right)f^{ace}f^{bde}\bigr]
g^4\phibar^a\phibar^b\lambdabar^c\lambdabar^d\cr
&+g^3d^{abc}\sqrt{2N}\phibar^a\lambdabar^b\left(3g_0\phibar^c\lambdabar^0
+g\phibar^0\lambdabar^c\right)
+4g^3g_0\phibar^a\phibar^0\lambdabar^a\lambdabar^0\Bigr\}.\cr}}
(Note that the contributions from Figs.~9(h--m) cancel those from 
Figs.~9(u--z); this is analogous to the situation with
Figs.~7 and 8, and is a consequence of our choice of coefficient for
the last term in Eq.~\lagmass.) 
\bigskip
\vbox{
\begintable   
Fig.|$X_9^{ABCD}$\cr
9a|$\frak12\alpha \tr[\Ftil^A\Ftil^B\Dtil^C\Dtil^D]$\cr
9b|$\frak12\tr[\Ftil^A\Ftil^B\Dtil^C\Dtil^D]$\cr
9c|$\frak12(3+\alpha)\tr[\Ftil^A\Ftil^B\Dtil^C\Dtil^D]$\cr
9d|$-\alpha \tr[\Ftil^A\Ftil^B\Dtil^C\Dtil^D]$\cr
9e|$Nd^{ABE}d^{CDE}c^Ac^Bc^Cc^Dc^E-2c^Cc^D\tr[\Ftil^A\Ftil^B\Dtil^C\Dtil^D]
+\frak4Nf^{ACE}f^{BDE}$
\crnorule
|$+\frak{2g^2}{g_0^2}\left(c^Ac^Bc^Cc^D\delta^{AB}\delta^{CD}
-\frak2Nf^{ACE}f^{BDE}\right)$\cr
9f|$\frak12\alpha \tr[\Ftil^A\Ftil^B\Ftil^C\Ftil^D]$\cr
9g|$\frak12\tr[\Ftil^A\Ftil^B\Ftil^C\Ftil^D]$\cr
9h|$-\tr[\Ftil^A\Ftil^B\Ftil^C\Ftil^D]-\frak12Nf^{ACE}f^{BDE}$\cr
9i|$-\frak12\alpha \left(\tr[\Ftil^A\Ftil^C\Ftil^B\Ftil^D]
-\frak12Nf^{ACE}f^{BDE}\right)$\cr
9j|$-2\alpha\tr[\Ftil^C\Ftil^A\Dtil^B\Dtil^D]$\cr
9k|$-4\tr[\Ftil^C\Ftil^A\Dtil^D\Dtil^B]$\cr
9l|$\frak12\alpha Nc^Ad^Bc^Ed^{ABE}d^{CDE}$\cr
9m|$2\tr[\Ftil^A\Dtil^C\Ftil^D\Dtil^B]$\cr
9n|$0$\cr
9o|$0$\cr
9p|$-\frak12\alpha Nd^{ABE}d^{CDE}d^Cc^Dc^E$\cr
9q|$\frak12(1+\alpha)Nd^{ABE}d^{CDE}c^E$\cr
9r|$-\frak14(3+\alpha)\tr[\Ftil^A\Ftil^B\Ftil^C\Ftil^D]$\cr
9s|$-\frak14\alpha \tr[\Ftil^A\Ftil^B\Ftil^C\Ftil^D]$\cr
9t|$-\frak14\tr[\Ftil^A\Ftil^B\Ftil^C\Ftil^D]$\cr
9u|$\tr[\Ftil^A\Ftil^B\Ftil^C\Ftil^D]+\frak12Nf^{ACE}f^{BDE}$\cr
9v|$\frak12\alpha\left(\tr[\Ftil^A\Ftil^C\Ftil^B\Ftil^D]
-\frak12Nf^{ACE}f^{BDE}\right)$\cr
9w|$2\alpha\tr[\Ftil^C\Ftil^A\Dtil^B\Dtil^D]$\cr
9x|$4\tr[\Ftil^C\Ftil^A\Dtil^D\Dtil^B]$
\endtable}
\centerline{{\it Table~9:\/} Contributions from Fig.~9}
\vfill
\eject
\vbox{
\begintable
Fig.|$X_9^{ABCD}$\cr
9y|$-\frak12\alpha Nc^Ad^Bc^Ed^{ABE}d^{CDE}$\cr
9z|$-2\tr[\Ftil^A\Dtil^C\Ftil^D\Dtil^B]$\cr
9aa|$0$\cr
9bb|$0$
\endtable}
\centerline{{\it Table~9:\/} Contributions from Fig.~9 (continued)}
\bigskip
The results from Fig.~10 are of the form
\eqn\formnine{
Ny g^2LX_{10}C^{\mu\nu}f^{abc}\pa_{\mu}\phibar^a\pa_{\nu}\phibar^b\phibar^c}
and the contributions from the individual graphs to $X_{10}$ are given in 
Table~10.
\bigskip  
\vbox{
\begintable
Fig.|$X_{10}$\cr
10a|$\frak12\alpha$\cr
10b|$0$\cr
10c|$2$\cr
10d|$0$\cr
10e|$-\frak32$
\endtable}
\centerline{{\it Table~10:\/} Contributions from Fig.~10}
\bigskip   
The results in Table~10 add to
\eqn\sumnine{\eqalign{
\Gamma_{10\rm{1PI}}^{(1)\rm{pole}}=&
\frak12 Ny g^2LC^{\mu\nu}(1+\alpha)
f^{abc}\pa_{\mu}\phibar^a\pa_{\nu}\phibar^b\phibar^c
\cr}}
We have not explicitly drawn most of 
the diagrams (labelled Fig.~(11a,b$\ldots$)) giving contributions of the form
\eqn\formten{
iC^{\mu\nu}y g^2g_D
L\left(X_{11}^{ABCD}\pa_{\mu}\phibar^A\phibar^B\phibar^CA_{\nu}^D
+Y_{11}^{ABCD}\phibar^A\phibar^B\phibar^C\pa_{\mu}A_{\nu}^D\right),}
since they can be obtained by adding external scalar lines to the diagrams 
of Fig.~7. Thus Figs.~11(e-o) are obtained from Figs.~7(a-k) by adding an 
external scalar ($\phibar$) line at the position of the
cross. Figs.~11(p-v) are obtained from Figs.~7(l-r) by adding an 
external scalar ($\phibar$) line at the position of the crossed circle. 
The remaining
Figs.~11(a-d) are depicted in Fig.~11. The 
individual contributions to $X_{11}^{ABCD}$ and 
$Y_{11}^{ABCD}$ in Eq.~\formten\
are given in Table~11.
\bigskip
\vbox{
\begintable
Fig.|$X_{11}^{ABCD}$|$Y_{11}^{ABCD}$\cr
11a|$-\alpha\tr[\Ftil^A\Ftil^B\Ftil^C\Ftil^D]$|$0$\cr
11b|$-\tr[\Ftil^A\Ftil^B\Ftil^C\Ftil^D]$|$0$\cr
11c|$\frak12\alpha(\tr[\Ftil^A\Ftil^B\Ftil^C\Ftil^D]-\frak12
Nf^{ABE}f^{CDE})$|$0$\cr
11d|$0$|$0$\cr
11e|$-4\tr[\Dtil^B\Dtil^A\Ftil^C\Ftil^D]$|$0$\cr
11f|$2\tr[\Dtil^A\Dtil^B\Ftil^C\Ftil^D]$|$0$\cr
11g|$2\tr[\Dtil^A\Ftil^B\Dtil^C\Ftil^D]$|$0$\cr
11h|$-2Nf^{ABE}f^{CDE}$|$0$\cr
11i|$0$|$-4\tr[\Dtil^B\Ftil^C\Dtil^D\Ftil^A]$\cr
11j|$0$|$-2\alpha\tr[\Ftil^B\Ftil^C\Dtil^D\Dtil^A]$\cr
11k|$0$|$-2\tr[\Ftil^B\Ftil^C\Dtil^D\Dtil^A]$\cr
11l|$\tr[\Dtil^{(A}\Dtil^{B)}\Ftil^C\Ftil^D]$|$0$\cr
11m|$2\tr[\Ftil^D\Ftil^A\Dtil^B\Dtil^C]$|$0$\cr
11n|$2\tr[\Dtil^B\Dtil^C\Ftil^D\Ftil^A]$|$-2\alpha\tr[\Dtil^B\Dtil^C\Ftil^D\Ftil^A]$\cr
11o|$-3\tr[\Dtil^B\Dtil^C\Ftil^D\Ftil^A]$|$0$\cr
11p|$0$|$\frak16(5+\alpha)\left(\tr[\Ftil^A\Ftil^B\Ftil^C\Ftil^D]
-Nc^Dd^{ABE}d^{CDE}\right)$\cr
11q|$\frak23\alpha\left(\tr[\Ftil^{[A}\Ftil^{B]}\Ftil^C\Ftil^D]
-\frak12Nf^{ABE}f^{CDE}\right)$|
$\frak13\alpha\left(4\tr[\Ftil^A\Ftil^B\Dtil^C\Dtil^D]
+Nc^Bc^Cd^Ed^{ADE}d^{BCE}\right)$\cr
11r|$\tr[\Ftil^A\Ftil^B\Ftil^C\Ftil^D]$
|$\frak13\tr[\Ftil^A\Ftil^B\Ftil^C\Ftil^D]$\crnorule
|$$|$+\frak13\left(4\tr[\Ftil^A\Ftil^B\Dtil^C\Dtil^D]
+Nc^Bc^Cd^Ed^{ADE}d^{BCE}\right)$\cr
11s|$\frak13\left((3+\alpha)Nf^{ABE}f^{CDE}-\tr[\Ftil^A\Ftil^B\Ftil^C\Ftil^D]
\right)$
|$\frak16(1+\alpha)\tr[\Ftil^A\Ftil^B\Ftil^C\Ftil^D]$\crnorule
|$+\frak16\left(4\tr[\Ftil^A\Ftil^D\Dtil^B\Dtil^C]+c^Ac^Dd^ENd^{ADE}d^{BCE}\right)$
|$+\frak16\left(4\tr[\Ftil^A\Ftil^D\Dtil^B\Dtil^C]+c^Ac^Dd^ENd^{ADE}d^{BCE}\right)$\cr
11t|$\frak16\left(4\tr[\Ftil^A\Ftil^B\Ftil^C\Ftil^D]-\alpha Nf^{ABE}f^{CDE}
\right)$|$-\frak13\alpha\tr[\Ftil^A\Ftil^B\Ftil^C\Ftil^D]$\crnorule
|$-\frak13\left(4\tr[\Ftil^D\Ftil^A\Dtil^B\Dtil^C]
+Nc^Ac^Dd^Ed^{ADE}d^{BCE}\right)$|
\endtable}
\centerline{{\it Table~11:\/} Contributions from Fig.~11}
\bigskip
\vbox{
\begintable
Fig.|$X_{11}^{ABCD}$|$Y_{11}^{ABCD}$\cr
11u|$\tr[\frak16\left\{(4+3\alpha)\Ftil^A\Ftil^B-
3\alpha\Ftil^B\Ftil^A\right\}\Ftil^C\Ftil^D]
$|$-\frak16\alpha\tr[\Ftil^A\Ftil^B\Ftil^C\Ftil^D]$\crnorule
|$-\frak14\alpha Nf^{ABE}f^{CDE}$|
$+\frak13\alpha
\left(4\tr[\Ftil^D\Ftil^A\Dtil^B\Dtil^C]+Nc^Ac^Dd^Ed^{ADE}d^{BCE}\right)$\crnorule
|$-\frak13\left(4\tr[\Ftil^D\Ftil^A\Dtil^B\Dtil^C]+Nc^Ac^Dd^Ed^{ADE}d^{BCE}\right)$|\cr
11v|$-\tr[\Ftil^A\Ftil^B\Ftil^C\Ftil^D]+\frak32Nf^{ABE}f^{CDE}$|$0$\crnorule
|$+\frak12\left(4\tr[\Ftil^D\Ftil^A\Dtil^B\Dtil^C]+Nc^Ac^Dd^Ed^{ADE}d^{BCE}\right)$|
\endtable}
\centerline{{\it Table~11:\/} Contributions from Fig.~11 (continued)}
\bigskip
The results sum to
\eqn\sumten{\eqalign{
\Gamma_{11\rm{1PI}}^{(1)\rm{pole}}=&iC^{\mu\nu}y g^2L\Bigl(
-\frak12g\left(3+\frak73\alpha\right)Nf^{abe}f^{cde}\pa_{\mu}\phibar^a
\phibar^b\phibar^cA_{\nu}^d\cr
&+\left[-\left(\frak54-\frak16\alpha\right)Nd^{abe}d^{cde}
+\left(3+\frak73\alpha\right)\delta^{ab}\delta^{cd}\right]
g\phibar^a\phibar^b\phibar^c\pa_{\mu}A_{\nu}^d\cr
&-\frak12(9+\alpha)g\sqrt{2N}d^{abc}\phibar^0\phibar^a\phibar^b
\pa_{\mu}A_{\nu}^c-(5+\alpha)g\phibar^0\phibar^0\phibar^a 
\pa_{\mu}A_{\nu}^a\cr
&-2g_0\sqrt{2N}d^{abc}\phibar^a\phibar^b\phibar^c\pa_{\mu}A_{\nu}^0
-8g_0\phibar^a\phibar^a\phibar^0\pa_{\mu}A_{\nu}^0\Bigr).\cr
}}

The divergent contributions
to the effective action from the graphs in Fig. 12 are of the form
\eqn\formtwelve{
iLNg^2yX_{12}f^{abc}F^a                      
(C\psi)^b\psi^c}                       
where the contributions from the
individual graphs to $X_{12}$ are given in Table 12:                            
\bigskip
\vbox{  
\begintable
Fig.|$X_{12}$\cr
12a|$1$\cr
12b|$1$
\endtable}
\centerline{{\it Table~12:\/} Contributions from Fig.~12}
\bigskip

The contributions from Table~12 add to
\eqn\sumtwelve{
\Gamma_{12\rm{1PI}}^{(1)\rm{pole}}
=2iLNg^2yf^{abc}F^a
(C\psi)^b\psi^c.}

The divergent contributions
to the effective action from the graphs in Fig. 13 are of the form
\eqn\formthirteen{
iLNg^2myX_{13}C^{\mu\nu}F^a_{\mu\nu}F^a,}
where the contributions from the
individual graphs to $X_{13}$ are given in Table 13:
\bigskip
\vbox{
\begintable
Fig.|$X_{13}$\cr
13a|$-1$\cr
13b|$\frak12$
\endtable} 
\centerline{{\it Table~13:\/} Contributions from Fig.~13}
\bigskip

The contributions from Table~13 add to
\eqn\sumthirteen{
\Gamma_{13\rm{1PI}}^{(1)\rm{pole}}
=-\frak12iLNg^2myC^{\mu\nu}F^a_{\mu\nu}F^a.}

The divergent contribution from Fig.~14 is
\eqn\figfourteen{
-{1\over{\sqrt2}}LNg^2myC^{\mu\nu}\lambdabar^a\sigmabar_{\nu}\pa_{\mu}\psi^a.}

The divergent contributions
to the effective action from the graphs in Fig. 15 are of the form
\eqn\formfifteen{
iLNg^2myX_{15}f^{abc}\phibar^a
(C\psi)^b\psi^c}
where the contributions from the
individual graphs to $X_{15}$ are given in Table 15:
\bigskip
\vbox{
\begintable
Fig.|$X_{15}$\cr
15a|$-1$\cr
15b|$-1$\cr
15c|$1$\cr
15d|$-1$\endtable}
\centerline{{\it Table~15:\/} Contributions from Fig.~15}
\bigskip

The contributions from Table~15(a)--(c) add to
\eqn\sumfifteen{
\Gamma_{15\rm{1PI}}^{(1)\rm{pole}}
=-iLNg^2myf^{abc}\phibar^a                      
(C\psi)^b\psi^c.}
Fig.~15(d) is only present in the eliminated case and is discussed 
separately in Section 5.  
The divergent contributions
to the effective action from the graphs in Fig. 16 are of the form
\eqn\formsixteen{
LNg^2y^2X_{16}f^{abe}d^{cde}(C\psi)^a\psi^b\phibar^c\phibar^d}
where the contributions from the
individual graphs to $X_{16}$ are given in Table 16:
\bigskip
\vbox{
\begintable
Fig.|$X_{16}$\cr
16a|$-\frak14$\cr
16b|$-\frak14$\cr
16c|$\frak14$\cr
16d|$0$\cr
16e|$-\frak14$
\endtable}
\centerline{{\it Table~16:\/} Contributions from Fig.~16}
\bigskip

The contributions from Table~16(a)--(d) add to
\eqn\sumsixteen{
\Gamma_{16\rm{1PI}}^{(1)\rm{pole}}
=-\frak14LNg^2y^2f^{abe}d^{cde}(C\psi)^a\psi^b\phibar^c\phibar^d.}
Fig.~16(e) is only present in the eliminated case and again is discussed 
separately in Section 5.  

Finally the contribution from Fig.~17 is 
\eqn\seventeen{
\Gamma_{17\rm{1PI}}^{(1)\rm{pole}}=\frak12im^2ygLC^{\mu\nu}F_{\mu\nu}^a
\phibar^a.}
This is only present in the eliminated case.

\appendix{B}{Results for one-loop diagrams in the fundamental case}
In this Appendix we list the divergent contributions from the 
one-loop diagrams for the fundamental case. 

The divergent contributions
to the effective action from the graphs in Fig. 7 are of the form
\eqn\formoneb{
imLg^AC^{\mu\nu}\pa_{\mu}A_{\nu}^A\phibar R^AX_{B7}^A\phitilbar}
where the contributions to $X_{B7}^A$ from the
individual graphs are given in Table 17:
\bigskip
\vbox{
\begintable
Graph|$X_{B7}^A$\cr
7a|$-4(2\Chat_2-Ng^2c^A)$\cr
7b|$-2Ng^2c^A$\cr
7c|$2(2\Chat_2-Ng^2c^A)$\cr
7d|$-4\Chat_2$\cr
7e|$-32\left(2-\frak{\Delta}{g_0^2}\delta^{A0}\right)\Chat_2
+(24-8\delta^{A0})Ng^2$\cr
7f|$-4\alpha(2\Chat_2-Ng^2c^A)$\cr
7g|$-4(2\Chat_2-Ng^2c^A)$\cr
7h|$4\Chat_2-Ng^2c^A$\cr
7i|$2(2\Chat_2-Ng^2c^A)$\cr
7j|$-2(1+2\alpha)Ng^2c^A$\cr
7k|$3Ng^2c^A$\cr
7l|$-(5+\alpha)Ng^2c^A$\cr
7m|$2\alpha(2\Chat_2-Ng^2c^A)$\cr
7n|$2(2\Chat_2-Ng^2c^A)$\cr
7o|$-(4\Chat_2-Ng^2c^A)$\cr
7p|$-2(2\Chat_2-Ng^2c^A)$\cr
7q|$2(1+2\alpha)Ng^2c^A$\cr
7r|$-3Ng^2c^A$
\endtable}
\centerline{{\it Table~17:\/} Contributions from Fig.~7 (fundamental case)}
\bigskip
These results add to
\eqn\sumoneb{\eqalign{
\Gamma_{B7\rm{1PI}}^{(1)\rm{pole}}=&
imLgC^{\mu\nu}\pa_{\mu}A_{\nu}^A\phibar R^A\Bigl[\left\{-(76+4\alpha)
+32\frak{\Delta}{g_0^2}\delta^{A0}\right\}
\Chat_2\cr
&+\left\{(21+\alpha)c^A+16\delta^{A0}\right\}Ng^2\Bigr]
\phitilbar.\cr}}
(Note that the contributions from Figs.~7(h-k) cancel those from
Figs.~7(o-r) respectively.)

The divergent contributions
to the effective action from the graphs in Fig. 8 are of the form
\eqn\formtwob{
imLg^2C^{\mu\nu}X_{B8}f^{abc}A_{\mu}^aA_{\nu}^b\phibar R^c 
\phitilbar} 
where the contributions to $X_{B8}$ from the
individual graphs are given in Table 18:
\bigskip 
\vbox{   
\begintable
Graph|$X_{B8}$\cr
8a|$4(2\Chat_2-Ng^2)$\cr
8b|$2Ng^2$\cr
8c|$-2(2\Chat_2-Ng^2)$\cr
8d|$32\Chat_2-12Ng^2$\cr
8e|$2\alpha(2\Chat_2-Ng^2)$\cr
8f|$2(2\Chat_2-Ng^2)$\cr
8g|$-\frak12(3+\alpha)Ng^2$\cr
8h|$0$\cr
8i|$0$\cr
8j|$-2(2\Chat_2-Ng^2)$\cr
8k|$\frak32\alpha Ng^2$\cr
8l|$-\alpha Ng^2$\cr
8m|$\frak32\alpha Ng^2$\cr
8n|$\frak12(2+\alpha)Ng^2$\cr
8o|$0$\cr
8p|$0$\cr
8q|$-\frak32\alpha Ng^2$\cr
8r|$\frak32(1+\alpha)Ng^2$\cr
8s|$-\alpha(2\Chat_2-Ng^2)$\cr
8t|$-(2\Chat_2-Ng^2)$\cr
8u|$\frak12(3+\alpha)Ng^2$\cr
8v|$0$\cr
8w|$0$\cr
8x|$2(2\Chat_2-Ng^2)$\cr
8y|$-\frak32\alpha Ng^2$
\endtable}
\centerline{{\it Table~18:\/} Contributions from Fig.~8 (fundamental case)}
\vbox{
\begintable
Graph|$X_{B8}$\cr
8z|$\alpha Ng^2$\cr
8aa|$-\frak32\alpha Ng^2$\cr
8bb|$-\frak12(2+\alpha)Ng^2$\cr
8cc|$0$\cr
8dd|$0$
\endtable}
\centerline{{\it Table~18:\/} Contributions from Fig.~8 (fundamental case) 
(continued)}
\bigskip 
These results add to
\eqn\sumtwob{
\Gamma_{B8\rm{1PI}}^{(1)\rm{pole}}=
imLNg^2C^{\mu\nu}f^{abc}A_{\mu}^aA_{\nu}^b\phibar 
\left[2(19+\alpha)\Chat_2-(\frak{23}{2}+\alpha)Ng^2\right]R^c\phitilbar}
(Note that the contributions from Figs.~8(g-p) cancel those from
Figs.~8(u-dd) respectively.)

The divergent contributions
to the effective action from the graphs in Fig. 9 are of the form
\eqn\formthreeb{
mL|C|^2g^Ag^B\lambdabar^A\lambdabar^B\phibar X_{B9}^{AB}
\phitilbar}
where the contributions to $X_{B9}^{AB}$ from the
individual graphs are given in Table 19.
The results from Table 19 add to 
\eqn\sumthreeb{\eqalign{
\Gamma_{B9\rm{1PI}}^{(1)\rm{pole}}=&mL|C|^2\phibar\Bigl(
\left\{\left[\frak14(2-\alpha)-4\frak{g^2}{g_0^2}\right]Ng^2
+\left[\frak14(9+\alpha)+\frak{8g^2}{g_0^2}
\right]\Chat_2\right\}g^2d^{abc}R^c\lambdabar^a\lambdabar^b\cr
+&\left\{-\left[\frak14(11+\alpha)+4\frak{g^2}{g_0^2}\right]Ng^2
+\left[\frak14(9+\alpha)+\frak{16g^2}{g_0^2}
\right]\Chat_2\right\}
\frak1Ng^2\lambdabar^a\lambdabar^a\cr
&+\left\{(9+\alpha)\Chat_2-\frak12(1+\alpha)Ng^2\right\}gg_0R^aR^0
\lambdabar^a\lambdabar^0\cr
&+\frak12(9+\alpha)g_0^2\Chat_2R^0R^0\lambdabar^0\lambdabar^0\Bigr)
\phitilbar.\cr}}
(Note that the contributions from Figs.~9(h--m) cancel those from 
Figs.~9(u--z), in analogy to the situation with Figs.~7 and 8; this is a 
consequence of our choice of coefficient for the last term in Eq.~\lagmassb.)
\bigskip
\vbox{
\begintable
Graph|$X_{B9}^{ab}$|$X_{B9}^{a0}$|$X_{B9}^{00}$\cr
9a|$\alpha\left(\frak{1}{2N}\Delta R^aR^b+\frak14g^2\delta^{ab}\right)$|
$\alpha\frak{1}{2N}\Delta R^aR^0$|$\alpha \Chat_2R^0R^0$\cr
9b|$\left(\frak{1}{2N}\Delta R^aR^b+\frak14g^2\delta^{ab}\right)$|
$\frak{1}{2N}\Delta R^aR^0$|$\Chat_2R^0R^0$\cr
9c|$(3+\alpha)\left(\frak{1}{2N}\Delta R^aR^b+\frak14g^2\delta^{ab}\right)$|
$(3+\alpha)\frak{1}{2N}\Delta R^aR^0$|$(3+\alpha) \Chat_2R^0R^0$\cr
9d|$-2\alpha\left(\frak{1}{2N}\Delta R^aR^b+\frak14g^2\delta^{ab}\right)$|
$-2\alpha\frak{1}{2N}\Delta R^aR^0$|$-2\alpha \Chat_2R^0R^0$\cr
9e|$\left(Ng^2+4\frak{g^2\Delta}{g_0^2N}\right)d^{abc}R^c+\frak{2g^2}{g_0^2}
\left(2g^2-g_0^2+4\frak{\Delta}{N^2}\right)\delta^{ab}$|$0$|$0$\cr
9f|$\alpha\left(\frak{1}{2N}\Delta R^aR^b+\frak14g^2\delta^{ab}\right)$|
$\alpha\frak{1}{2N}\Delta R^aR^0$|$\alpha \Chat_2R^0R^0$\cr
9g|$\left(\frak{1}{2N}\Delta R^aR^b+\frak14g^2\delta^{ab}\right)$|
$\frak{1}{2N}\Delta R^aR^0$|$\Chat_2R^0R^0$\cr 
9h|$-2\left(\frak{1}{N}\Delta R^aR^b+\frak14g^2\delta^{ab}\right)$|
$-\frak12\left(2\Chat_2+\frak3N\Delta\right)R^aR^0$|$-4\Chat_2R^0R^0$\cr
9i|$\frak14\alpha g^2 Nd^{abc}R^c$|$\frak12\alpha NR^aR^0$|$0$\cr
9j|$-\alpha g^2Nd^{abc}R^c$|$-2\alpha NR^aR^0$|$0$\cr      
9k|$8\left(\frak{1}{N}\Delta R^aR^b+\frak14g^2\delta^{ab}\right)$|
$2\left(2\Chat_2+\frak3N\Delta\right)R^aR^0$|$16\Chat_2R_0R_0$\cr
9l|$\alpha g^2 Nd^{abc}R^c$|$2\alpha NR^aR^0$|$0$\cr
9m|$-4\left(\frak{1}{2N}\Delta R^aR^b+\frak14g^2\delta^{ab}\right)$|     
$-\frak{2}{N}\Delta R^aR^0$|$-4\Chat_2R^0R^0$\cr             
9n|$0$|$0$|$0$\cr
9o|$0$|$0$|$0$\cr   
9p|$-\alpha g^2 Nd^{abc}R^c$|$-2\alpha NR^aR^0$|$0$\cr
9q|$(1+\alpha) g^2 Nd^{abc}R^c$|$2(1+\alpha) NR^aR^0$|$0$\cr
9r|$-\frak18(3+\alpha)g^2(Nd^{abc}R^c+2g^2\delta^{ab})$|$0$|$0$\cr
9s|$-\frak12\alpha\left(\frak{1}{2N}\Delta R^aR^b+\frak14g^2\delta^{ab}\right)$|
$-\frak12\alpha\frak{1}{2N}\Delta R^aR^0$|$-\frak12\alpha \Chat_2R^0R^0$\cr
9t|$-\frak12\left(\frak{1}{2N}\Delta R^aR^b+\frak14g^2\delta^{ab}\right)$|
$-\frak12\frak{1}{2N}\Delta R^aR^0$|$-\frak12 \Chat_2R^0R^0$\cr
9u|$2\left(\frak{1}{N}\Delta R^aR^b+\frak14g^2\delta^{ab}\right)$|
$\frak12\left(2\Chat_2+\frak3N\Delta\right)R^aR^0$|$4\Chat_2R^0R^0$\cr
9v|$-\frak14\alpha g^2 Nd^{abc}R^c$|$-\frak12\alpha NR^aR^0$|$0$\cr
9w|$\alpha g^2Nd^{abc}R^c$|$2\alpha NR^aR^0$|$0$\cr
9x|$-8\left(\frak{1}{N}\Delta R^aR^b+\frak14g^2\delta^{ab}\right)$|
$-2\left(2\Chat_2+\frak3N\Delta\right)R^aR^0$|$-16\Chat_2R_0R_0$
\endtable}
\centerline{{\it Table~19:\/} Contributions from Fig.~9 (fundamental case)}

\vfill
\eject
\vbox{
\begintable
Graph|$X_{B9}^{ab}$|$X_{B9}^{a0}$|$X_{B9}^{00}$\cr
9y|$-\alpha g^2 Nd^{abc}R^c$|$-2\alpha NR^aR^0$|$0$\cr
9z|$4\left(\frak{1}{2N}\Delta R^aR^b+\frak14g^2\delta^{ab}\right)$|
$\frak{2}{N}\Delta R^aR^0$|$4\Chat_2R^0R^0$\cr
9aa|$0$|$0$|$0$\cr
9bb|$0$|$0$|$0$
\endtable}
\centerline{{\it Table~19:\/} Contributions from Fig.~9 (fundamental case)
(continued)}

\appendix{C}{Analysis of terms required for renormalisability}
In this Appendix we perform a systematic analysis of the terms which can be 
generated by radiative corrections, focussing on the adjoint case. 
We follow the broad outlines of the analysis of Lunin and Rey\lunin, with
modifications to accommodate the presence of Yukawa couplings.
 
We therefore start by assuming that the couplings for the antichiral
fields, $\ybar$ and $\mbar$ are distinct from those for the chiral fields,
$y$ and $m$.    
The most general operator which can appear through radiative corrections
may be written schematically as 
\eqn\appb{
O=\Lambda^{\beta}y^{\delta}\ybar^{\deltabar}m^{\mu}\mbar^{\mubar}
C_{\mu\nu}^{\alpha}\pa^{\alpha_0}\phi^{\alpha_1}
\phibar^{\alphabar_1}F^{\alpha_2}\Fbar^{\alphabar_2}\psi^{\alpha_3}
\psibar^{\alphabar_3}A_{\mu}^{\alpha_4}\lambda^{\alpha_5}
\lambdabar^{\alphabar_5}D^{\alpha_6},}
where $\Lambda$ is an ultraviolet cutoff scale and $\alpha$, $\beta$,
$\delta$, $\deltabar$, $\mu$, $\mubar$, $\alpha_i$, 
$\alphabar_i$ are non-negative integers.
For a dimension four operator
\eqn\appc{
\beta-\alpha+\alpha_0+\alpha_1+\alphabar_1+\alpha_4+2(\alpha_2+\alphabar_2
+\alpha_6)+\frak32(\alpha_3+\alphabar_3+\alpha_5+\alphabar_5)+\mu+\mubar=4.}
There is also a pseudo-R-parity which acts as
\eqn\appd{\eqalign{
\phi\rightarrow e^{-i\rho}\phi, \quad F\rightarrow& e^{i\rho}F, \quad
\lambda\rightarrow e^{-i\rho}\lambda,\cr
C^{\mu\nu}\rightarrow e^{-2i\rho}C^{\mu\nu},\quad & y\rightarrow
e^{i\rho}y,\cr}}
$\phibar$, $\Fbar$, $\lambdabar$ and $\ybar$ transforming with opposite
charge and all other fields being neutral. This entails
\eqn\appe{
-2\alpha+\alpha_2+\alphabar_1+\alphabar_5-\alphabar_2-\alpha_1-\alpha_5
+\delta-\deltabar=0.}
Finally there is the pseudo $U_A(1)$ chiral symmetry which acts as
\eqn\appf{
\phi\rightarrow e^{i\gamma}\phi,\quad m\rightarrow e^{-2i\gamma}m,\quad
y\rightarrow e^{-3i\gamma} y,}
and acts in the same fashion on $\psi$ and $F$ as on $\phi$; the barred
fields transform with opposite charge. This leads to
\eqn\appg{
\alpha_1+\alpha_2+\alpha_3-\alphabar_1-\alphabar_2-\alphabar_3
+3(\deltabar-\delta)+2(\mubar-\mu)=0.}
Combining Eqs.~\appe, \appg\ we have
\eqn\apph{
\alphabar_1=3\alpha+\alpha_1-2(\alpha_2-\alphabar_2)-\frak12(\alpha_3
-\alphabar_3)+\frak32(\alpha_5-\alphabar_5)+\mu-\mubar}
and substituting in Eq.~\appc\ we find 
\eqn\appi{
\beta+2\alpha+\alpha_0+2\alpha_1+4\alphabar_2+
\alpha_3+2\alphabar_3+\alpha_4+3\alpha_5
+2\alpha_6+2\mu=4.}
For simplicity we shall start by analysing the massless case, i.e.
$\mu=\mubar=0$. 

For $\alpha=2$ we find  from Eq.~\appi\
\eqn\appj{
\beta=\alpha_0=\alpha_1=\alphabar_2=\alpha_3=\alphabar_3=\alpha_4=\alpha_5
=\alpha_6=0}
and hence Eqs.~\appg\ and \apph\ become
\eqn\appk{\eqalign{
\alphabar_1=&6-2\alpha_2-\frak32\alphabar_5,\cr
\alphabar_1-\alpha_2-3\deltabar+3\delta=&0.\cr}} 
The only solutions to Eq.~\appk\ are
\eqn\appl{\eqalign{
\alphabar_5=4,\quad\alpha_2=0,\quad&\alphabar_1=0,\quad\deltabar-\delta=0,\cr
\alphabar_5=2,\quad\alpha_2=0,\quad&\alphabar_1=3,\quad\deltabar-\delta=1,\cr
\alphabar_5=2,\quad\alpha_2=1,\quad&\alphabar_1=1,\quad\deltabar-\delta=0,\cr
\alphabar_5=0,\quad\alpha_2=0,\quad&\alphabar_1=6,\quad\deltabar-\delta=2,\cr
\alphabar_5=0,\quad\alpha_2=1,\quad&\alphabar_1=4,\quad\deltabar-\delta=1,\cr
\alphabar_5=0,\quad\alpha_2=2,\quad&\alphabar_1=2,\quad\deltabar-\delta=0,\cr
\alphabar_5=0,\quad\alpha_2=3,\quad&\alphabar_1=0,\quad\deltabar-\delta=-1.
\cr}}
These solutions correspond to terms
\eqn\appm{
|C|^2(\lambdabar\lambdabar)^2,\quad
\ybar|C|^2\lambdabar\lambdabar\phibar^3,\quad \ybar^2|C|^2\phibar^6,\quad
|C|^2\phibar \lambdabar\lambdabar F,\quad |C|^2\phibar^2F^2,\quad
\ybar|C|^2\phibar^4F,\quad y|C|^2F^3.}
Here arbitrary powers of $y\ybar$ and of course $g$ are suppressed. 
The terms in Eq.~\appm\ without a $y$ or $\ybar$ are already in the Lagrangian,
with the exception of the $\phibar^2F^2$ term. As stated earlier, in the
case with superpotential we have not computed terms purely involving 
$\phibar$, $\lambdabar$ and/or $F$, which are individually 
$\Ncal=\frak12$ invariant.
 
For $\alpha=1$ we need either a $\sigmabar_{\mu\nu}$, a $\sigma_{\mu\nu}$,
a $\sigma_{\mu}D_{\nu}$, a $\sigmabar_{\mu}D_{\nu}$ or a $D_{\mu}D_{\nu}$
to contract with the $C^{\mu\nu}$. However, $C^{\mu\nu}\sigmabar_{\mu\nu}=0$
due to the self-duality of $C^{\mu\nu}$, and we see from Eq.~\appi\ that 
$\alphabar_2=\alpha_5=0$ so no $\lambda$ may appear. So $\sigma_{\mu\nu}$
could only appear in the form $\psi\sigma_{\mu\nu}\psi$ and we need 
$\alpha_3=2$. Eqs.~\appg\ and \apph\ then give
\eqn\appn{\eqalign{
\alphabar_1=&2-2\alpha_2-\frak32\alphabar_5,\cr
\alphabar_1-\alpha_2+3(\delta-\deltabar)=&2,\cr}}
whose solutions are
\eqn\appna{\eqalign{
\alphabar_5=0,\quad\alpha_2=1,\quad&\alphabar_1=0,\quad\deltabar-\delta=-1,\cr
\alphabar_5=0,\quad\alpha_2=0,\quad&\alphabar_1=2,\quad\deltabar-\delta=0.
\cr}}
These solutions may be written in the form
\eqn\appnb{
yF\psi C\psi,\quad \phibar^2\psi C\psi}
where $(C\psi)^{\alpha}=C^{\alpha}{}_{\beta}\psi^{\beta}$. These terms are not
in the original Lagrangian but can be generated at one loop (see 
Figs.~6 and 12); in fact the $F\psi C\psi$term is individually $\Ncal=\frak12$ 
invariant. They are both discussed in detail in the main text. 

If we have one 
derivative then $\alpha_0+\alpha_4=1$ which implies from Eq.~\appi\
that $\alpha_1=\alphabar_3=\alpha_6=0$ and $\alpha_3=1$. So
$\lambdabar\sigmabar_{\mu}\psi$ is the only possibility for the fermion fields,
i.e. $\alphabar_5=1$. Eqs.~\appg\ and \apph\ then give
\eqn\appo{\eqalign{
\alphabar_1=&1-2\alpha_2,\cr
\alphabar_1-\alpha_2+3(\delta-\deltabar)=&1\cr}}
whose only solution is $\alphabar_1=1$, $\alpha_2=\delta-\deltabar=0$
which corresponds to $C^{\mu\nu}D_{\mu}\phibar\lambdabar\sigmabar_{\nu}\psi$
which is already in the Lagrangian.

Finally $D_{\mu}D_{\nu}$ corresponds to $\alpha_0+\alpha_4 =2$ which from
Eq.~\appi\ gives $\alpha_1=\alpha_3=\alphabar_3=\alpha_6=0$ .
Eqs.~\appg\ and \apph\ then give
\eqn\appp{\eqalign{
\alphabar_1=&3-\frak32\alphabar_5-2\alpha_2,\cr
\alphabar_1-\alpha_2+3(\delta-\deltabar)=&0,\cr}}
whose solutions are
\eqn\appq{\eqalign{
\alphabar_5=0,\quad \alpha_2=1,\quad \alphabar_1=&1,\quad\deltabar-\delta=0,\cr
\alphabar_5=0,\quad \alpha_2=0,\quad \alphabar_1=&3,\quad\deltabar-\delta=1,\cr 
\alphabar_5=2,\quad \alpha_2=0,\quad \alphabar_1=&0,\quad\deltabar-\delta=0.\cr
}}
These solutions correspond to
\eqn\apppa{
C^{\mu\nu}\phibar F_{\mu\nu}F,\quad C^{\mu\nu}F_{\mu\nu}\lambdabar\lambdabar,
\quad\ybar C^{\mu\nu}\phibar D_{\mu}\phibar D_{\nu}\phibar,\quad
\ybar C^{\mu\nu}\phibar^3 F_{\mu\nu},}
all of which are already in the
Lagrangian.

The inclusion of mass terms is straightforward but laborious and we shall
confine ourselves to stating the results. Firstly we invariably find $\mu=0$.
For $\alpha=2$ we find 
\eqn\appr{\eqalign{
\mubar=1,\quad \alphabar_5=0,\quad\alpha_2=&0,\quad\alphabar_1=5,\quad
\deltabar-\delta=1,\cr
\mubar=1,\quad \alphabar_5=0,\quad\alpha_2=&1,\quad\alphabar_1=3,\quad
\deltabar-\delta=0,\cr
\mubar=1,\quad \alphabar_5=0,\quad\alpha_2=&2,\quad\alphabar_1=1,\quad
\deltabar-\delta=-1,\cr
\mubar=1,\quad \alphabar_5=2,\quad\alpha_2=&0,\quad\alphabar_1=2,\quad
\deltabar-\delta=0,\cr
\mubar=1,\quad \alphabar_5=2,\quad\alpha_2=&1,\quad\alphabar_1=0,\quad
\deltabar-\delta=-1,\cr
\mubar=2,\quad \alphabar_5=0,\quad\alpha_2=&0,\quad\alphabar_1=4,\quad
\deltabar-\delta=0,\cr
\mubar=2,\quad \alphabar_5=0,\quad\alpha_2=&1,\quad\alphabar_1=2,\quad
\deltabar-\delta=-1,\cr
\mubar=2,\quad \alphabar_5=0,\quad\alpha_2=&2,\quad\alphabar_1=0,\quad  
\deltabar-\delta=-2,\cr
\mubar=2,\quad \alphabar_5=2,\quad\alpha_2=&0,\quad\alphabar_1=1,\quad
\deltabar-\delta=-1,\cr
\mubar=3,\quad \alphabar_5=0,\quad\alpha_2=&0,\quad\alphabar_1=3,\quad
\deltabar-\delta=-1,\cr
\mubar=3,\quad \alphabar_5=0,\quad\alpha_2=&1,\quad\alphabar_1=1,\quad
\deltabar-\delta=-2,\cr
\mubar=3,\quad \alphabar_5=2,\quad\alpha_2=&0,\quad\alphabar_1=0,\quad
\deltabar-\delta=-2,\cr
\mubar=4,\quad \alphabar_5=0,\quad\alpha_2=&0,\quad\alphabar_1=2,\quad
\deltabar-\delta=-2,\cr
\mubar=4,\quad \alphabar_5=0,\quad\alpha_2=&1,\quad\alphabar_1=0,\quad
\deltabar-\delta=-3,\cr
\mubar=5,\quad \alphabar_5=0,\quad\alpha_2=&0,\quad\alphabar_1=1,\quad
\deltabar-\delta=-3.\cr}}
These solutions correspond to terms 
\eqn\apps{\eqalign{
\ybar\mbar|C|^2\phibar^5,\quad \mbar|C|^2\phibar^3F,\quad
y\mbar|C|^2\phibar F^2,\quad & \mbar |C|^2\phibar^2\lambdabar\lambdabar,\quad
y\mbar |C|^2F\lambdabar\lambdabar,\quad
\mbar^2|C|^2\phibar^4,\cr 
y\mbar^2|C|^2\phibar^2F,\quad
y^2\mbar^2|C|^2F^2,\quad
y\mbar^2|C|^2&\phibar\lambdabar\lambdabar,\quad 
y\mbar^3|C|^2\phibar^3,\quad y^2\mbar^3|C|^2\phibar F,\cr
y^2\mbar^3|C|^2\lambdabar\lambdabar,\quad
y^2\mbar^4|C|^2\phibar^2,\quad&
y^3\mbar^4|C|^2F,\quad
y^3\mbar^5|C|^2\phibar.\cr
}}
The possible terms without gauge fields in Eq.~\apps\ have
already been given in Ref.~\gris. In any case all these terms are
individually $\Ncal=\frak12$ invariant and we have not 
calculated their counterterms (with the exception of 
$\mbar |C|^2\phibar^2\lambdabar\lambdabar$ which was in the classical
lagrangian).

For the case $\alpha=1$ we find solutions with
$\mubar=\alphabar=1$, $\alpha_3=2$, $\deltabar-\delta=-1$ and 
$\mubar=2$, $\alpha_3=2$, $\deltabar-\delta=-2$, all the other parameters zero,
 corresponding to $y\mbar\phibar\psi C\psi$ and $y^2\mbar^2\psi C\psi$ 
respectively. The $y\mbar\phibar\psi C\psi$ term has one loop 
contributions shown in Fig.~15, which are cancelled by our choice of 
$\rho_7$; we could find no one-loop diagrams contributing to the 
$y^2\mbar^2\psi C\psi$ term. There is also  
a solution with $\alphabar_5=\alpha_3=
\alpha_0+\alpha_4=\mubar=\delta-\deltabar=1$, all the other parameters zero,
corresponding to 
$y\mbar C^{\mu\nu}\lambdabar\sigmabar_{\mu}D_{\nu}\psi$. This has a one-loop
contribution shown in Fig.~14, which is cancelled by $\gamma_4$ in 
Eq.~\lagmass.
We also find solutions with $\alpha_0+\alpha_4=2$: 
\eqn\appt{\eqalign{
\mubar=1,\quad \alphabar_5=0,\quad\alpha_2=&0,\quad\alphabar_1=2,\quad
\deltabar-\delta=0,\cr  
\mubar=1,\quad \alphabar_5=0,\quad\alpha_2=&1,\quad\alphabar_1=0,\quad
\deltabar-\delta=-1,\cr 
\mubar=2,\quad \alphabar_5=0,\quad\alpha_2=&0,\quad\alphabar_1=1,\quad
\deltabar-\delta=-1,\cr 
\mubar=3,\quad \alphabar_5=0,\quad\alpha_2=&0,\quad\alphabar_1=0,\quad
\deltabar-\delta=-2.\cr 
}}
These correspond to terms
\eqn\appu{
C^{\mu\nu}\mbar\phibar^2F_{\mu\nu},\quad
C^{\mu\nu}y\mbar FF_{\mu\nu},\quad
C^{\mu\nu}y\mbar^2\phibar F_{\mu\nu}.}

The first term is already in the classical lagrangian;
the second gets contributions from Fig.~13 which again are cancelled by the 
$\gamma_4$ term in Eq.~\lagmass; and we have not been able to find 
any one-loop diagrams contributing to the third term.

This exhausts all the solutions.

We now turn to the case of the fundamental/anti-fundamental
representation. We can derive the possible terms by selecting terms above
with no $y$ or $\ybar$ (since there are now no Yukawa couplings)
and taking care to replace $\phi$s (for example) by $\phi$s and $\phitil$s 
in gauge-invariant combinations. This analysis was of course already performed 
by Lunin and Rey\lunin\ and we reproduce their results; however
with the addition of a $\mbar|C|^2(\phibar F)(\phibar\phitilbar)$ term,
a $\mbar|C|^2(\phitilbar \Ftil)(\phibar\phitilbar)$ term,
a $\mbar^2|C|^2(\phibar\phitilbar)^2$ term and a 
$C^{\mu\nu}\mbar\phibar F_{\mu\nu}\phitilbar$ term. It is straightforward to
construct one-loop diagrams which give divergent contributions to these terms 
and so we see no reason to omit them. 

\appendix{D} {Group identities for $U(N)$}
The basic
commutation relations for $U(N)$ are (for the fundamental representation):
\eqn\commrel{ [R^a,R^b]=if^{abc}R^c,\quad
\{R^A,R^B\}=d^{ABC}R^C,}
where $d^{ABC}$ is totally symmetric.
Defining matrices $\Ftil^A$, $\Dtil^A$ by $(\Ftil^A)^{BC}=if^{BAC}$, 
$(\Dtil^A)^{BC}=d^{ABC}$,
useful identities for $U(N)$ are
\eqn\sunidents{\eqalign{
\Tr[\Ftil^A\Ftil^B]=&N\delta^{AB},\qquad \Tr[\Dtil^A\Dtil^B]=N\delta^{AB},\cr
\Tr[\Ftil^A\Ftil^B\Dtil^C]=&\frak{N}{2}d^{ABC}c^Ac^Bd^C,\qquad
\Tr[\Ftil^A\Dtil^B\Dtil^C]=i\frak{N}{2}f^{ABC},\cr
f^{ABE}d^{CDE}+&f^{ACE}d^{DBE}+f^{ADE}d^{BCE}=0,\cr
f^{ABE}f^{CDE}=&d^{ACE}d^{BDE}-d^{ADE}d^{BCE},\cr}}
(where  $c^A=1-\delta^{A0}$ and $d^A=1+\delta^{A0}$),
and also
\eqn\sunidentsa{\eqalign{
\Tr[\Ftil^A\Ftil^B\Ftil^C\Ftil^D]=&c^Ac^Bc^Cc^D\Bigl[
\frak12\delta^{(AB}\delta^{CD)}\cr
&+\frak{N}{4}\left(d^{ABE}d^{CDE}+d^{ADE}d^{BCE}-d^{ACE}d^{BDE}\right)\Bigr],\cr
\Tr[\Ftil^A\Ftil^B\Ftil^C\Dtil^D]=&-\frak{N}{4}i(d^{ABE}f^{CDE}+f^{ABE}d^{CDE})
c^Ac^Bc^Cd^D,\cr
\Tr[\Ftil^A\Ftil^B\Dtil^C\Dtil^D]=&\Bigl[\frak12c^Ac^Bc^Cc^D\left(
\delta^{AB}\delta^{CD}-\delta^{AC}\delta^{BD}-\delta^{AD}\delta^{BC}\right)\cr
&+\frak{N}{4}c^Ac^Bd^Cd^D\left(
d^{ABE}d^{CDE}+d^{ADE}d^{BCE}-d^{ACE}d^{BDE}\right)\Bigr],\cr
\Tr[\Ftil^A\Dtil^B\Ftil^C\Dtil^D]=&
c^Ac^Bc^Cc^D\frak12\left(\delta^{AC}\delta^{BD}
-\delta^{AB}\delta^{CD}-\delta^{AD}\delta^{BC}\right)\cr
&+\frak{N}{4}c^Ad^Bc^Cd^D\left(
d^{ABE}d^{CDE}+d^{ADE}d^{BCE}-d^{ACE}d^{BDE}\right)\Bigr].\cr
}}
\vfill
\eject
\epsfysize= 5in
\centerline{\epsfbox{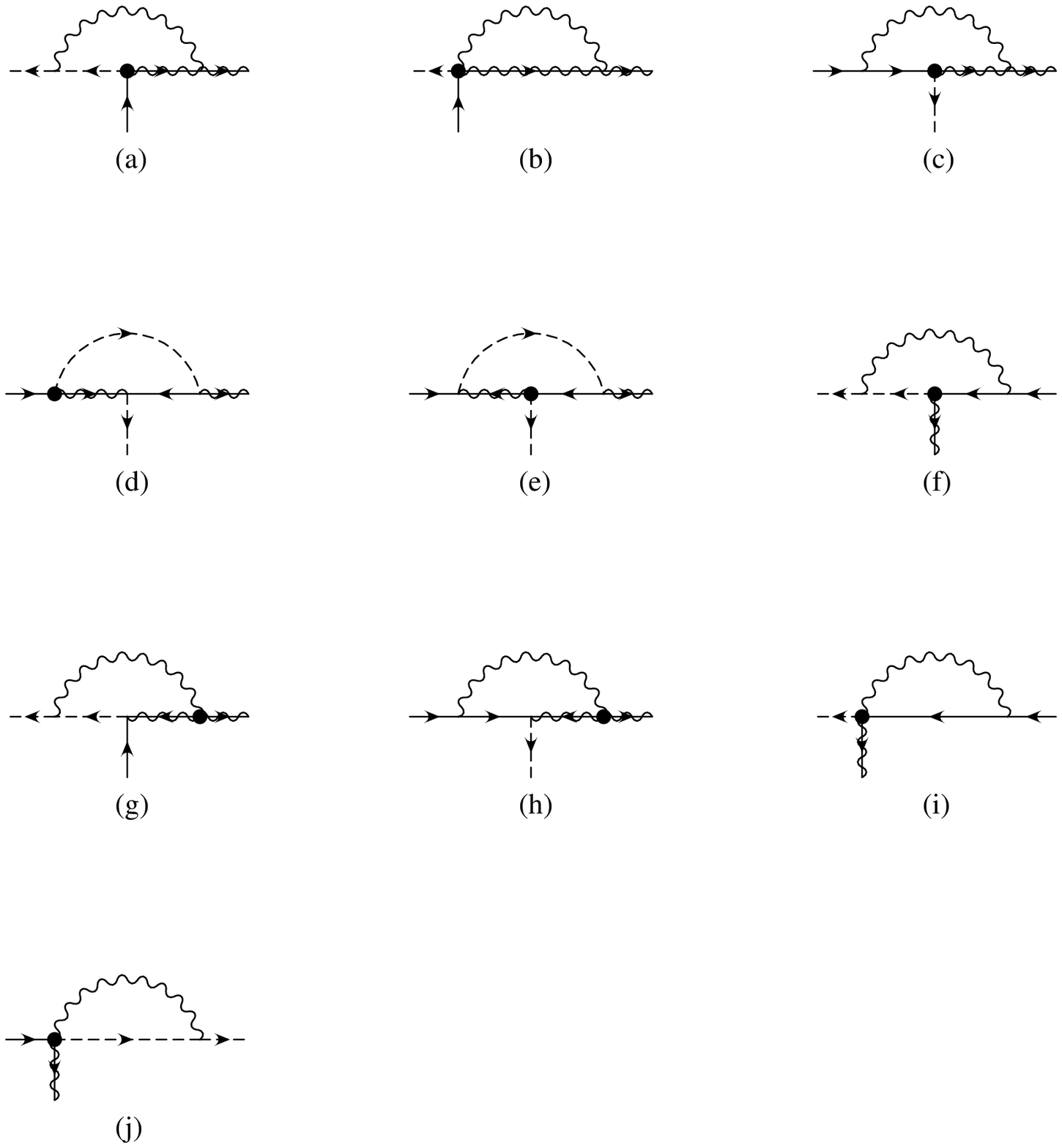}}
\inparg
{\it \noindent Fig. 1: Diagrams with one gaugino, one scalar and one
chiral fermion line; the dot represents the position of a $C$.} 
\medskip
\outparg
\vfill  
\eject  
\epsfysize= 7in
\centerline{\epsfbox{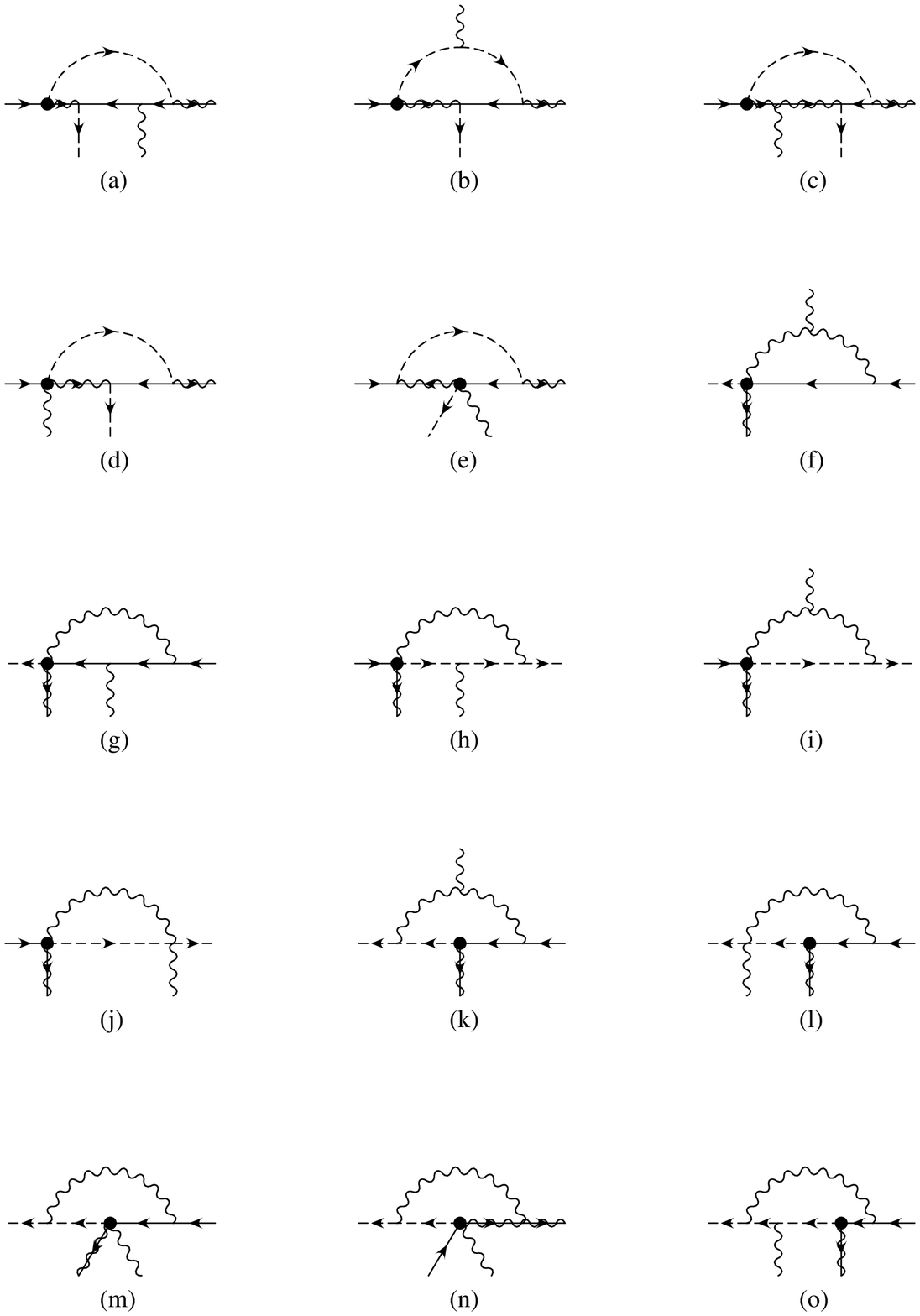}}
\inparg
{\it \noindent Fig. 2: Diagrams with one gaugino, one scalar, one
chiral fermion and one gauge line; the dot represents the position of a $C$.}
\medskip
\outparg

\bigskip
\epsfysize= 7in
\centerline{\epsfbox{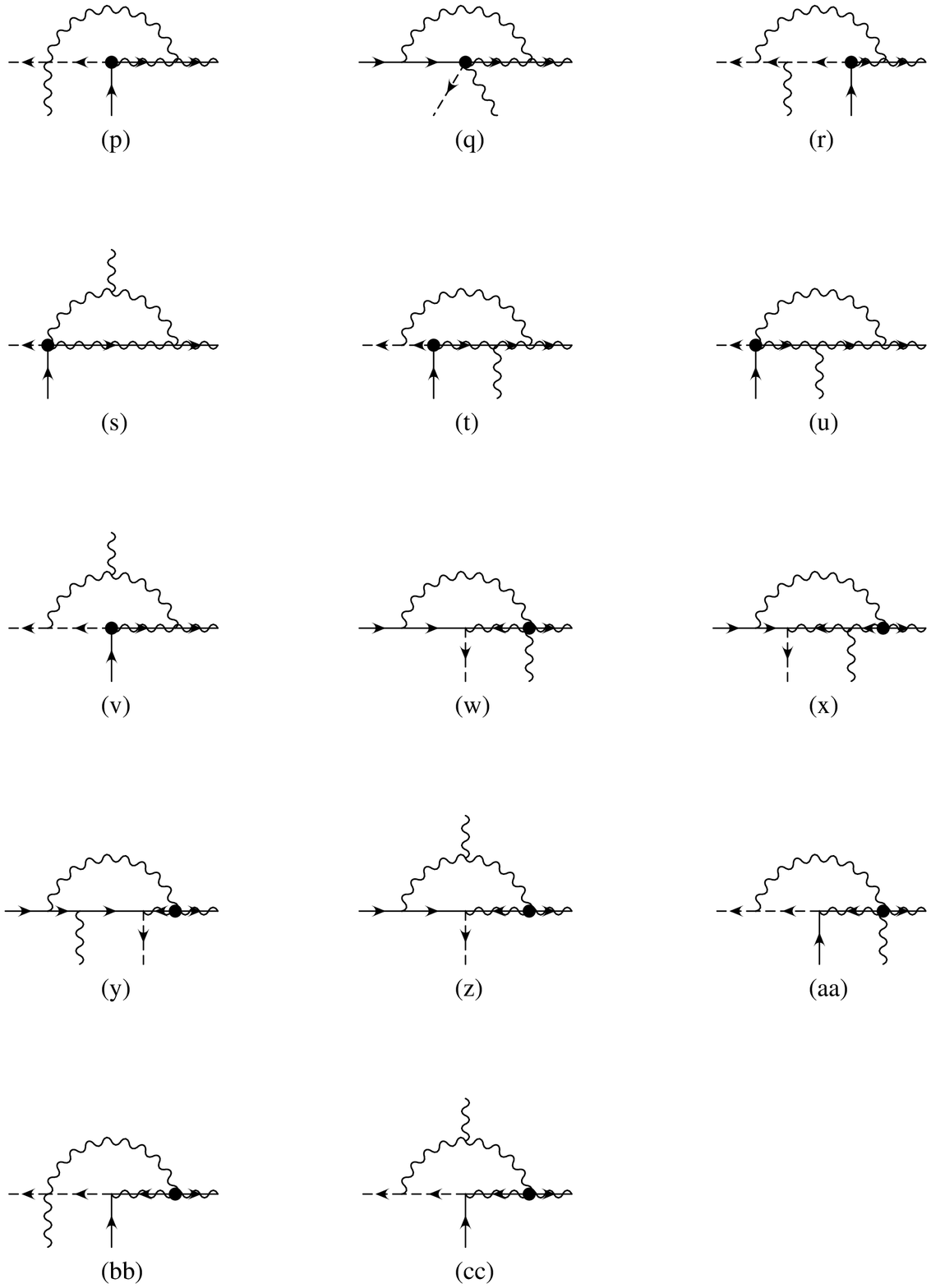}}
\inparg
{\it \noindent Fig. 2(continued).}   
\medskip
\outparg

\epsfysize= 2.5in
\centerline{\epsfbox{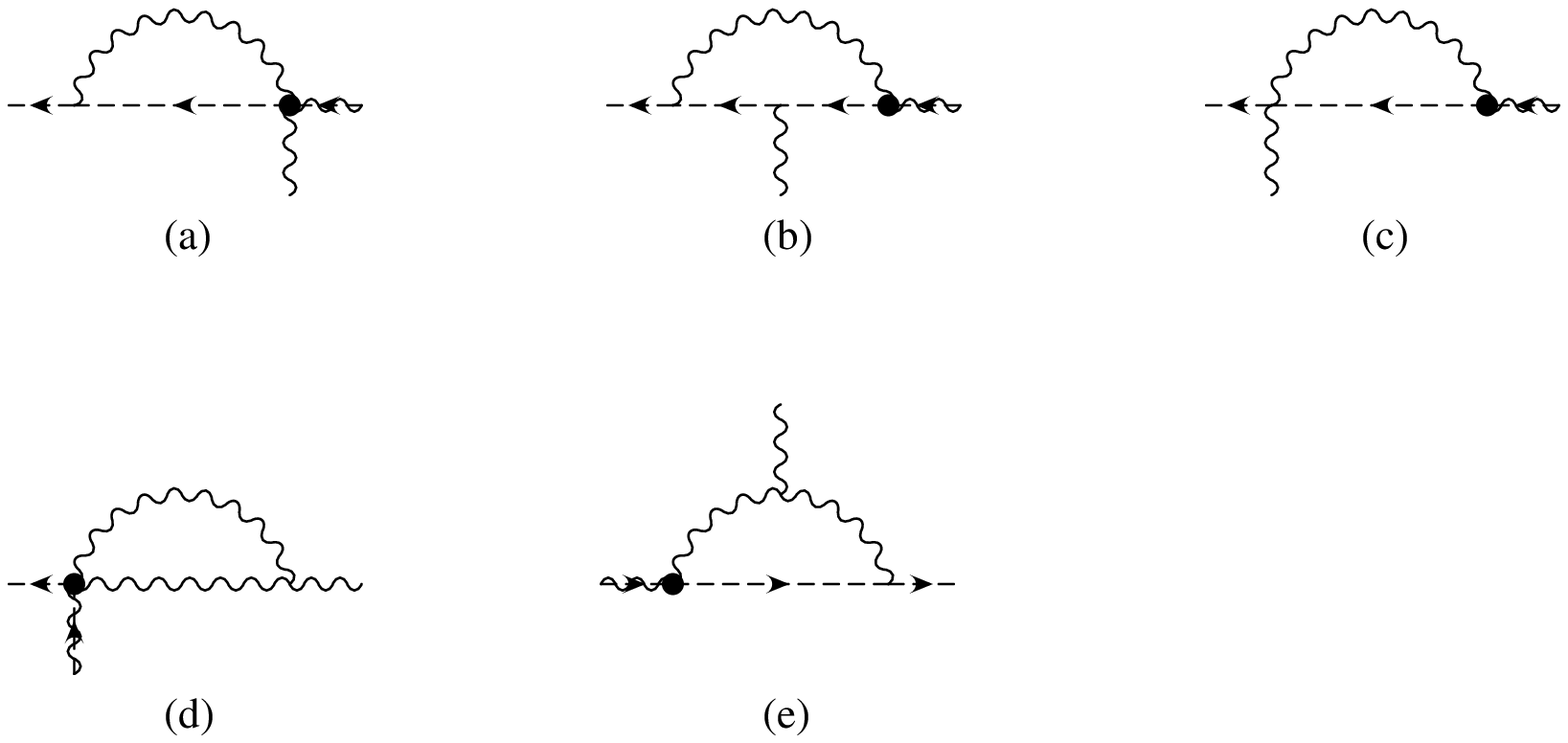}}
\inparg 
{\it \noindent Fig. 3: Diagrams with one gauge, one scalar and one
auxiliary line; the dot represents the position of a $C$.}
\medskip
\outparg
\vfill  
\eject  
\epsfysize= 5in
\centerline{\epsfbox{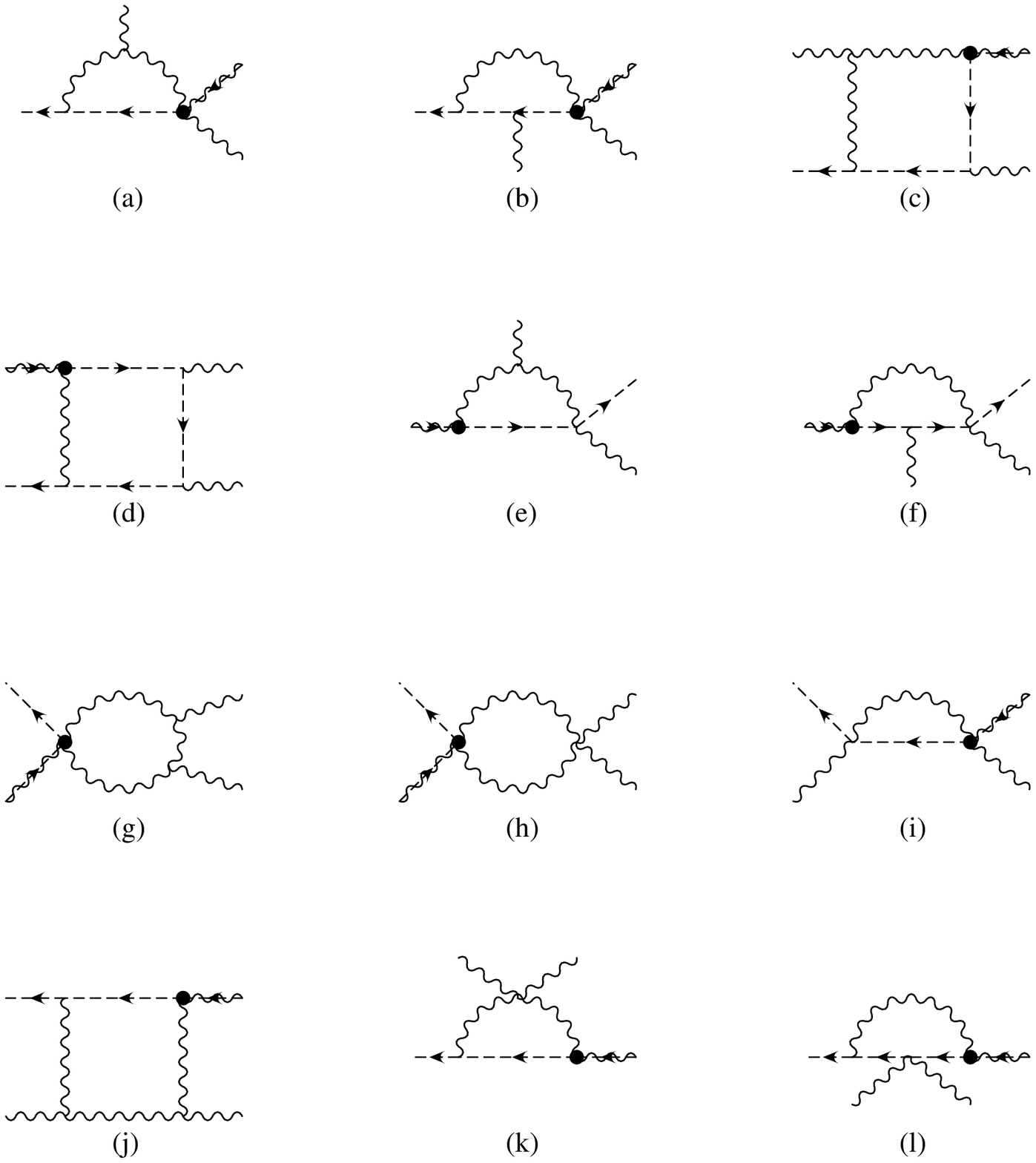}}
\inparg
{\it \noindent Fig. 4: Diagrams with two gauge, one scalar and one
auxiliary line; the dot represents the position of a $C$.}
\medskip
\outparg
\vfill  
\eject  
\bigskip
\epsfysize= 5in
\centerline{\epsfbox{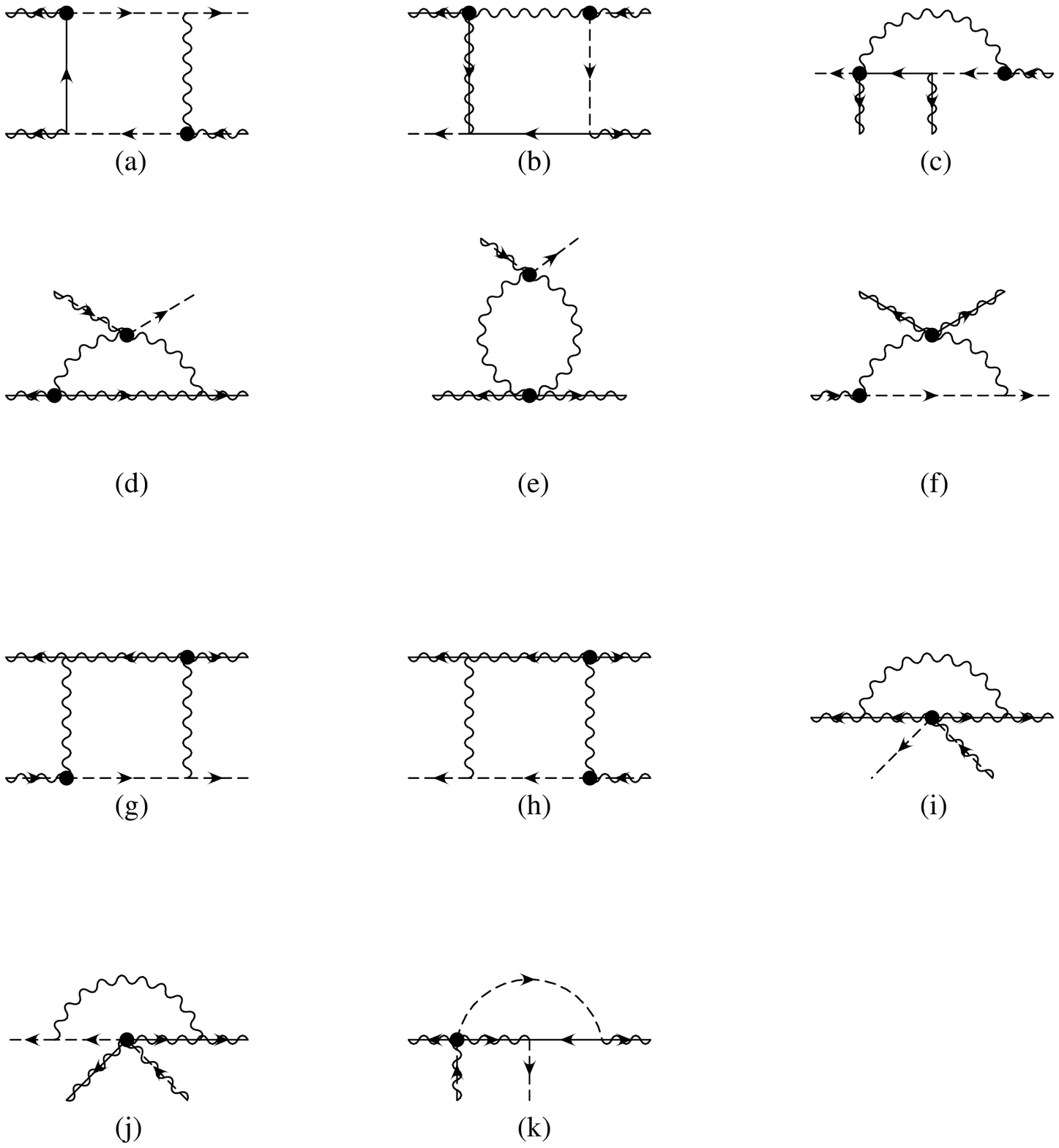}}
\inparg
{\it \noindent Fig. 5: Diagrams with two gaugino, one scalar and one
auxiliary line; the dot represents the position of a $C$ or a $|C|^2$.}
\medskip
\outparg
\vfill
\eject

\bigskip
\epsfysize= 2.5in
\centerline{\epsfbox{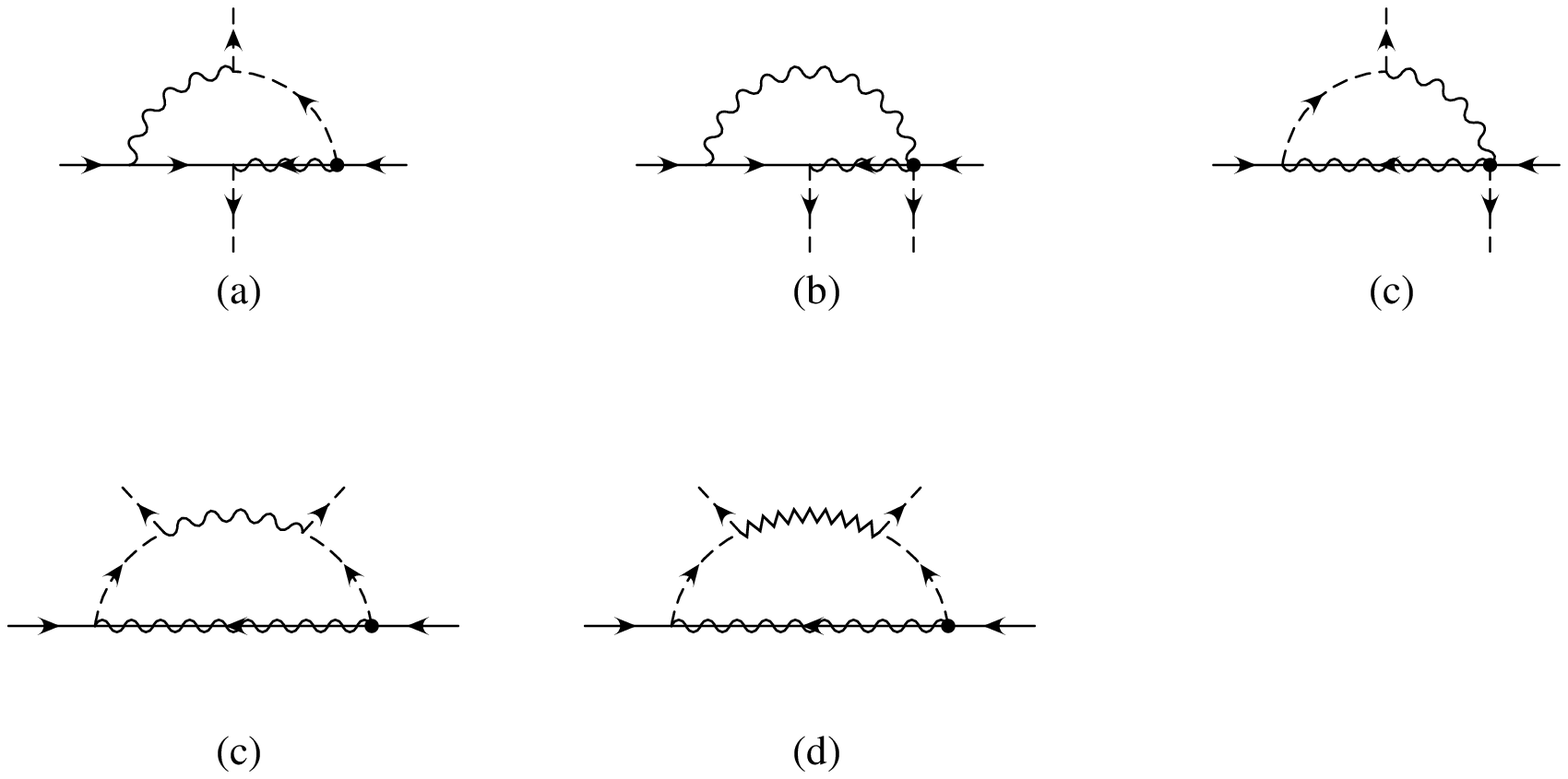}}
\inparg
{\it \noindent Fig. 6: Diagrams with two scalar
and two chiral fermion lines; the dot represents the position of a $C$.}
\medskip
\outparg

\vfill  
\eject  
\epsfysize= 7in
\centerline{\epsfbox{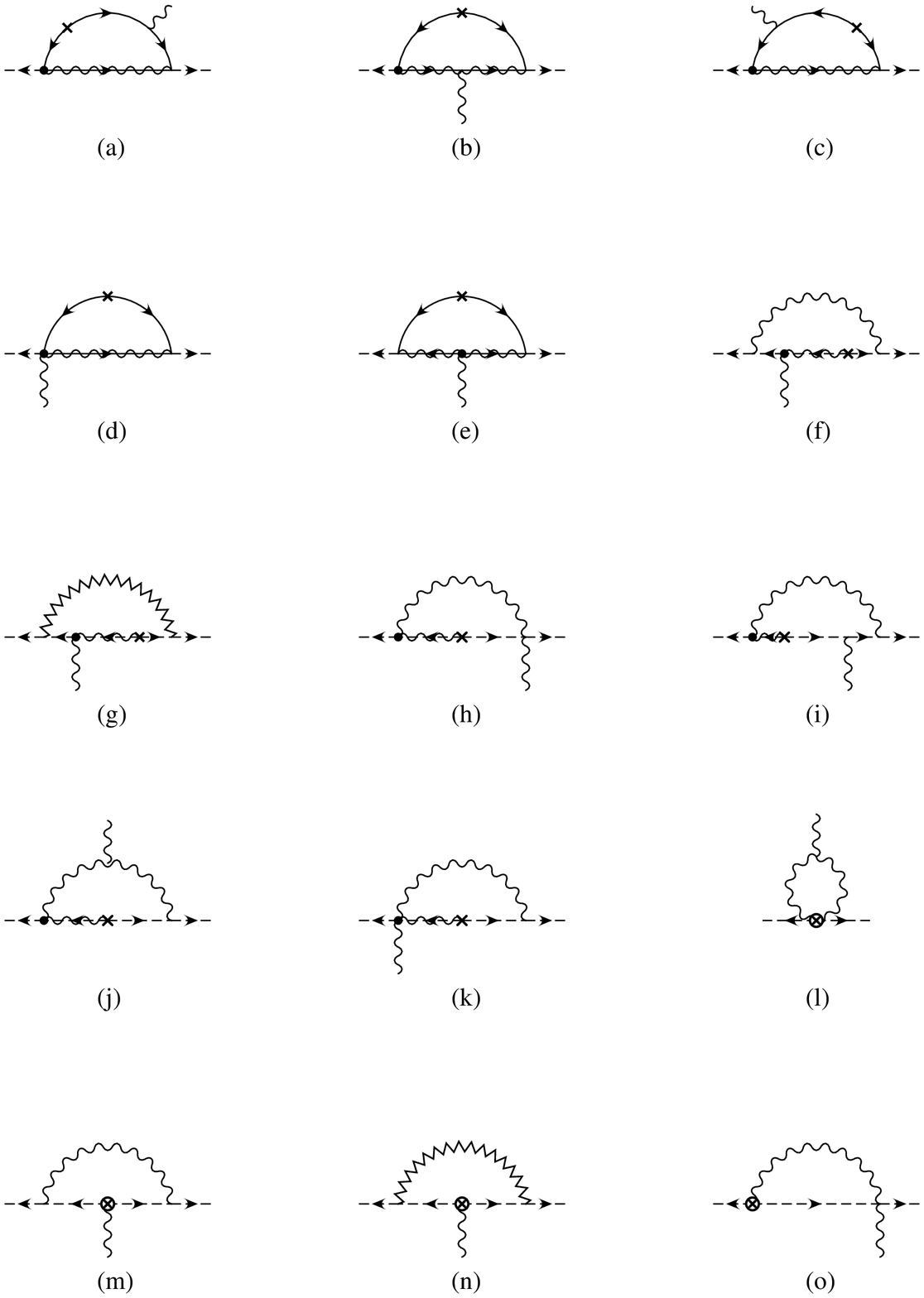}}
\inparg
{\it \noindent Fig. 7: Diagrams with two scalar,
one gauge line; a dot denotes a $C$, a cross a mass
and a crossed circle a vertex with both a mass and a $C$.}
\medskip
\outparg
\bigskip
\epsfysize= 1.2in
\centerline{\epsfbox{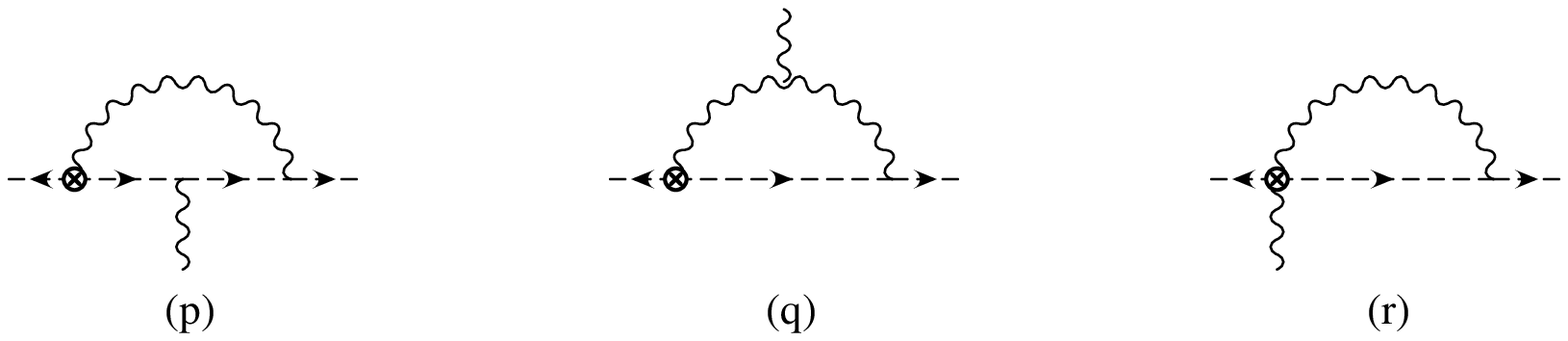}}
\inparg
{\it \noindent Fig. 7(continued)}   
\medskip
\outparg

\vfill
\eject 
\epsfysize= 7in
\centerline{\epsfbox{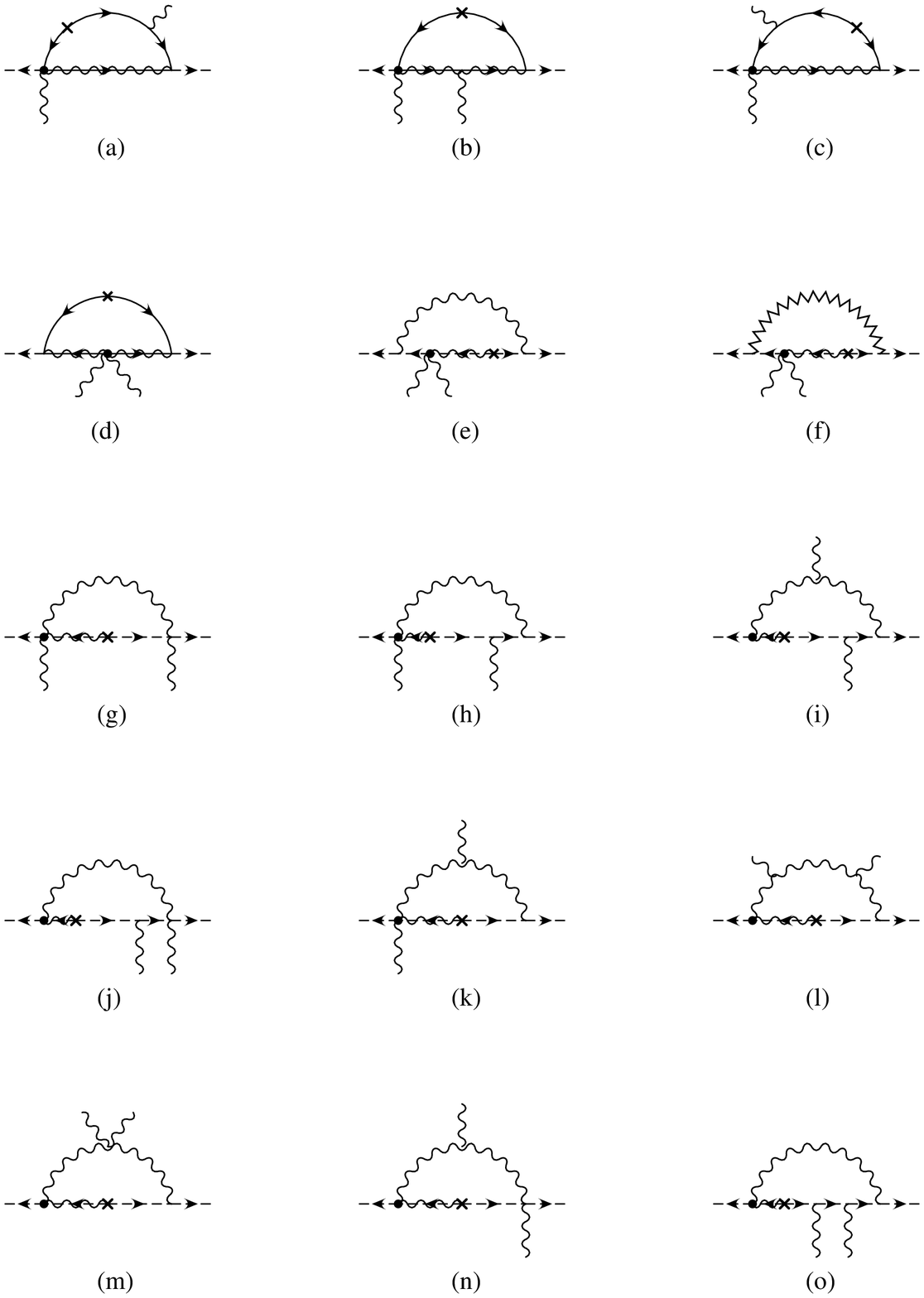}}
\inparg 
{\it \noindent Fig. 8: Diagrams with two scalar, two
gauge lines; a dot denotes a $C$, a cross a mass
and a crossed circle a vertex with both a mass and a $C$.}
\medskip
\outparg
\vfill
\eject 
\epsfysize= 7in
\centerline{\epsfbox{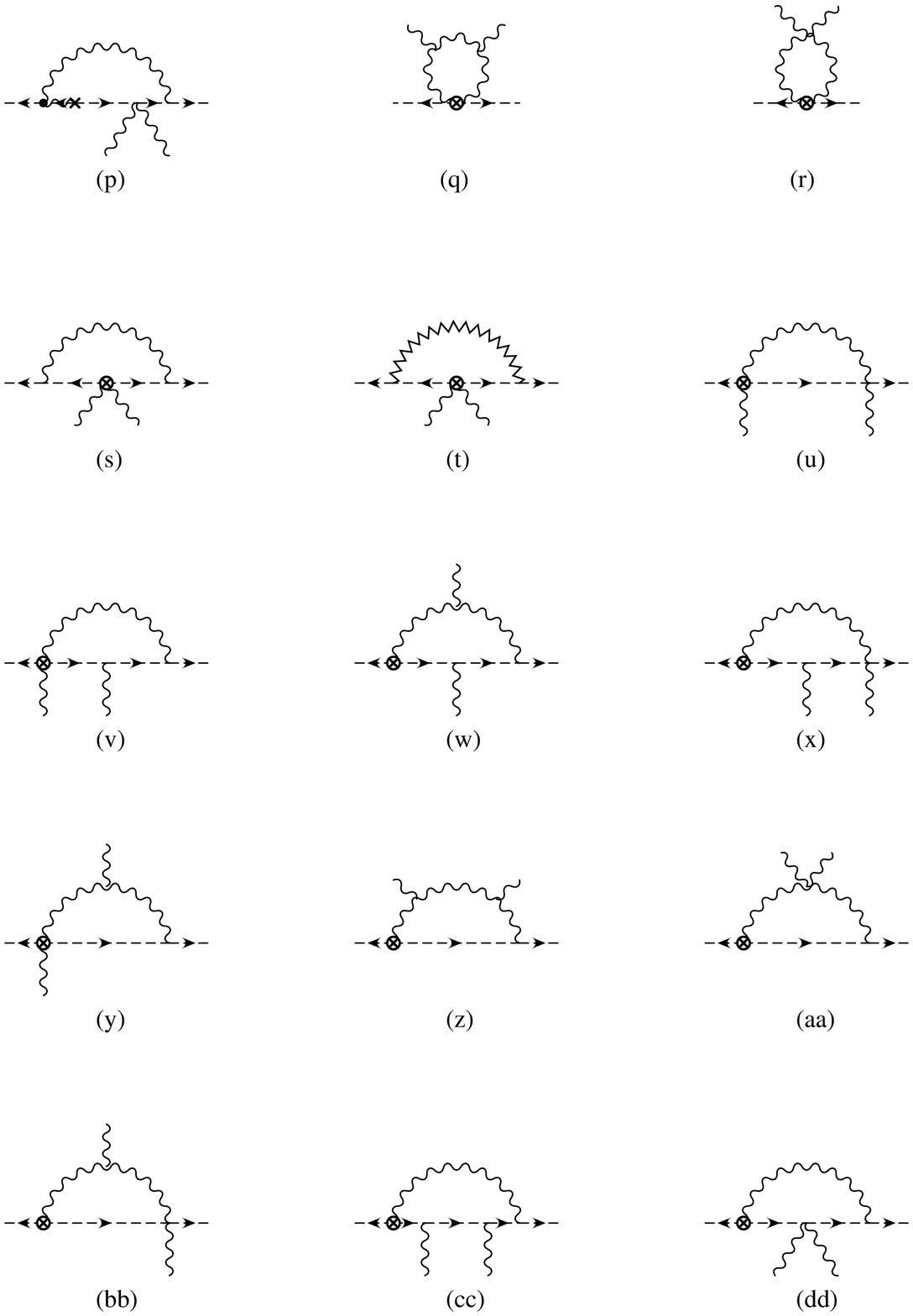}}
\inparg 
{\it \noindent Fig. 8(continued)}
\medskip
\outparg

\vfill
\eject  
\epsfysize= 7in
\centerline{\epsfbox{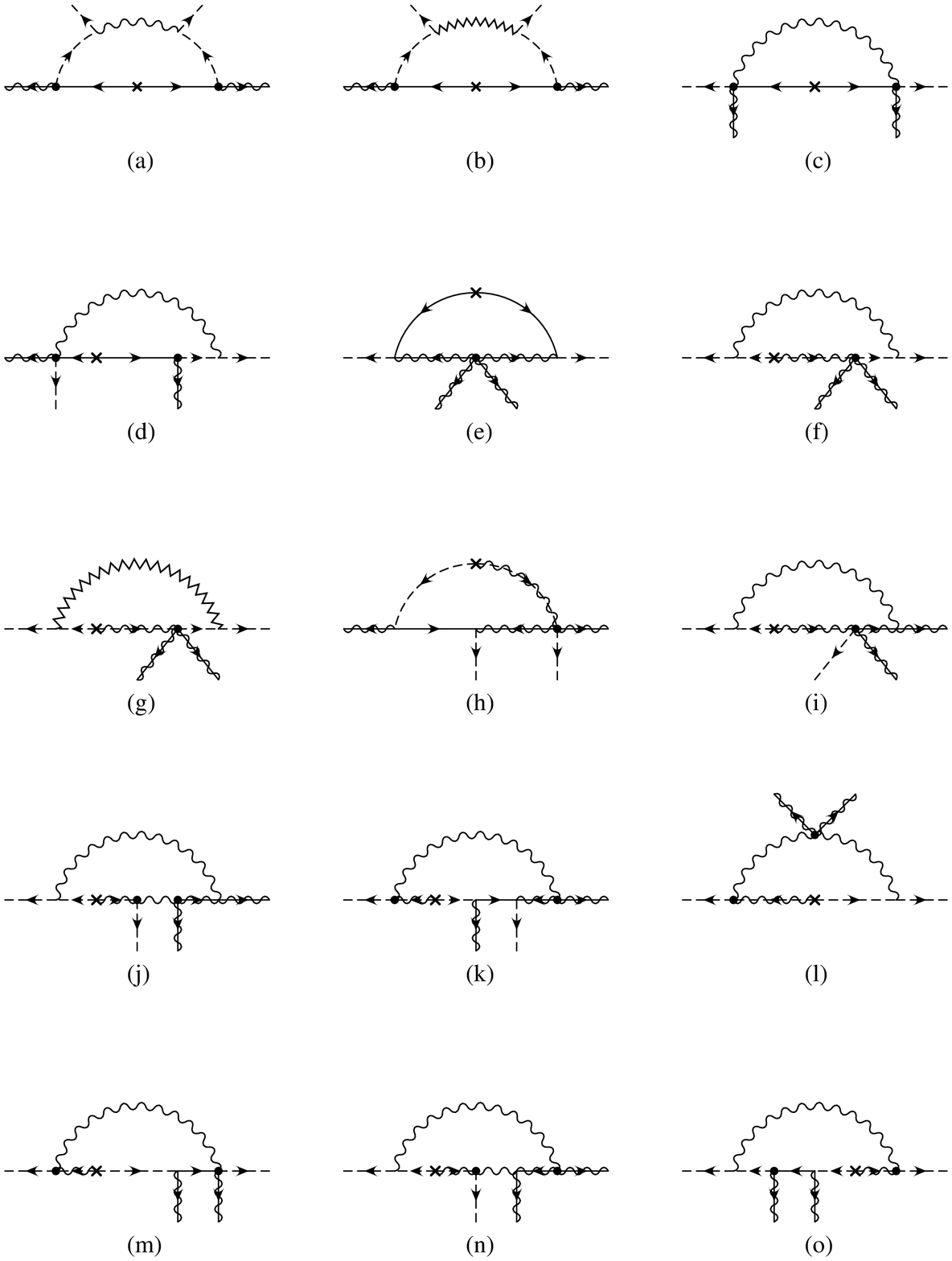}}
\inparg 
{\it \noindent Fig. 9: Diagrams with two scalar, two gaugino lines; 
a dot denotes a $C$, a cross a mass  
and a crossed circle a vertex with both a mass and a $C$.}
\medskip
\outparg
\vfill 
\eject
\epsfysize= 7in
\centerline{\epsfbox{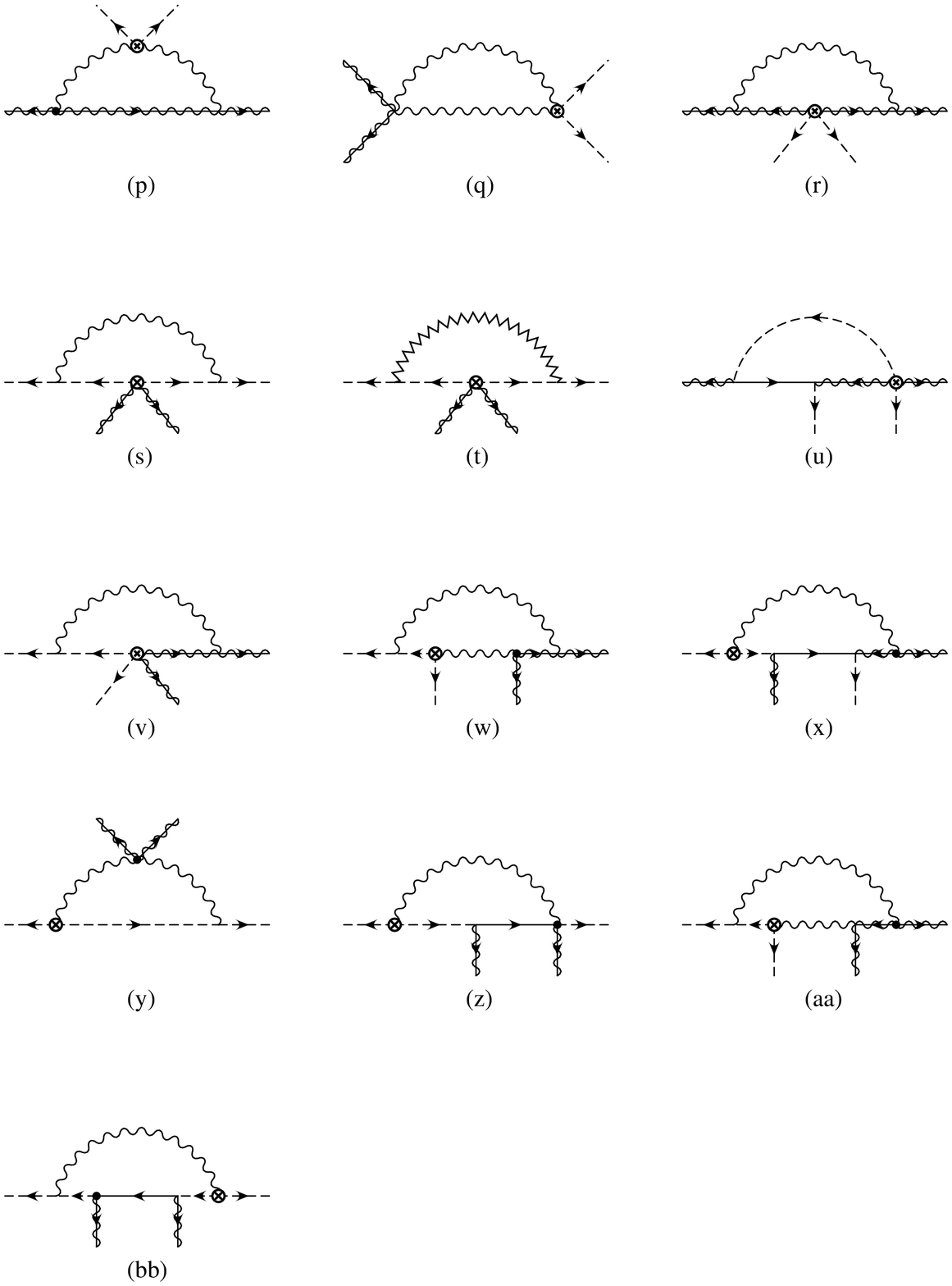}}
\inparg
{\it \noindent Fig. 9 (continued)}
\medskip
\outparg
\vfill
\eject
\epsfysize= 2.5in
\centerline{\epsfbox{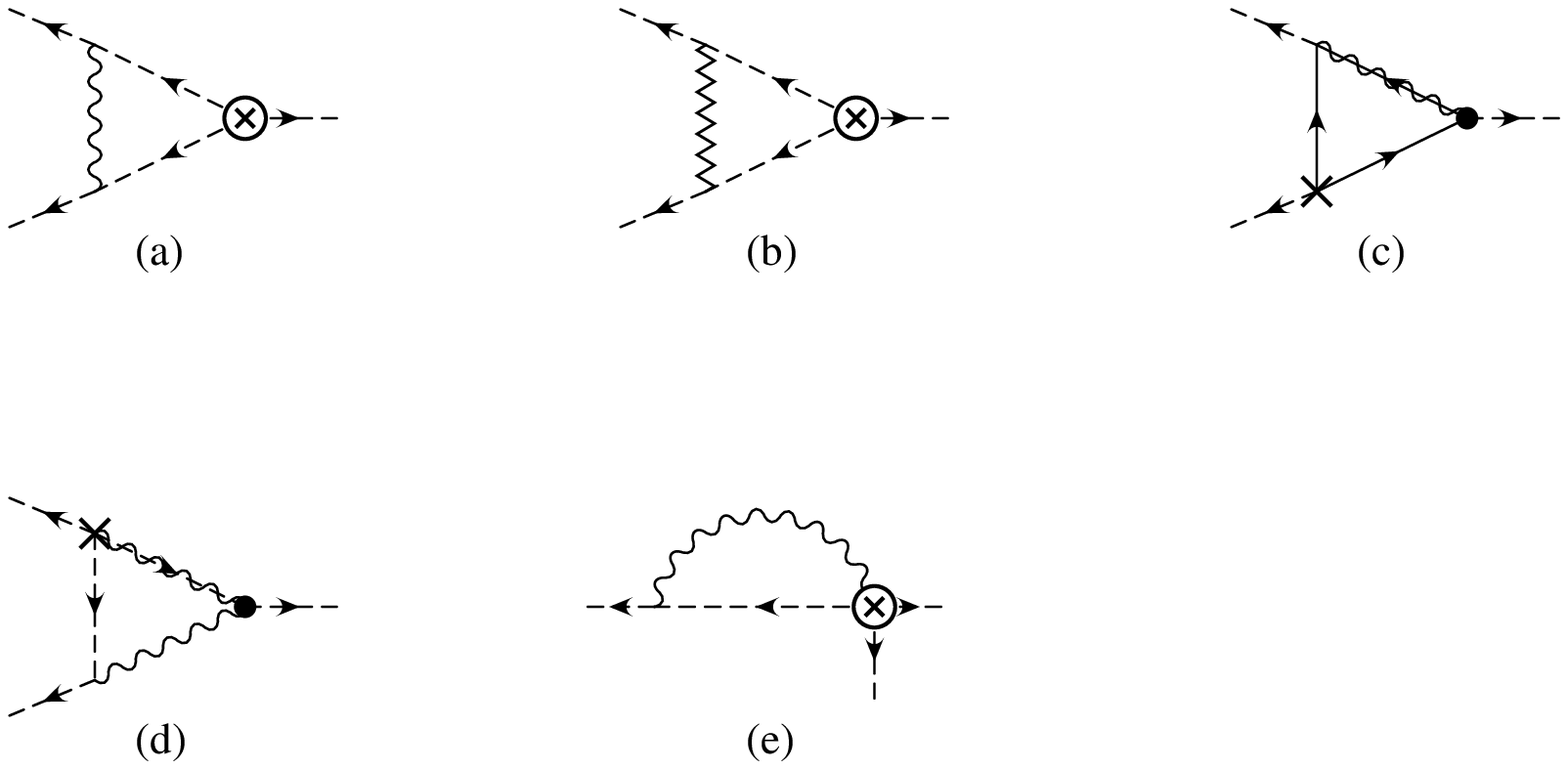}}
\inparg 
{\it \noindent Fig. 10: Diagrams with three scalar lines; a dot represents 
the position of a $C$, a cross a Yukawa vertex
and a crossed circle a Yukawa vertex with a $C$.}
\medskip
\outparg

\bigskip
\epsfysize= 2.5in
\centerline{\epsfbox{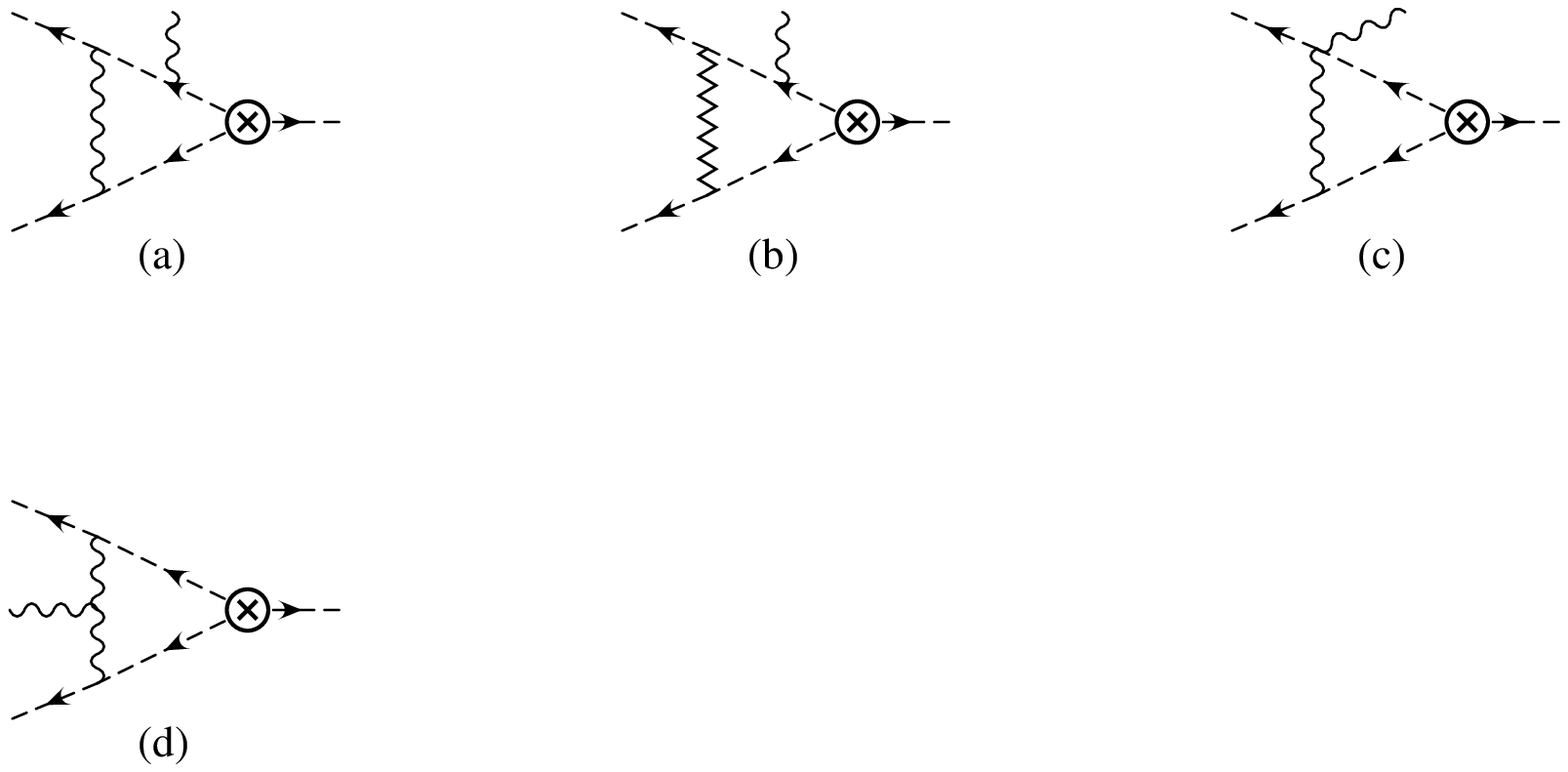}}
\inparg
{\it \noindent Fig. 11: Diagrams with three scalar lines and one gauge
line; a crossed circle represents a Yukawa vertex with a $C$.
Figs.~11(e-v) are not depicted explicitly.
Figs.~11(e-o) are obtained from Figs.~7(a-k) by adding an
external scalar ($\phibar$) line at the position of the
cross. Figs.~11(p-v) are obtained from Figs.~7(l-r) by adding an
external scalar ($\phibar$) line at the position of the crossed circle.}
\medskip
\outparg

\bigskip
\epsfysize= 1.2in
\centerline{\epsfbox{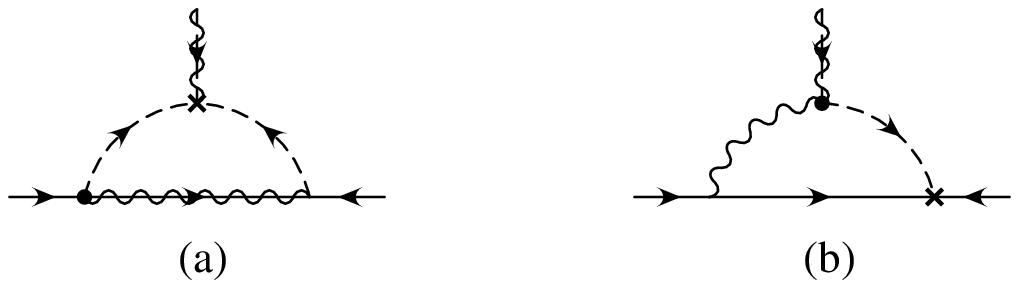}}
\inparg
{\it \noindent Fig. 12: Diagrams with two chiral fermion lines and one
auxiliary line; the dot represents the position of a $C$ and the cross 
a Yukawa vertex.} 
\medskip
\outparg

\bigskip
\epsfysize= .8in
\centerline{\epsfbox{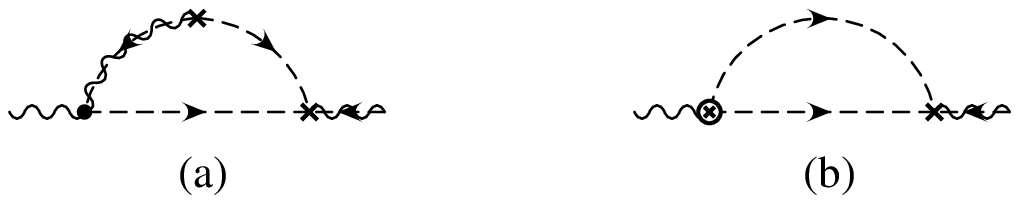}}
\inparg
{\it \noindent Fig. 13: Diagrams with one gauge line and one
auxiliary line; a cross represents a mass insertion or Yukawa vertex and
a crossed circle a vertex with both a mass and a $C$.}
\medskip
\outparg

\bigskip
\epsfysize= .5in
\centerline{\epsfbox{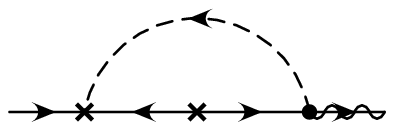}}
\inparg
{\it \noindent Fig. 14: A diagram with one chiral fermion lines and one
gaugino line; a cross represents a mass insertion or Yukawa vertex.}
\medskip
\outparg

\vfill
\eject
\bigskip
\epsfysize= 2.5in
\centerline{\epsfbox{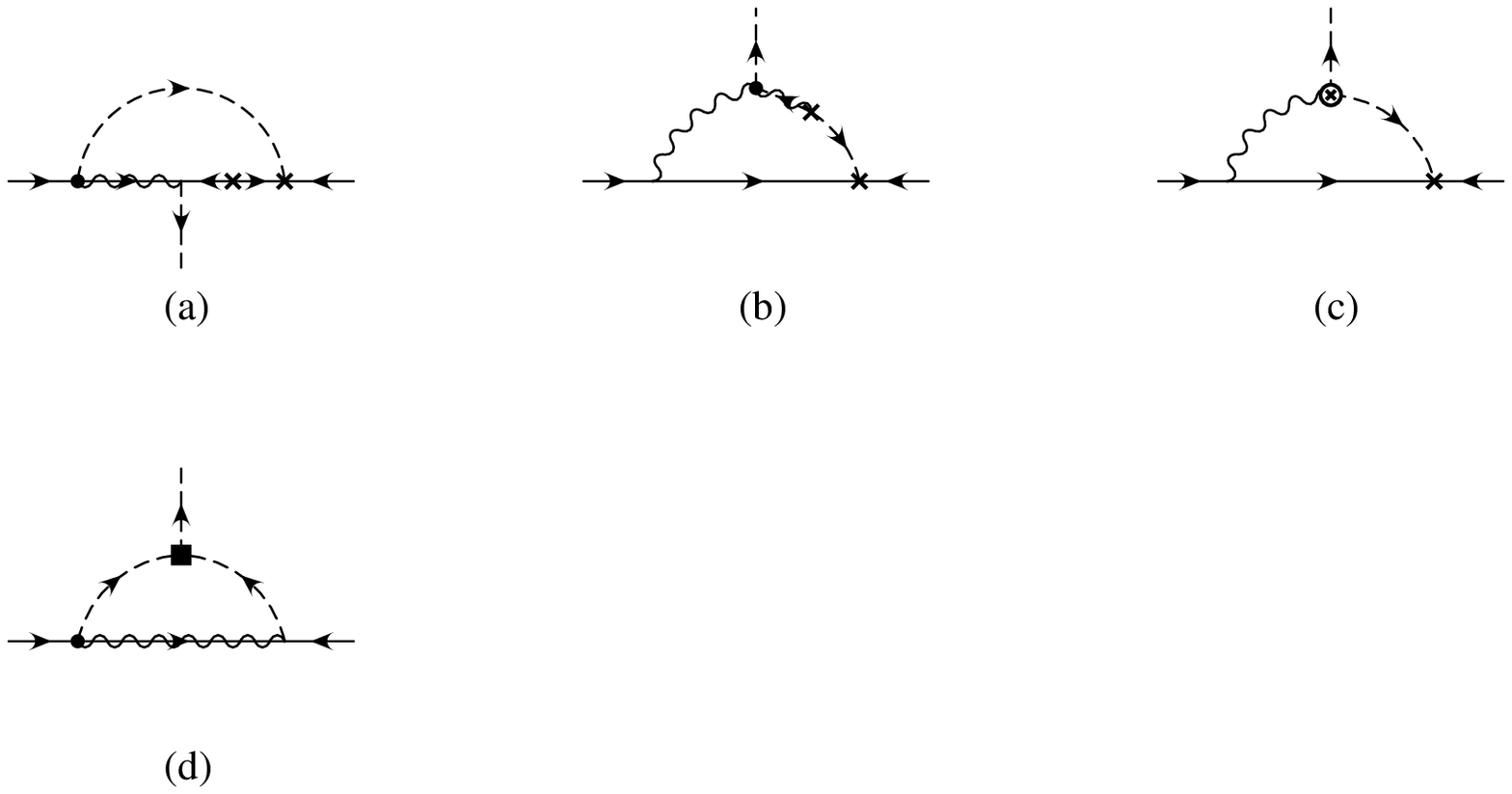}}
\inparg
{\it \noindent Fig. 15: Diagrams with two chiral fermion lines and one 
scalar; a dot represents the position of a $C$, a cross represents a 
mass insertion or Yukawa vertex, a crossed circle a vertex with 
both a mass and a $C$, and a box a mass-Yukawa vertex.}
\medskip
\outparg

\bigskip
\epsfysize= 2.5in
\centerline{\epsfbox{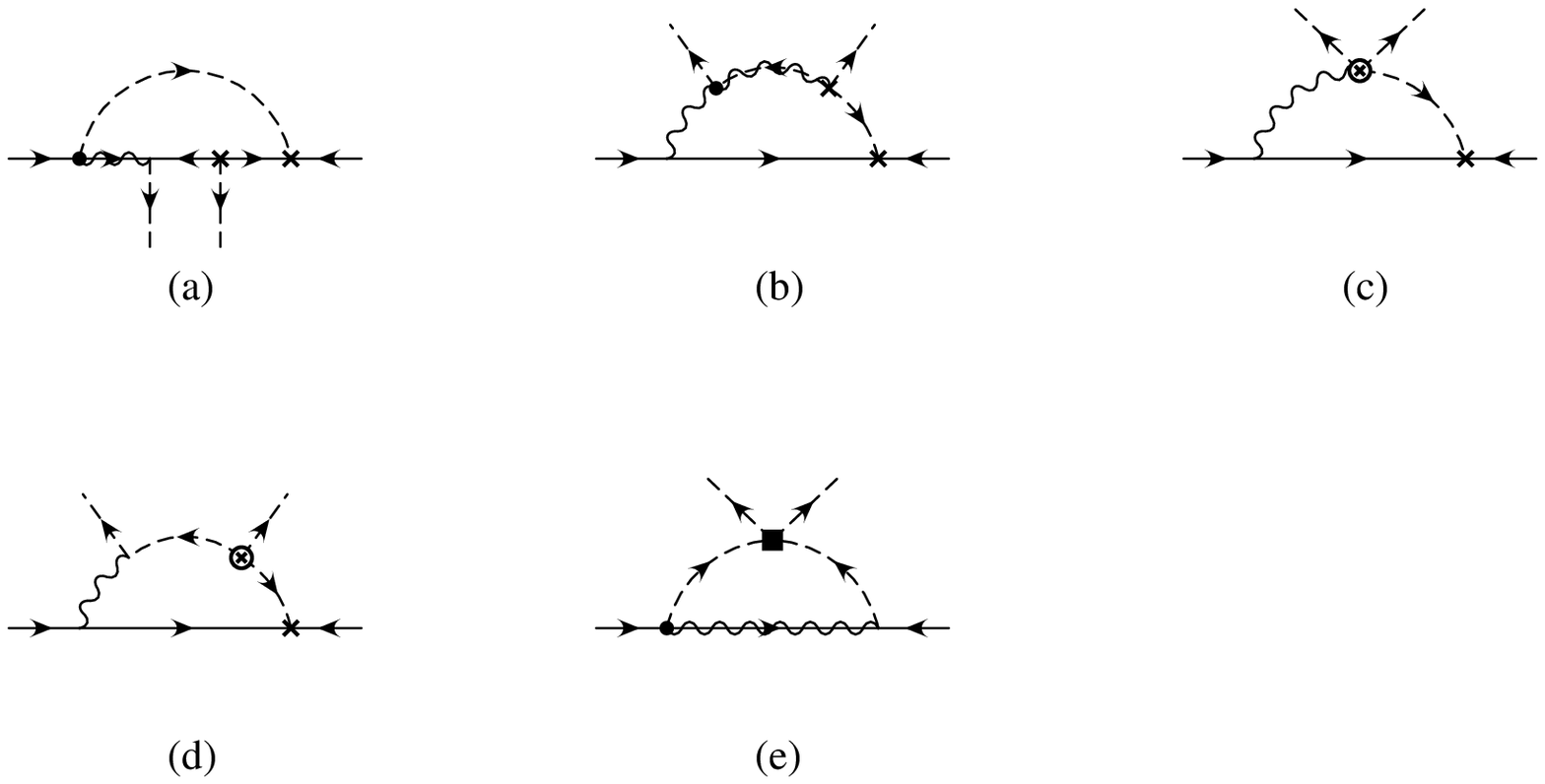}}
\inparg
{\it \noindent Fig. 16: Diagrams with two chiral fermion lines and two
scalars; a dot represents
the position of a $C$, a cross a Yukawa vertex
and a crossed circle a Yukawa vertex with a $C$.}
\medskip
\outparg

\bigskip
\epsfysize= .7in
\centerline{\epsfbox{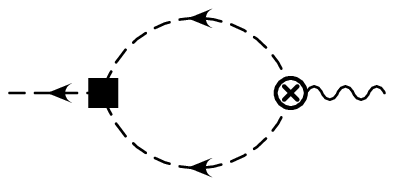}}
\inparg
{\it \noindent Fig. 17: A diagram with one scalar and one gauge line; 
a crossed circle represents a vertex with
both a mass and a $C$, and a box a mass-Yukawa vertex.}
\medskip
\outparg

\listrefs
\bye